%% file: main.tex
\documentclass{article}
\usepackage[english]{babel}

\usepackage[round,authoryear,semicolon,compress]{natbib}
\usepackage[letterpaper,top=2.7cm,bottom=2cm,left=3cm,right=3cm,marginparwidth=1.75cm]{geometry}
\usepackage{amsmath}
\usepackage{amssymb}
\usepackage{graphicx}
\usepackage[colorlinks=true, allcolors=blue]{hyperref}
\usepackage{aas_macros}
\usepackage{multirow}
\usepackage{url}
\usepackage{ulem}
\usepackage{xcolor}
\usepackage{rotating}
\usepackage{enumerate}

\usepackage{titlesec}

\titleformat{\paragraph}
{\normalfont\normalsize\bfseries}{\theparagraph}{1em}{}
\titlespacing*{\paragraph}
{0pt}{3.25ex plus 1ex minus .2ex}{1.5ex plus .2ex}

\titleformat{\subparagraph}
    {\normalfont\normalsize\bfseries}{\thesubparagraph}{1em}{}
\titlespacing*{\subparagraph}{\parindent}{3.25ex plus 1ex minus .2ex}{.75ex plus .1ex}

\definecolor{accentcolor}{RGB}{26, 137, 201}
\definecolor{lightaccent}{RGB}{89, 181, 235}
\definecolor{lighteraccent}{RGB}{181, 227, 255}

\usepackage{titlesec} 

\usepackage{graphicx} 
\graphicspath{{Pictures/}} 

\usepackage{lipsum} 

\usepackage{tikz} 

\usepackage[english]{babel} 

\usepackage{enumitem} 
\setlist{nolistsep} 

\usepackage{booktabs} 

\usepackage{eso-pic} 

\usepackage{xcolor} 
\usepackage{colortbl}

\usepackage{booktabs}

\usepackage{tabularx}
\usepackage{array}

\usepackage{floatrow}
\usepackage{sidecap}
\sidecaptionvpos{figure}{t}

\usepackage[version=4]{mhchem}
\usepackage{enumerate}
\usepackage{enumitem}

\usepackage{lato}

\usepackage{parskip}
\usepackage{rotating}
\usepackage[euler]{textgreek}

\usepackage{microtype} 
\usepackage[utf8]{inputenc} 
\usepackage[T1]{fontenc} 

\usepackage{afterpage}

\usepackage{multicol}

\usepackage[labelfont={color=accentcolor,bf}]{caption}

\usepackage[all]{nowidow}
\usepackage{latexsym}

\usepackage{amsmath}

 \usepackage{lipsum}
 \usepackage{microtype}
 \usepackage{vwcol}  

\usepackage{hyperref}
\usepackage{cleveref}
\crefname{section}{§}{§§}
\Crefname{section}{§}{§§}

\usepackage{amssymb}

\usepackage{textcomp}

\usepackage{xpatch}
\xpatchcmd{\citeauthor}{\begingroup}{\begingroup\em}{}{}

\makeatletter
\newcommand{\unchapter}[1]{%
  \begingroup
  \let\@makechapterhead\@gobble 
  \chapter{#1}
  \endgroup
}
\makeatother

\newcommand{\tabitem}
\newcommand\textalign[2][]{%
\ifx#1l\relax
  \makebox[0pt][l]{#2}%
\else
  \ifx#1r\relax
    \makebox[0pt][r]{#2}%
  \else  
    \ifx#1c\relax
      \makebox[0pt][c]{#2}%
    \fi\fi\fi
}
	
\usepackage{multirow}
\usepackage{multicol}
\usepackage{xfrac}

\usepackage{xurl}

\usepackage{breqn}

\usepackage[T1]{fontenc}
\usepackage{tipa}

\usepackage{flafter}

\usepackage{fancyhdr} 

\fancypagestyle{plain}{\fancyhead{}} 

\makeatletter
\renewcommand{\cleardoublepage}{
\clearpage\ifodd\c@page\else
\hbox{}
\vspace*{\fill}
\thispagestyle{empty}
\newpage
\fi}

\usepackage{amsmath,amsfonts,amssymb,amsthm} 

\newtheoremstyle{accentcolornumbox}
{0pt}
{0pt}
{\normalfont}
{}
{\small\bf\sffamily\color{accentcolor}}
{\;}
{0.25em}
{\small\sffamily\color{accentcolor}\thmname{#1}\nobreakspace\thmnumber{\@ifnotempty{#1}{}\@upn{#2}}
\thmnote{\nobreakspace\the\thm@notefont\sffamily\bfseries\color{black}---\nobreakspace#3.}} 

\newtheoremstyle{blacknumex}
{5pt}
{5pt}
{\normalfont}
{} 
{\small\bf\sffamily}
{\;}
{0.25em}
{\small\sffamily{\tiny\ensuremath{\blacksquare}}\nobreakspace\thmname{#1}\nobreakspace\thmnumber{\@ifnotempty{#1}{}\@upn{#2}}
\thmnote{\nobreakspace\the\thm@notefont\sffamily\bfseries---\nobreakspace#3.}}

\newtheoremstyle{blacknumbox} 
{0pt}
{0pt}
{\normalfont}
{}
{\small\bf\sffamily}
{\;}
{0.25em}
{\small\sffamily\thmname{#1}\nobreakspace\thmnumber{\@ifnotempty{#1}{}\@upn{#2}}
\thmnote{\nobreakspace\the\thm@notefont\sffamily\bfseries---\nobreakspace#3.}}

\newtheoremstyle{accentcolornum}
{5pt}
{5pt}
{\normalfont}
{}
{\small\bf\sffamily\color{accentcolor}}
{\;}
{0.25em}
{\small\sffamily\color{accentcolor}\thmname{#1}\nobreakspace\thmnumber{\@ifnotempty{#1}{}\@upn{#2}}
\thmnote{\nobreakspace\the\thm@notefont\sffamily\bfseries\color{black}---\nobreakspace#3.}} 
\makeatother

\RequirePackage[framemethod=default]{mdframed} 

\newmdenv[skipabove=7pt,
skipbelow=7pt,
backgroundcolor=black!5,
linecolor=accentcolor,
innerleftmargin=5pt,
innerrightmargin=5pt,
innertopmargin=5pt,
leftmargin=0cm,
rightmargin=0cm,
innerbottommargin=5pt]{tBox}

\newmdenv[skipabove=7pt,
skipbelow=7pt,
rightline=false,
leftline=true,
topline=false,
bottomline=false,
backgroundcolor=accentcolor!10,
linecolor=accentcolor,
innerleftmargin=5pt,
innerrightmargin=5pt,
innertopmargin=5pt,
innerbottommargin=5pt,
leftmargin=0cm,
rightmargin=0cm,
linewidth=4pt]{eBox}	

\newmdenv[skipabove=28pt,
skipbelow=7pt,
rightline=false,
leftline=false,
topline=false,
bottomline=false,
linecolor=accentcolor,
innerleftmargin=30pt,
innerrightmargin=15pt,
innertopmargin=0pt,
leftmargin=0cm,
rightmargin=0cm,
linewidth=4pt,
innerbottommargin=0pt]{dBox}	

\newmdenv[skipabove=7pt,
skipbelow=7pt,
rightline=false,
leftline=true,
topline=false,
bottomline=false,
linecolor=gray,
backgroundcolor=black!5,
innerleftmargin=5pt,
innerrightmargin=5pt,
innertopmargin=5pt,
leftmargin=0cm,
rightmargin=0cm,
linewidth=4pt,
innerbottommargin=5pt]{cBox}

\makeatletter
\renewcommand{\@seccntformat}[1]{\llap{\textcolor{accentcolor}{\csname the#1\endcsname}\hspace{1em}}}                    
\renewcommand{\section}{\@startsection{section}{1}{\z@}
{-4ex \@plus -1ex \@minus -.4ex}
{1ex \@plus.2ex }
{\normalfont\large\sffamily\bfseries}}
\renewcommand{\subsection}{\@startsection {subsection}{2}{\z@}
{-3ex \@plus -0.1ex \@minus -.4ex}
{0.5ex \@plus.2ex }
{\normalfont\sffamily\bfseries}}
\renewcommand{\subsubsection}{\@startsection {subsubsection}{3}{\z@}
{-2ex \@plus -0.1ex \@minus -.2ex}
{.2ex \@plus.2ex }
{\normalfont\small\sffamily\bfseries}}                        
\renewcommand\paragraph{\@startsection{paragraph}{4}{\z@}
{-2ex \@plus-.2ex \@minus .2ex}
{.1ex}
{\normalfont\small\sffamily\bfseries}}

\usepackage{hyperref}
\hypersetup{hidelinks,pagebackref=true,hyperindex=true,colorlinks=false,breaklinks=true,urlcolor= accentcolor,bookmarks=true,bookmarksopen=false,pdftitle={Title},pdfauthor={Author}}

\newcommand{\thechapterimage}{}

\def\@makechapterhead#1{
\thispagestyle{empty}
{\centering \normalfont\sffamily
\ifnum \c@secnumdepth >\m@ne
\if@mainmatter
\startcontents
\begin{tikzpicture}[remember picture,overlay]
\node at (current page.north west)
{\begin{tikzpicture}[remember picture,overlay]
\node[anchor=north west,inner sep=0pt] at (0,0) {\includegraphics[width=\paperwidth]{\thechapterimage}};
\end{tikzpicture}};
\end{tikzpicture}}
\par\vspace*{230\p@}
\fi
\fi}

\makeatother

\title{Uranus Study Report}

\begin{document}
\maketitle

\newpage
.
\vfill

\noindent Study Report prepared for the W. M. Keck Institute for Space Studies (KISS)\\
\par

\noindent Team Leads: Mark Hofstadter, Ravit Helled, and David Stevenson\\

\noindent Parts of this research were carried out at the Jet Propulsion Laboratory, California Institute of Technology, under a contract with the National Aeronautics and Space Administration (80NM0018D0004).\\

\noindent Pre-Decisional Information — For Planning and Discussion Purposes Only\\

\noindent DOI:  \\

\vspace{10mm}

\noindent Director: Prof. Bethany Ehlmann\\
\noindent Executive Director: Harriet Brettle\\
\noindent Editing and Formatting: Jessica Law\\
\noindent Cover Image: Keck Institute for Space Studies/Chuck Carter\\
\noindent © 2024. All rights reserved.

\newpage
\begin{figure}[h!]
    \centering
    \includegraphics[width=0.8\linewidth]{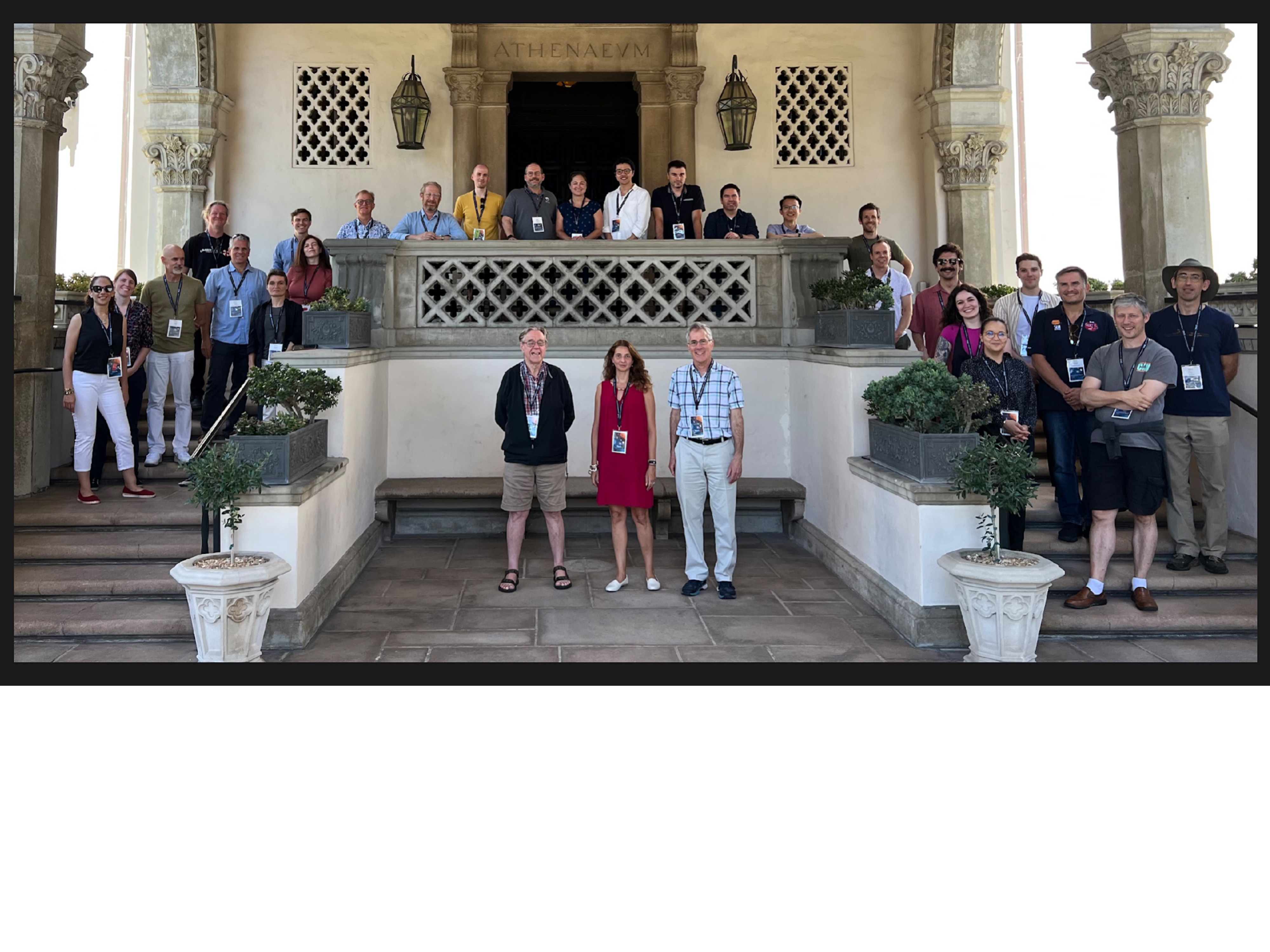}
\end{figure}

\vspace*{-2cm}

\centerline{KISS Uranus Interior Workshop Participants}

\noindent Mandy Bethkenhagen -- Institute of Science and Technology Austria \\
Hao Cao -- University of California, Los Angeles\\
Junjie Dong -- California Institute of Technology \\
Maryame El Moutamid -- Southwest Research Institute\\
Anton Ermakov -- University of California, Berkeley (now at Stanford University)\\
Jim Fuller -- California Institute of Technology \\
Tristan Guillot -- Observatoire de la Côte d'Azur\\
Ravit Helled -- University of Zürich \\
Mark Hofstadter -- Jet Propulsion Laboratory, California Institute of Technology\\
Benjamin Idini -- University of California, Santa Cruz \\
Andre Izidoro -- Rice University \\
Yohai Kaspi -- Weizmann Institute \\
Tanja Kovacevic -- University of California, Berkeley \\
Valéry Lainey -- Paris Observatory\\
Steve Levin -- Jet Propulsion Laboratory, California Institute of Technology \\
Jonathan Lunine -- Cornell University\\
Christopher Mankovich -- Jet Propulsion Laboratory, California Institute of Technology \\
Stephen Markham -- Observatoire de la Côte d'Azur\\
Marius Millot -- Lawrence Livermore National Laboratory\\
Olivier Mousis -- Aix-Marseille University  \\
Simon Müller -- University of Zürich\\
Nadine Nettelmann -- University of Zürich\\
Francis Nimmo -- University of California, Santa Cruz\\\
Marzia Parisi -- Jet Propulsion Laboratory, California Institute of Technology\\
Krista Soderland -- University of Texas at Austin\\
Dave Stevenson -- California Institute of Technology \\
Lars Stixrude -- University of California, Los Angeles\\
Nick Teanby -- University of Bristol\\
Allona Vazan -- The Open University of Israel \\
\newpage


\tableofcontents

\newpage

\section{Executive Summary}
Determining the internal structure of Uranus is a key objective for planetary science. Knowledge of Uranus's bulk composition and the distribution of elements is crucial to understanding its origin and evolutionary path. In addition, Uranus represents a poorly understood class of ``intermediate-mass planets" (intermediate in size between the relatively well studied terrestrial and gas giant planets), which appear to be very common in the Galaxy. As a result, a better characterization of Uranus will also help us to better understand exoplanets in this mass and size regime.\\

Recognizing the importance of Uranus, a Keck Institute for Space Studies (KISS) workshop was held in September 2023 to investigate how we can improve our knowledge of Uranus's internal structure in the context of a future Uranus mission that includes an orbiter and a probe. The scientific goals and objectives of the recently released Planetary Science and Astrobiology Decadal Survey were taken as our starting point. We reviewed our current knowledge of Uranus's interior and identified measurement and other mission requirements for a future Uranus spacecraft, providing more detail than was possible in the Decadal Survey's mission study and including new insights into the measurements to be made.  We also identified important knowledge gaps to be closed with Earth-based efforts in the near term that will help guide the design of the mission and interpret the data returned.\\

Our report focuses on the following topics: 
\begin{itemize}
    
\item Uranus's atmosphere (noble gas, elemental, and molecular abundances and the pressure-temperature profile).
\item Its gravity field and interior models.
\item The planet's magnetic field, heat flux, and energy balance.
\item The potential for Uranian seismology.
\end{itemize}

In the words of the Decadal Survey, the Uranus Flagship mission can produce ``transformative, breakthrough science across a broad range of topics." We argue that to fulfill its promise, great progress should be made in terms of mission design, theoretical calculations, numerical modeling, experiments, ground-based observations, and bringing communities together to plan the mission and interpret the data it returns.


\input{sections/introduction}



\newpage
\section{Uranus's Atmospheric Composition and Temperature}
\label{sec.atmos.comp.temp}

\subsection{Science Motivation}

\begin{figure*}[h!]
    \centering
    \includegraphics[width = 0.65\textwidth]{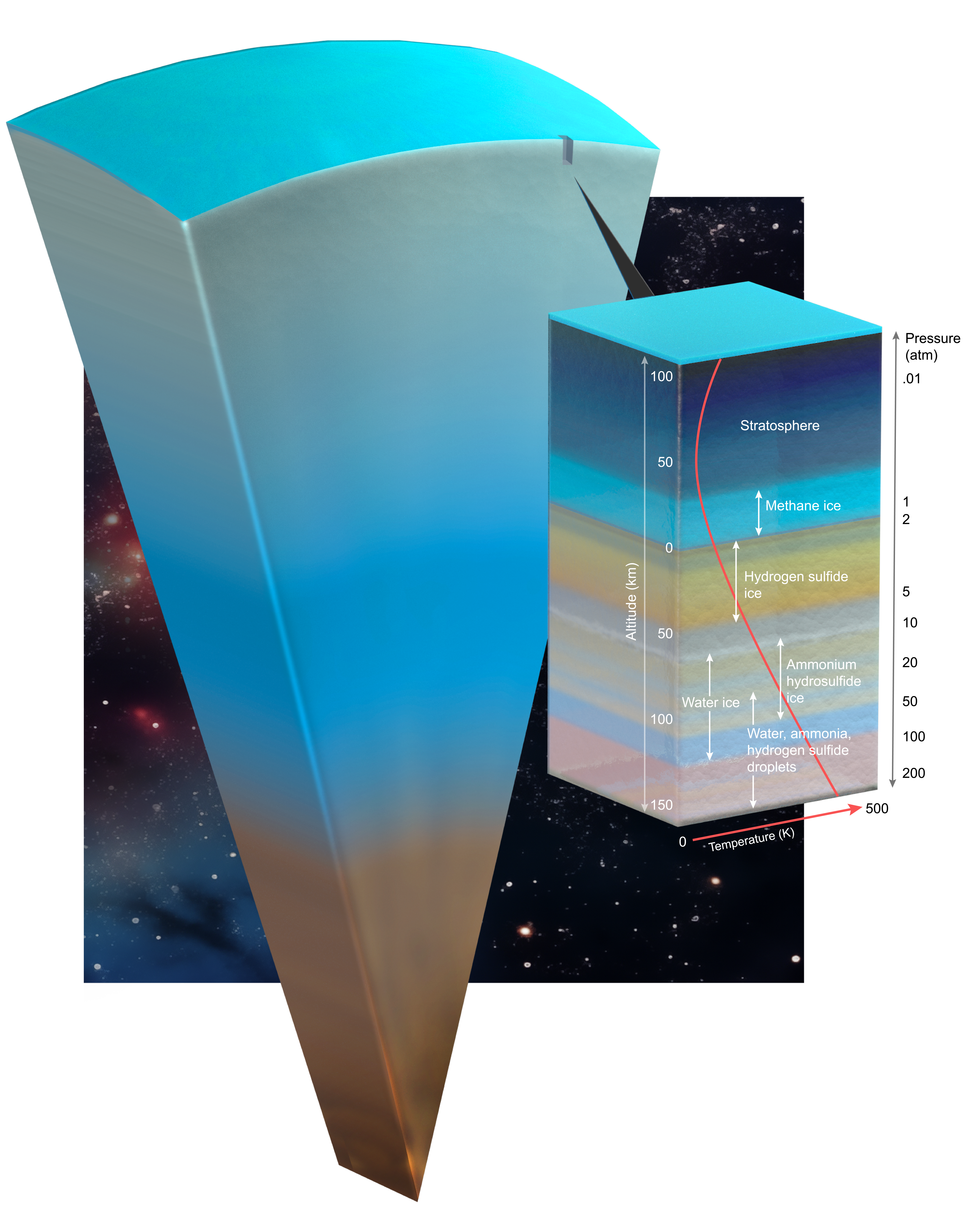}%
    \caption{A sketch of Uranus's internal structure and a zoom-in on the vertical structure of its atmosphere. In a planet like Uranus, there is no obvious boundary between the atmosphere and the deeper interior. The internal structure of Uranus remains unknown and although the atmosphere corresponds to a tiny fraction of the planet it is important for Uranus's characterization as we discuss in this report.  The zoom-in on Uranus's atmosphere presents the atmospheric temperature and pressure profiles as well as the expected cloud layers. }
    \label{fig:UranusInteriorAtmosphere.png}
\end{figure*}

While Uranus's atmosphere is small in mass (and size) in comparison to the entire planet, it offers an important window for studying Uranus's origin and internal structure. The atmosphere sets a critical boundary condition for interior models and also reflects physical and chemical processes occurring in the deeper interior. The expected atmospheric structure is presented in Figure \ref{fig:UranusInteriorAtmosphere.png}. 

Measuring atmospheric elemental abundances in Uranus provides clues as to what the planet is made of and how it formed, and comparing its abundances to the other planets advances our understanding of the formation and evolution of the solar system as a whole.  Studying how elements are transported from the deeper interior to the atmosphere by detecting disequilibrium species, the location and composition of clouds, and measuring the atmospheric temperature profile will provide us with direct constraints on interior-atmosphere interactions.  Note that very little is known about the temperature profile in deep hydrogen atmospheres in part because of our poor understanding of moist convection and the inhibition of convection in the presence of molecular weight gradients.  The methane cloud on Uranus, near 1 bar, is the most accessible region of all the giant planets where moist convection is thought to dominate, so direct measurement of the temperature profile in that region via an entry probe and indirect determination in the deeper atmosphere will provide us with the means to understand how all planets with hydrogen atmospheres cool and contract. 

In the remainder of Section \ref{sec.atmos.comp.temp} we discuss some of the important aspects of measuring noble gases, isotopic ratios, the planet's temperature profile, and the questions and opportunities regarding the determination of the abundances of condensable and disequilibrium species.  
\input{sections/subsections/atmospheric_pt_profile}

\input{sections/subsections/condensable_and_disequilibrium_species}
\input{sections/subsections/noble_gasses_and_isotopes}
\label{sec:noble_gasses_and_isotopes}

\newpage
\input{sections/subsections/gravity_and_winds}

\newpage
\input{sections/subsections/magnetic_field}

\newpage
\input{sections/subsections/atmospheric_energy_balance}

\newpage
\input{sections/subsections/seismology}
\newpage
\input{sections/recommendations}

\newpage
\input{sections/summary}
\newpage
\bibliography{sources}

\input{sections/appendix}

\end{document}

%% file: sections/introduction.tex
\newpage
\rhead{\begin{picture}(0,0) \put(0,0){\includegraphics[width=1cm]{example-image-a}} \end{picture}}

\section{Introduction}\label{sec:introduction}

Uranus is a member of the least understood class of planets in the Solar System, the Ice Giants. There are many fundamental unknowns about Uranus itself including its origin, evolutionary path, internal structure and bulk composition, rotation, atmosphere, and interior dynamics. Similar unknowns exist regarding its moons, rings, and magnetosphere. In addition to being the most mysterious planet in our solar system, most of the exoplanets discovered around other stars appear to be similar in size to Uranus and Neptune, indicating that this class of planet is common in our galaxy and much more common than the better-understood Gas Giant planets such as Jupiter and Saturn.  For these reasons, among others, the recently completed Planetary Science and Astrobiology Decadal Survey (National Academies 2022) selected a Uranus Orbiter and Probe (UOP) as the top priority new Flagship mission for the coming decade ``primarily for its ability to produce transformative, breakthrough science across a broad range of topics" (ibid p. 582).\\

The Decadal Survey, titled Origins, Worlds, and Life (OWL), highlights the importance of determining the interior structure and composition of Uranus. OWL identifies that investigation as one of the central objectives of the UOP mission. It is not our intent to re-argue the science goals and objectives presented in OWL. Instead we take them as a starting point for our study and focus in more detail than was possible for OWL on identifying the measurements to make and the knowledge gaps to be filled in order to best address those objectives. The key relevant questions presented in the OWL Decadal Survey and its UOP mission study are:

\begin{itemize}
    \item When, where, and how did Uranus form and evolve?
    \item  What is the bulk composition of Uranus and its depth dependence?
    \item Does Uranus have discrete layers or a fuzzy core?
    \item What is the true rotation rate of Uranus, does it rotate uniformly, and how deep are the winds?
    \item What mechanisms are transporting heat and energy in the planet today?
\end{itemize} 

\noindent Collectively, these questions get at the higher level goal of understanding how our solar system and exoplanetary systems form and evolve.  

This report reviews the key open questions linked to Uranus's internal structure, identifies a desirable instrument suite and spacecraft architecture for addressing those questions, and recommends Earth-based activities to support the Uranus Flagship mission. To do this, our Keck Institute for Space Studies (KISS) team includes experts from various disciplines such as modeling the interior of Uranus, high-pressure physics and equations of state, measurement techniques, and mission design. A list of study participants and other information is available on the website \href{https://kiss.caltech.edu/workshops/uranus/uranus.html}{https://kiss.caltech.edu/workshops/uranus/uranus.html}.\\

Sections \ref{sec.atmos.comp.temp} through \ref{sec:seismology} of this report are organized around the measurements we have identified as necessary to understand Uranus's current interior structure. In each case we discuss some of the science driving the measurements, how the measurements will be interpreted, and -- perhaps most importantly -- knowledge gaps that currently exist that limit our ability to interpret the measurements. Filling those knowledge gaps is an important task to be performed before the arrival of UOP in the Uranus system. We note that all of the interior-related measurements called for by OWL's UOP study appear in our list of measurements. In addition to those in OWL, we call for the detection of tides raised in Uranus by its moons (Section \ref{sec:gravity_and_winds}) and for detecting normal-mode oscillations within Uranus by their perturbations to a spacecraft's orbit (Section \ref{section.seismology.detection}) in addition to the ring seismology called for by OWL.  We also emphasize the importance of measuring the atmospheric pressure-temperature profile for studies of the interior (Section \ref{sec:atmospheric_pressure_temperature_profile}), whereas OWL calls for these measurements primarily in the context of atmospheric dynamics and dynamo generation. Finally, our discussions suggest that a microwave radiometer be elevated in importance from its status in the UOP study.\\
\par

Section \ref{sec:recommendations} of this document presents our recommendations for the UOP mission and for Earth-based work that can fill important knowledge gaps in advance of the mission.  Some of the Earth-based work is best performed now, when it can guide the design of the basic mission architecture.  Section \ref{sec:summary} presents a summary of the entire report. After an extensive list of references, five appendices provide background information on radio science, gravity science, giant-planet seismology, high-pressure material properties, and magnetic fields.
\par

%% file: sections/subsections/atmospheric_pt_profile.tex
\subsection{Atmospheric Pressure-Temperature Profile}\label{sec:atmospheric_pressure_temperature_profile}

Atmospheric pressure-temperature (P-T) profiles are essential to characterize a planetary atmosphere: They provide the thermodynamic conditions that govern the phases of the chemical species involved. Their gradient (vertically and also horizontally) is directly related to energy transport, and they thus crucially inform us on transport mechanism and heat loss to outer space. For giant planets, this is thus key to understanding planetary cooling and planet formation. \\
\par 

For planets other than the Earth, atmospheric P-T profiles are generally inferred \textit{indirectly} from radiative transfer models. The strong limitation of the technique is that, instead of constraining pressure directly, it relies on the optical depth, which is a function of both the quantity of column atmosphere and opacity. The opacity, in turn, depends on composition. Opacities being generally very uncertain, pressures cannot be estimated with high accuracy. Another strong limitation of spectroscopy, at least at optical and infrared wavelengths, is that it cannot penetrate the atmosphere deeper than a few bars. Observations at millimeter to centimeter wavelength can penetrate the deeper atmosphere, down to pressures of hundreds of bars in giant planets, but at the expense of limited spectral information and again, a highly model-dependent P-T profile determination. Limb darkening, measured with microwave observations at multiple emission angles, can help break this degeneracy. The microwave brightness temperature is an integral along the line of sight of the product of temperature and opacity.  Comparing measurements at different emission angles effectively measures the vertical gradient of temperature times opacity, while different microwave frequencies penetrate to different pressure levels, in a composition-dependent way. Combining limb darkening with absolute microwave brightness at multiple frequencies constrains both the composition and the vertical temperature profile.
\par 

\begin{figure}[bt]
    \centering
    \includegraphics[width = 10cm]{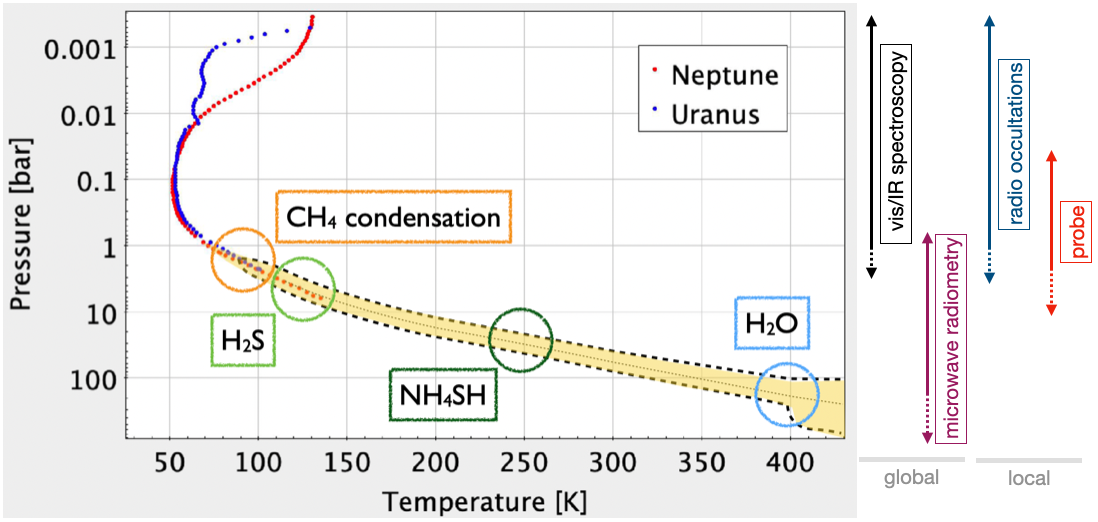}
    \caption{P-T profiles of Uranus and Neptune as inferred from radio-occultation measurements \citep[see][]{Lindal1992} and extrapolated adiabatically to deep levels. The possible inhibition of moist convection in hydrogen atmospheres \citep{Guillot1995} implies that the temperature profile might be either sub- or super-adiabatic in the region of condensation of abundant species as shown here. The pressure levels probed by different instruments are shown on the right.}
    \label{fig:deepPT}
\end{figure}

Another approach relies on occultations, in particular radio occultations from a spacecraft behind the planetary atmosphere. The atmospheric refractivity is then measured as a function of altitude, which allows us to retrieve a combination of the mean molecular weight and temperature. As shown in Figure~\ref{fig:deepPT}, the Voyager 2 spacecraft was able thus to probe Uranus's atmosphere down to 2.3\,bar and Neptune's down to 6.3\,bar \citep{Lindal1992}. As noted in that paper, the temperature profile retrieved highly depends on an assumed methane abundance profile and could in fact be either sub or super-adiabatic. In addition, the fact that condensates have a higher molar weight than hydrogen implies that moist convection might be inhibited locally, regardless of the assumed temperature gradient \citep{Guillot1995}. This implies that, as depicted in Figure~\ref{fig:deepPT}, the temperature profile is unknown in regions of condensation of abundant species such as methane and water, and is expected to become super-adiabatic \citep{Leconte+2017}. In fact, without such a determination, we cannot pretend to understand heat transport in planets with hydrogen atmospheres! As shown by \citet{Markham+2022}, this in turn has consequences for our ability (or lack thereof) to interpret measurements of exoplanets.

\begin{figure}
    \centering
    \includegraphics[width = \textwidth]{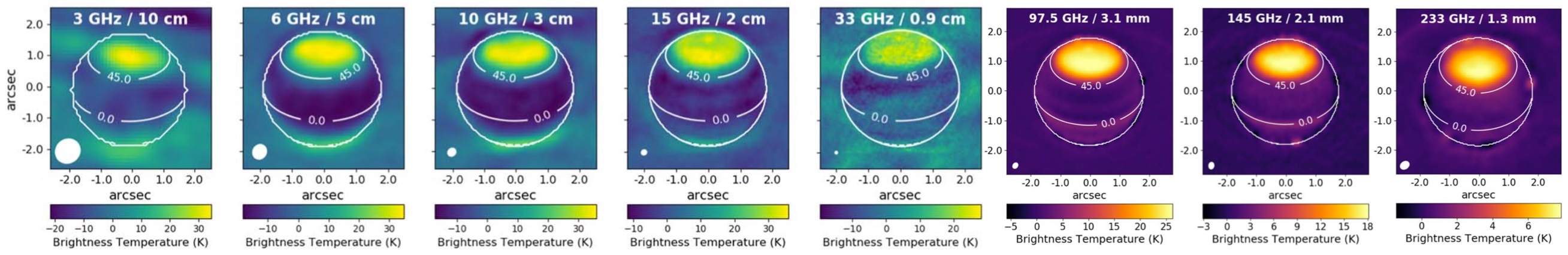}
    \caption{Maps of brightness temperature of Uranus as measured from the VLA (wavelengths 10--0.9 cm) on 2015-08-29 and ALMA (3.1--1.3 mm) from 2017-12-03 to 2018-09-13. Figure from \citet{Molter+2021}.}
    \label{fig:brightnessTmaps}
\end{figure}

It is therefore crucial to directly and unambiguously measure Uranus's P-T profile. However, this measurement faces another challenge, the known variability of the planet's atmosphere, which, as shown in Figure~\ref{fig:brightnessTmaps}, exhibits a strong latitudinal gradient in brightness at millimeter and centimeter wavelengths \citep{Molter+2021}. This gradient is interpreted to be mainly due to an equator-to-pole variation of the abundance of condensable species such as methane and hydrogen sulfide, but whether this also partly reveals (or hides) an equator-to-pole temperature variation is an open question. 

In order to lift the degeneracies and infer the pressure-temperature unambiguously, we envision the four types of measurements listed in Table~\ref{tab:atm_suite}. Optical and IR spectroscopy (including from the ground) have the capability to characterize the high levels (including the stratosphere) down to the medium troposphere at pressures of a few bars. Temperature-pressure maps can be derived, but they depend on models with assumed opacities, and the technique cannot be used below optically thick clouds. Microwave measurements are able to provide information even below the cloud decks down into the deep atmosphere (pressures of a hundred bars or more). Again, the measurements are ambiguous because they depend on the unknown abundances and profiles of opacity providers. As shown recently through the analysis of the Juno Microwave Radiometer (MWR) results at Jupiter \citep{Li+2024}, the ability to measure limb-darkening parameters is necessary to lift that degeneracy, at least partly. This is possible comprehensively only by using a microwave instrument on an orbiting spacecraft. Observations from the ground (e.g., with ngVLA or ALMA), however, provide extremely useful and complementary global and temporal coverage \citep{dePater+2021BAAS}. 
\par 

While spectroscopy and radiometry allow global mapping of the atmosphere and the potential to determine whether the temperature profile is adiabatic, sub-adiabatic, or super-adiabatic, the results remain model dependent. The unambiguous determination of a P-T profile at a specific location is therefore crucial to lift the degeneracy at that location and gain the ability to propagate the knowledge obtained to other regions of the planet. This can be done only through an in-situ sounding, preferably in a relatively quiet region that is representative of a wider region of the planet (i.e., avoiding the situation experienced with the Galileo probe, which fell into a Jupiter hot spot). The probe should go at least to a pressure scale height below the base of the methane cloud, that is, down to at least 5 bars, in order to fully characterize this crucial region \citep{Guillot2022ExA}. The MWR instrument should be calibrated properly in order to include wavelengths whose contribution functions peak at lower pressures (as well as others that peak at much higher pressures of course). In addition, using the Voyager experience, radio occultations at several latitudes (from equator to the poles) would provide another means to calibrate temperature structures. Reaching sufficiently deep levels might require the ability to steer the communication antenna. 

\begin{table}[hb]
    \centering
    \begin{tabular}{lccc}\hline
       Measurement  &  High levels & Medium troposphere & Deep atmosphere \\\hline
Optical+IR spectroscopy   &  yes & limited & no \\
MWR  & no & yes & yes \\
Radio occultations    & yes & limited & no \\
Probe      & yes & yes & no \\
\hline
    \end{tabular}
    \caption{Measurements capable of inferring temperatures in various parts of the atmosphere.}
    \label{tab:atm_suite}
\end{table}

%% file: sections/subsections/condensable_and_disequilibrium_species.tex
\subsection{Condensable and Disequilibrium Species}\label{sec:condensable_and_disequilibrium_species}



\begin{figure}
    \centering
    \includegraphics[height = 9cm]{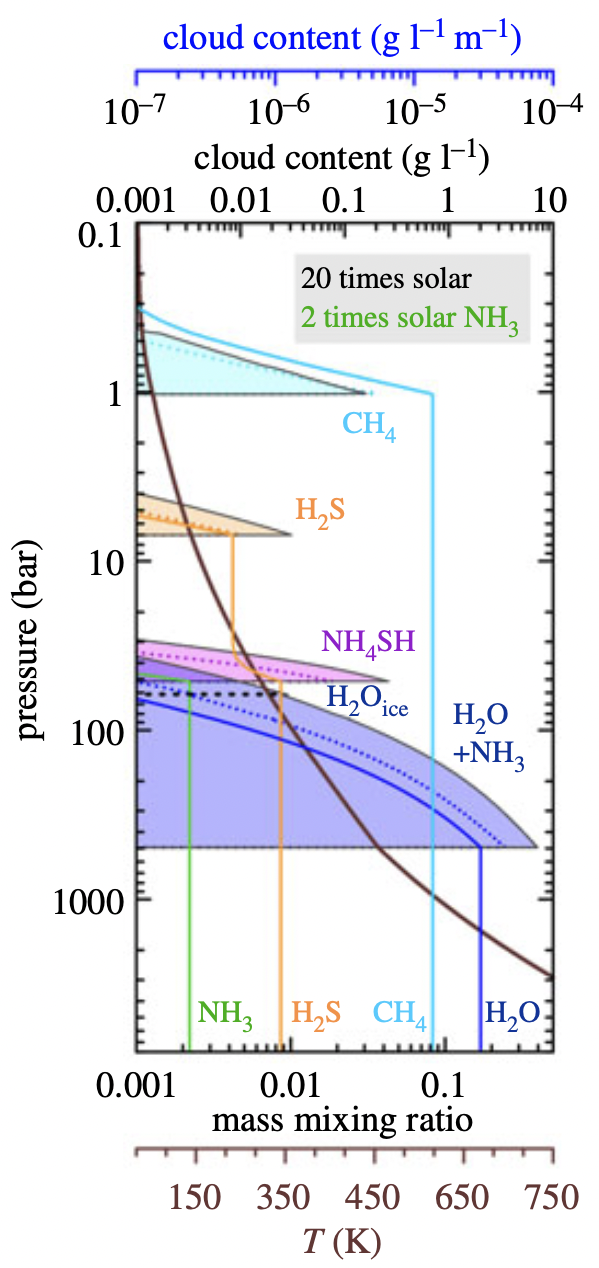}
    \caption{Cloud densities as a function of pressure in Uranus as obtained from an equilibrium condensation model. Figure from \citet{Hueso+2020}.}
    \label{fig:ECCM}
\end{figure}

Measuring the composition of any giant planet's atmosphere provides diagnostic clues as to the formation and evolution of that planet, as well as to the current interior structure.  As discussed in Section 2.4, the abundance of noble gases and the determination of isotopic ratios are particularly important because their nonreactive nature (for the former) and near-conservation during chemical reactions (for the latter) minimize our need to understand the dynamic and chemical processing that other species undergo in the deep atmosphere or interior. The noble gases and their isotopes cannot, however, provide a complete picture, and we need information on the abundances of so-called ices such as H$_2$O, CH$_4$, NH$_3$, and H$_2$S if we hope to determine the bulk composition of the giant planet and details of the formation process \citep{Atreya-etal:2019, guillot2022giant}.  Because, as shown in Figure~\ref{fig:ECCM}, all these species condense in Uranus's atmosphere, their abundances can vary spatially due to atmospheric dynamics and global circulation patterns. Hence it is important to measure certain disequilibrium species in the upper troposphere (species that are thought to form at high temperatures deep down but that are not in local thermodynamic equilibrium with conditions higher up) to also constrain vertical motions of the atmosphere. Examples of such species are CO, PH$_3$, and C$_2$H$_6$. The hydrogen molecule is also concerned, as it can be in either an ortho state in which both atoms have the same spin or into a para state with opposite spins. Measuring the ortho-para fraction has been shown to provide key information on mixing in giant planets' atmospheres, particularly Uranus \citep{Gierasch+Conrath1987}. 

Ground-based radio observations of Uranus at centimeter and longer wavelengths, as shown in Figure~\ref{fig:brightnessTmaps}, probing pressures from a few to tens of bars, indicate microwave absorbing species such as NH$_3$ and H$_2$S are a few to 100 times  more abundant in equatorial regions (latitudes roughly from +45 to -45 degrees) than they are over the polar regions \citep[e.g.,][]{HofstMuhl1989,HofstButler2003,Molter+2021}.  Observations in the near-IR \citep{KT2009,Sromovsky2011} suggest that near the 1 bar level, CH$_4$ is about 3 times more abundant over equatorial regions than it is over the poles. To explain these variations, it is suggested that low latitudes are areas of rising air that bring condensable species high into the troposphere. At the low temperatures encountered while rising, CH$_4$, NH$_3$, and H$_2$S condense out, leaving the air parcels ``dry" (i.e., depleted in those species).  These dry air parcels then move poleward at altitude and descend over the poles, creating a depleted region extending from ~1 bar down to ~50 bars. Sromovsky et al. further suggest that the latitudinal variations in CH$_4$ disappear by 5 bars, which, combined with the fact that NH$_3$ and H$_2$S variations persist deeper, could mean there are distinct circulation cells, similarly to what is suggested at Jupiter and Saturn \citep{fletcher2}. Alternatively, local storms may also lead to a depletion of condensable species because of the asymmetry between the ascending air (in which the concentration of condensing species is capped by condensation and particle growth) and the downdrafts powered by evaporative rainout, as indicated in Jupiter by the inverse correlation between ammonia abundance and lightning rate \citep{Guillot+2020b}.

A striking feature of the radio data at Uranus is that they requires a low NH$_3$ abundance, at least at the levels that are probed, down to pressures of 100 bars or so (e.g., Gulkis et al. 1978; de Pater et al. 1989; Molter et al. 2020). In addition, as shown in Figure~\ref{fig:ECCM} and contrary to Jupiter and Saturn, no NH$_3$ but instead an H$_2$S cloud was detected in the infrared \citep{Irwin+2018}, implying an N/S abundance ratio less than 1.0, when a solar mix would predict this value to be $\sim 7$. Specifically, infrared observations indicate that the tropospheric CH$_4$ abundance is 43--74 times larger, relative to H$_2$, than a solar abundance \citep{sromovsky2019methane, guillot2022giant}, ~16 times larger than what is seen in Jupiter, and ~8 times larger than Saturn's CH$_4$ abundance.  Constraints from radio observations indicate a similarly high S/H ratio between 29 and 47 times solar, but an N/H ratio of only 1--2 times solar \citep{Molter+2021, guillot2022giant}. It is not clear why Uranus would have formed with a low nitrogen abundance in the first place, and it seems more likely that NH$_3$ was been trapped in the deeper atmosphere, into water-ammonia ``mushballs" as invoked for Jupiter \citep{Guillot+2020a}, or deeper in a water ocean or into super-ionic water ice \citep[]{Atr20}. 

Probing Uranus's atmosphere for condensable and disequilibrium species requires a mixture of approaches. With the experience of ammonia on Jupiter, which is variable much deeper than its condensation level \citep{Bolton+2017, Li+2017ammonia}, we should be cautious of using equilibrium models such as those shown in Figure~\ref{fig:ECCM} at face value to set the levels that must be reached in order to get the bulk abundance of a condensable species. The combination of a probe reaching at least 5 bars at a unique location, of radio occultations, and of high spatial resolution infrared observations at many locations on the planet should be sufficient to fully characterize the methane condensation layer. However, we should be open to the possibility of variations to much deeper levels. Since methane accounts for about 10\% of the atmospheric mean molecular weight there, and directly impacts convection \citep{Guillot1995, Leconte+2017}, we should be able to rely on indirect analyses based on microwave observations that probe those levels, provided they include limb-darkening information in order to disentangle degeneracies between temperature and opacities. The determination of the abundances of other species such as H$_2$S and H$_2$O requires probing to between tens and hundreds of bars and relies almost solely on the information provided by microwave observations.   

Probing the abundance of disequilibrium species such as CO, CO$_2$, N$_2$, PH$_3$, GeH$_4$, and the hydrogen ortho-para ratio at several locations in the atmosphere is key to understanding mixing and atmospheric dynamics. With the possibility to constrain the vertical transport at several locations, it becomes possible to get from the measurement of the abundance of CO (and CO$_2$, if possible), an indirect constraint on the deep water abundance. Currently, an upper limit to the deep oxygen abundance of 250 times the solar abundance is derived from the tropospheric CO abundance \citep{Moses+2020}. Better constraints can be derived from (1) accurate measurements of tropospheric abundances of several disequilibrium species including CO, (2) an understanding of upward mixing, including the identification of possible stable regions in the deep atmosphere, if any, and (3) a constraint on the deep atmosphere's temperature profile. 

One point worth mentioning is that, in order to interpret the microwave radiometer data, it is necessary to know the microwave properties of the gases that provide significant opacity at the relevant pressures and temperatures. Water has been measured in the laboratory over a range of relevant conditions \citep{Steffes+2023}, but additional laboratory measurements are likely needed for H$_2$S, and possibly other species.

While our current understanding of the planets cannot explain the observations of condensable and disequilibrium species in Uranus's atmosphere, it is exciting that we see a mystery that is within our reach to resolve.  High spatial resolution visible and near-IR observations of the upper troposphere and moderate spatial resolution microwave sounding at $\sim$meter wavelengths, both of which are only possible from an orbiting spacecraft, when combined with measurements from an atmospheric entry probe, will constrain the dynamical state of the troposphere and abundances sufficiently to distinguish among the competing ideas discussed below.

%% file: sections/subsections/noble_gasses_and_isotopes.tex
\subsection{Noble gases and isotopes}

 The heavy noble gases are argon (Ar), krypton (Kr), and xenon (Xe). These differ from neon (Ne) in that they do not partition strongly into helium raindrops when the latter partitions out of metallic hydrogen at high pressures and temperatures. The strongly subsolar abundance of Ne measured by the Galileo Probe Mass Spectrometer (GPMS) in Jupiter is an indicator of this partitioning. It remains uncertain whether helium separates from hydrogen in Uranus (and Neptune). We discuss this in detail in the following sections. The heavy noble gases, in contrast, remain mixed in the gas phase except, importantly, xenon, which saturates in the gas and condenses to form a cloud in Uranus and Neptune in roughly the same pressure level regime as methane (see section II). Leaving this aside, the heavy noble gas abundance pattern is of high value because it tracks the patterns of some of the molecular species of the elements whose transport and chemical histories in the protoplanetary disk we seek to discern from present-day planetary properties. Also, the noble gas abundance pattern is known in meteorites and for now, one comet, allowing a direct comparison to relatively primitive reservoirs of condensed material from the period of, and prior to, planet formation.
\subsubsection{He/H and He/Ne abundance ratios} 

The He abundance was first measured by Voyager 2 in Uranus and Neptune during their respective flybys. \cite{Con87} and \cite{Con91} report He mass ratios of Y = 0.264 ± 0.048 and 0.32 ± 0.05 for Uranus and Neptune, respectively, for an H$_2$--He mixture. \cite{Lod09} give a protosolar He mass ratio of 0.278 when considering H$_2$ and He only, leading to the puzzling situation where He was nominally almost protosolar in Uranus and super-protosolar in Neptune. 

However, a recent re-analysis of the Voyager remote sending data in conjunction with HST spectra led to lower He/H ratios for both planets as a result of higher methane abundance in Uranus and higher N$_2$ abundance in Neptune than previously assumed. The higher abundances of heavier molecules at the altitude where the mean molecular weight of the atmosphere was inferred from the Voyager refractivity data leads to a reduction in the possible mass fraction of helium. The revised Y values for Uranus are 0.19--0.26 \citep{Sro11}, and at a protosolar level of $\sim$0.265 for Neptune \citep{Atr20}, indicating perhaps a moderate depletion in Uranus but a protosolar level for Neptune.

Measurements of the noble gases He and Ne in addition to Ar in the deep atmosphere at a few bars are important,as these species carry information on processes at much greater depths. They can inform us on the mixing between the atmosphere and the heavy-element-rich deep interior, and also about the deep composition. However, the interpretation of the measured values will be non-unique, and therefore must be coupled to other observational constraints. 

Both the $X_i$/H ratios and the $X_i$/$X_j$ ratios are important. The ratios with respect to H can be influenced by at least three processes. (1) Evaporation of light elements from the protosolar nebula, (2) rain-out to deeper levels in the evolved planet, and (3) chemical reactions that bind H. 

\paragraph*{\it {Preferential loss of hydrogen and helium in the protosolar disk}}
A sometimes-neglected pre-formation effect on composition is the following: the composition of the gas that was accreted by the ice giants could have been non-solar due to evaporation of the lightest elements hydrogen and helium \citep{Gui06}. If only this process took place, one would expect to see the protosolar H/He ratio in the atmospheres of the ice giants, and an enrichment in Ar, Kr, and Xe today. Two important important underlying assumptions here are that (1) the evaporation occurs hydrodynamically, meaning that the evaporation process removes gas from the disk without fractionation and independent of the individual atomic weight of the gas particles, and (2) fractionation in the gas did occur before removal due to condensation/sticking of the heavier noble gases (Ar, Kr, and Xe) to cold grains.

\paragraph*{\it Rain-out} 
Sedimentation of heavier material after phase separation can change the composition in the atmosphere. He/H is observed to be weakly depleted and Ne/H strongly depleted in the atmosphere of Jupiter. For Jupiter, this atmospheric He- and Ne-depletion is attributed to H/He phase separation at Megabar pressure levels at $\sim 0.85\ R_{\rm J}$. When He-rich droplets form, neon preferentially partitions in the He-rich phase, as the Gibbs free energy of the He-Ne system is lower than that of the H-Ne system \citep{Wil10}. This partitioning appears to be so efficient that when the He-rich droplets rain downward, they carry a large portion of the available Ne-budget with them so that the remaining He-poor material is more depleted in Ne than in He. Adiabatic models of the ice giants require a gas components at the Mbar levels, or else a much supersolar Ice-to-Rock ratio. If this gas component is from disk gas, perhaps once accreted together with porous planetesimals, He and Ne are present at depth together with H. As the ice giants have colder atmospheres than Jupiter, they might also have lower temperatures at the Mbar levels. Thus, H/He phase separation might occur and deplete the deep interior in He-Ne. If particle exchange with the atmosphere occurs, He-Ne depletion at depth can be communicated to the atmosphere. Observed moderate depletion in He/H and Ne/H could therefore indicate the presence of disk gas at depth. The level of He/H and He/N depletion would indicate inhibited large-scale convection and the temperature and fluid ices in the deep interior.

\paragraph*{\it CO dissociation}
Uranus might have primarily accumulated volatiles in the form of CO rather than as water (H$_2$O) or methane (CH$_4$) \citep{Ali14,Mou23}. CO has been identified in numerous protostellar disks, with its abundance notably measured highest among carbon-bearing species in the protostellar disk around DM Tau \citep{Hen13}. This compound can either condense as pure ice or become encapsulated in H$_2$O-clathrates within the cold midplane before being accreted by the ice giants in the form of pebbles and planetesimals \citep{Sch23}. Carbon can also be acquired in the form of methane pure ice or H$_2$O-clathrate. In the latter case (CH$_4$, H$_2$O), the accreted icy material would be hydrogen-rich, while in the former case (CO), it would be hydrogen-poor. When in contact with the disk gas, accreted CO undergoes chemical reactions to form CH$_4$ and H$_2$O. If CO were accreted in substantial amounts, this reaction would deplete the atmosphere of hydrogen, leading to apparent enrichments in He and Ne.

\vspace{0.5cm}

\subsubsection{$\rm ^{40}$Ar:  extent of mixing, original composition of the rock, volatilization of the rock during accretion }

The mass of silicates contained inside the interior of Uranus is essentially unknown. This is because the density of rock + H/He is the same as the density of ice, so there is a degeneracy \citep{Podol-etal:1995}. Condensation of solar nebular composition produces an ice:rock ratio of about 5:2 \citep{Lodde:2003}. If this ratio is representative of Uranus, about 3 Earth masses of rock should be present, but this is very uncertain (see Section \ref{sec:gravity_and_winds}).

One way of probing the rock abundance in Uranus, and the degree of transport between the interior and the gas envelope, is to use $^{40}$Ar measurements.
$\rm ^{40}$Ar is a noble gas that is produced by radioactive decay of $\rm ^{40}$K (half-life 1.25 Gyr) \cite{Kaula:1999}. Silicates contained within Uranus will produce $\rm ^{40}$Ar over the age of the solar system. The concentration of $\rm ^{40}$Ar in the atmosphere of Uranus therefore depends on the total silicate mass, the initial K concentration, and the degree to which $\rm ^{40}$Ar has been transported outwards. 

In principle $^{4}$He (which is produced by decay of U and Th) could be used in addition to $^{40}$Ar. However, there is also a primordial $^{4}$He component which will completely overwhelm the radiogenic helium contribution. There is also a primordial $^{40}$Ar component, which will partly obscure the radiogenic component, but its magnitude (based primarily on meteorite analyses) is uncertain \citep{Gobel-etal:1978}.

An example calculation, ignoring for the moment the primordial $^{40}$Ar issue, is as follows. We will assume chondritic K concentrations (546~ppm \citep{PalmeONeil:2014}). Assuming a three Earth mass chondritic core, the mass of $\rm ^{40}$Ar produced over 4 Gyr is about $\rm 1.4\times10^{18}$~kg. If this were all dispersed within a one Earth mass envelope with a mean molecular weight of 2.5, the resulting $\rm ^{40}$Ar number concentration (atoms/atom) is about $\rm 1.5\times10^{-8}$. This is well above the detection limit for modern mass spectrometers. Thus, in the case of efficient transport from the interior to the surface, the $^{40}$Ar concentration may be used to infer the silicate mass present. If transport of $\rm ^{40}$Ar from the deep interior to the surface is impeded, the envelope concentration will be lower.

How easy it is to transport $^{40}$Ar from the interior to the surface is unclear, mainly because the physical properties of Ar at the relevant temperatures and pressures are not known. Existing Ar phase diagrams \citep{VanNg-etal:2022} require extrapolation, and whether Ar in the deep interior would be solid or not depends on the poorly known temperature structure of Uranus (Section \ref{sec:atmospheric_pressure_temperature_profile}). Depending on the nature of the silicates in the interior (e.g., monolithic vs. dispersed) and the presence or absence of grain boundaries and/or fluids, Ar diffusion through silicates may be fast or slow. Whether Ar is buoyant relative to the surrounding material is unclear, but significantly affects transport upwards, as does the vigor of any interior convection. Stably stratified layers \citep{VazanHelle:2020} impede the upwards transport of such particles. Such layers could in principle be detected using other techniques, especially seismology or measurement of the tidal response (Sections \ref{section.seismology.detection} and \ref{sec:gravity_and_winds}).

Another way of interrupting $^{40}$Ar transport would be if $^{40}$Ar partitions preferentially into non-gas species. For instance, $^{40}$Ar might partition into superionic water rather than H/He gas. The presence of such a water layer might be inferred from other techniques, including those mentioned above and also magnetic observations \cite{Redme-etal:2011}.

One further but highly speculative complication is that K might also be partitioned into an iron core, if one exists. At Earth-like pressures, such partitioning does not appear to be important \cite{Xiong-etal:2018}, but the partitioning behaviour of K at higher pressures is currently unknown.

The analysis above assumes that all the silicates reside in the deep interior of Uranus. In reality, it is possible that the envelope contains a non-negligible fraction of heavy elements \citep{Haseg:2022}, deposited during accretion and depending strongly on the details of the accretion process. Detection of rock-related elements (e.g., K and Na) in the envelope by an atmospheric probe would thus be of great interest. Observations of exoplanets can also help resolve this issue. Sodium is a moderately volatile element that is chemically very similar to potassium, and is readily detectable by spectrosopy \cite[e.g.][]{Nikol-etal:2018}. If young Uranus-class exoplanets do not show strong Na spectral features, that would suggest that envelope pollution is not important, and vice versa.

The use of $^{40}$Ar as a proxy for silicate abundance and interior-atmospheric communication is novel. However, since a low atmospheric $^{40}$Ar concentration could imply either low rock abundance or low transport efficiency, this measurement would need to be combined with other observations to yield a unique answer. Further, the primordial $^{40}$Ar might be large enough to give a lower-bound concentration that limits our ability to diagnose limited mixing or low rock abundance. Research that could be carried out in the near-future to address specific knowledge gaps include: modeling of Uranus envelope pollution during accretion and comparison with exoplanet spectra; determining the physical properties and partitioning behavior of $^{40}$Ar at relevant temperatures and pressures; and a better determination of the primordial $\rm ^{36}Ar/^{40}Ar$ ratio.

\subsubsection*{D/H: A possible way to constrain the rock abundance in Uranus using existing data}

From lines of HD in the far-infrared detected by the Herschel mission, D/H in Uranus was found to be $4.4\pm0.4\times10^{-5}$, a number that is intermediate between that in the Sun and in comets \citep{Feuchtgruber-etal:2013}. Given these data, it is possible, under a few assumptions, to use this value to derive a ratio of rock-to-ice in Uranus. These assumptions are as follows:
\begin{itemize}
    \item The ice in Uranus is primarily water ice.
    \item Jupiter-family comets represent the source material of the ices in Uranus.
    \item We ignore deuterium and hydrogen in the rock phases in favor of the ice. \item D/H between water and hydrogen in Uranus is fully equilibrated.
    \item The envelope of Uranus is fully mixed. 
\end{itemize}
The equations are given and worked out in \cite {Feuchtgruber-etal:2013}. For a cometary D/H in water of $15-30\times10^{-5}$ and a value in the protosolar gas of D/H$\sim2\times10^{-5}$ in hydrogen, the ice mass fraction relative to total rock and ice is only $15-30$\%, implying that Uranus might be rock-rich rather than ice-rich. 

However, another alternative is that Uranus's interior was not fully mixed. In that case, only a limited amount of ices equilibriated with H$_2$ in the envelope. Uranus might then be ice-rich while at the same time having a relatively low atmospheric D/H value. 
A third alternative could be that Uranus formed near or beyond the CO iceline. Its constituents might then have become enriched in CO at the expense of H$_2$O \citep{Ali14, Mou23}. 
Disentangling these different possibilities would requires a better understanding of the planet's deep atmosphere in terms of transport of chemical species and temperature structure. This in turn requires measuring abundances of disequilibrium species, constraining the temperature profile in regions of condensation of abundant species, and understanding the phenomenon of convection inhibition globally (see Sections~\ref{sec:atmospheric_pressure_temperature_profile} and \ref{sec:condensable_and_disequilibrium_species}). 


\subsubsection{Stable isotopes of krypton and xenon}

There are multiple stable isotopes of xenon and krypton. The ratios among them for each noble gas allow comparison between the atmospheric values and candidate primitive reservoirs, such as meteorites and comets. Section 2 discusses potential condensation effects in Jupiter's atmosphere, which could alter the gas phase ratios, though this appears to be limited to xenon and in a narrow pressure range. Xenon isotopes were measured with sufficient precision in Jupiter's atmosphere by the GPMS to allow meaningful comparison with other solar system bodies, although it is important to use the error bars given in Figure 10 of \cite {Mahaffy-etal:2000} only. Table 1 of the same paper and Table 2.2 of \cite{Atreya-etal:2019} give incorrect values for the error bars. The krypton isotopes in Jupiter's atmosphere were not measured with sufficient precision to allow comparison with other sources \citep{Mahaffy-etal:2000}.  

Noble gas isotopes are available for the Jupiter family comet, 67P/Churyumov-Gerasimenko (\cite{Rubin-etal:2018} for krypton; \cite{Marty-etal:2017} for xenon). Comparison between Galileo-measured xenon isotopes in the atmosphere of Jupiter and comet 67P indicates a disagreement in the abundance pattern between the two, specifically that the cometary abundances decline with increasing neutron number while the Jovian remain constant. Interestingly, the Jovian values agree better with the Q-phase xenon thought to be the remains of solar nebula gas at the time xenon was trapped \citep{Lewis-etal:1975}. These are strong indicators that, if 67P is representative of Jupiter-family comets in its noble gas isotopic composition, such comets do not represent dominant contributors of heavy elements in Jupiter's envelope--whether during accretion or post-formation episode(s) of contamination. If xenon values show Q-phase dominance in Jupiter, it indicates that the heavy elements in its atmosphere were either introduced during the presence of the gaseous disk or originated from the parent bodies of chondrites. These parent bodies could have contributed to Jupiter's heavy element content either early or late in its formation process. 

This is not the place to debate these alternatives, as intriguing as they are. But it is evident from the discussion how valuable a similar set of data from Uranus would be, both in terms of sampling a body that resides well outward of Jupiter and one that has been less diluted by hydrogen and helium.

\vspace{0.5cm}

\subsubsection*{$^{15}$N/$^{14}$N and $^{13}$C/$^{12}$C}
Comparison of these isotopic ratios with those in comets provides an additional test of whether comets contributed significant amounts of material to the atmosphere of Uranus and (perhaps) by extension, to the other giant planets as well. The disadvantage of these isotopic ratios is that they might differ from one molecular carrier to another (i.e., ammonia vs. nitrogen), and hence modeling must be done to determine what the measured ratio in the dominant species in the atmosphere means in terms of correspondence with primitive carriers. The advantage is that remote sensing allows these ratios to be determined in a number of comets. This comparison has been done for nitrogen isotopes in Jupiter \citep{Atreya-etal:2019}.


The oxygen isotopic ratios constitute important measurements to be made in Uranus atmosphere. The terrestrial $^{16}$O/$^{18}$O and $^{16}$O/$^{17}$O isotopic ratios are 499 and 2632, respectively \citep{Asplund2009}. At the high accuracy levels achievable with meteorite analysis, these ratios present some small variations (expressed in $\delta$ units, which are deviations in part per thousand). Measurements performed in comets \citep{Bock2012}, far less accurate, match the terrestrial $^{16}$O/$^{18}$O value. The $^{16}$O/$^{18}$O ratio has been found to be $\sim$380 in Titan's atmosphere from Herschel SPIRE observations, but this value may be due to some fractionation process \citep{Courtin2011,Loison2017}. On the other hand, \citet{Serigano2016} found values consistent with the terrestrial ratios in CO with ALMA. The only $^{16}$O/$^{18}$O measurement made so far in a giant planet was obtained from ground-based infrared observations in Jupiter's atmosphere and had too large an uncertainty to be interpreted \citep[1--3 times the terrestrial value;][]{Noll1995}.

\vspace{0.5cm}

\subsubsection*{Challenges of interpretation} 

{\it Partitioning into phases}\\         
Generally we expect only liquid, and not solid, to be important for partitioning of the species considered here into condensates. For example, krypton and xenon can partition into liquid methane at cryogenic temperatures to an extent that could significantly alter the gas phase abundances. However, Uranus’s atmosphere is very cold. At the base of the stratosphere the temperature is below 60 K. The dominant cloud forming species in the upper troposphere is methane, which condenses as an ice. The base of the methane cloud, for an observed methane enrichment of 100 times solar, is at 2 bars \citep{Atr20}. For lower values allowed by the observations, the base will be higher but around a bar. Below this level, condensation is made complex by the apparent depletion of ammonia, leading to a supersolar S/N ratio. Under these conditions, clouds of frozen H$_2$S dominate with a base at 11 bars \citep{Atr20}. Deeper down, the extent of the clouds is very uncertain, because of the depletion of nitrogen and unknown abundance of oxygen. Pure ammonia condensates would be solid \citep{Atr20}. Water-ammonia clouds would be partially liquid, but might be thin or absent. This leaves water as the only potential liquid condensate into which noble gases might partition, with a cloud base dependent on the unknown enrichment of O relative to solar. (The deep ionic ocean of water is a separate problem that is part of the general question of the relationship between the atmosphere and deep interior, and so is covered below.)

The low solubility of the noble gases allows Henry’s law to be used to calculate the partitioning into the water droplets. With the constants provided from the NIST Webbook (\url{https://webbook.nist.gov/ chemistry/}) and an assumed water cloud base at 100 bar pressure, we find that only a tiny fraction of the gases partition into the water droplets in the cloud (the largest being argon at a ppm level) , so that the mixing ratios to be observed by a probe are not affected. \\

\noindent{\it Direct condensation}\\
Although the partial pressures of the noble gases are very small compared to the total pressure, the low temperatures near Uranus’s tropopause allow xenon to condense as a cloud between 0.02 and 0.5 bars, for a 100 fold enrichment in xenon relative to solar \citep{Zahnle2024}. Clouds occupying a narrower region of pressure (altitude) range are possible down to an enrichment of twice solar. The other noble gases do not condense. Therefore, mass spectrometric measurements of the xenon abundance intended to determine the bulk abundance should be made deeper than 0.5 bar. Since the methane cloud base is at about 1--2 bars, satisfying the requirement for a subcloud methane measurement will satisfy the same for xenon.

\subsubsection*{Measuring noble gases and isotopes}
The noble gas and isotope abundances can be determined using various measurements. Two examples are a mass spectrometer (MS) and a tunable laser spectrometer (TLS). 
The GPMS validated the concept of measuring key elemental and isotopic abundances in the atmospheres of the giant planets by mass spectrometry. For molecular carriers, these measurements are complementary to remote sensing determinations; for the noble gases and isotopes the measurements are unique. The GPMS itself is a product of 1970s technology and has been superseded in mass resolution, sensitivity, and dynamic range by mass spectrometers deployed at Saturn, Titan, comets, and Mars, among other targets. 

Concepts for the Saturn probe mission provide the set of parameters needed to determine the abundance patterns described in this chapter. The Hera concept \citep{MOUSIS201680} includes a mass spectrometer with a mass resolution of 1100 and both a much larger mass range and better sensitivity than the GPMS. Others have proposed something closer in capability to the GPMS, in particular in the most recent US Decadal Survey for Planetary Science and Astrophysics Saturn Probe study \url{https://nap.nationalacademies.org/read/26522/chapter/1}. Multiple Saturn probe concepts include a TLS to provide very high sensitivity determinations of selected species, and these are now being proposed specifically for a Uranus probe as well \citep{Webster2023TunableLS}. The challenge for Uranus is that only a subset of desirable species do not condense, do not partition into condensed phases, or have cloud bases at relatively shallow levels, making it unlikely that a probe that descends to $\sim$20 bars (the depth reached by the Galileo probe) will measure the bulk atmospheric abundance ratios of most condensables. 
\par 

It is also possible to make the measurements from Earth, near-Earth space, and satellites. For example, the Herschel satellite has demonstrated the ability to determine an important isotopic ratio, namely HD/H$_2$, in the atmosphere of Uranus \citep{Feuchtgruber-etal:2013}. JWST might be able to provide some ratios in species with infrared spectral features, as will ALMA, but in general noble gases are invisible to spectroscopic sensors. 


%% file: sections/subsections/gravity_and_winds.tex
\section{Gravitational Field  and Winds}\label{sec:gravity_and_winds}


%
 
 
 

\subsection{Science Motivation}
Constraining Uranus's internal structure and composition is essential to understanding the planet’s origin and evolution. However, nearly the entire mass of Uranus lies beneath the depths that can be observed using remote sensing techniques or from an entry probe. For Jupiter, results from the Juno mission suggest microwaves (0.6--22 GHz) sense down to the $\approx$1 kbar level, which is at 0.1\% of Uranus's mass, while an entry probe down to a $\approx$5 bars would scratch only the outermost 0.04\% of Uranus's mass. On the other hand, measurements of the planetary gravity field and tidal response are sensitive to the entire planetary volume and therefore provide integral constraints on the deep internal structure. Appendix \ref{section.radio.science} has a description of how radio science techniques are used to measure the gravity field. (See Section \ref{sec:seismology} for a discussion of seismology, which also has the potential to sample the deep interior.)
\par 

The accuracy of gravity measurements critically depends on the spacecraft's orbits, namely, the amount of time the spacecraft interacts with the planet at close range. At the moment, interior models of Uranus are constrained using the gravity data collected by the single flyby of Voyager 2, augmented by Voyager and ground-based measurements of the rings (e.g., \citet{jacob3}. Currently, only the $J_2$ and $J_4$ spherical harmonics of the gravity field have been measured, and the relatively large uncertainty on their values introduce uncertainty in structure models \citep[see][and references therein]{Helled2020a}. Additional uncertainties arise from poor knowledge of the temperature profile and the rotation rate of the interior (see Appendix \ref{gravitydetails} for additional details.)

\begin{figure*}[h!]
    \centering
    \includegraphics[width = 0.85\textwidth]{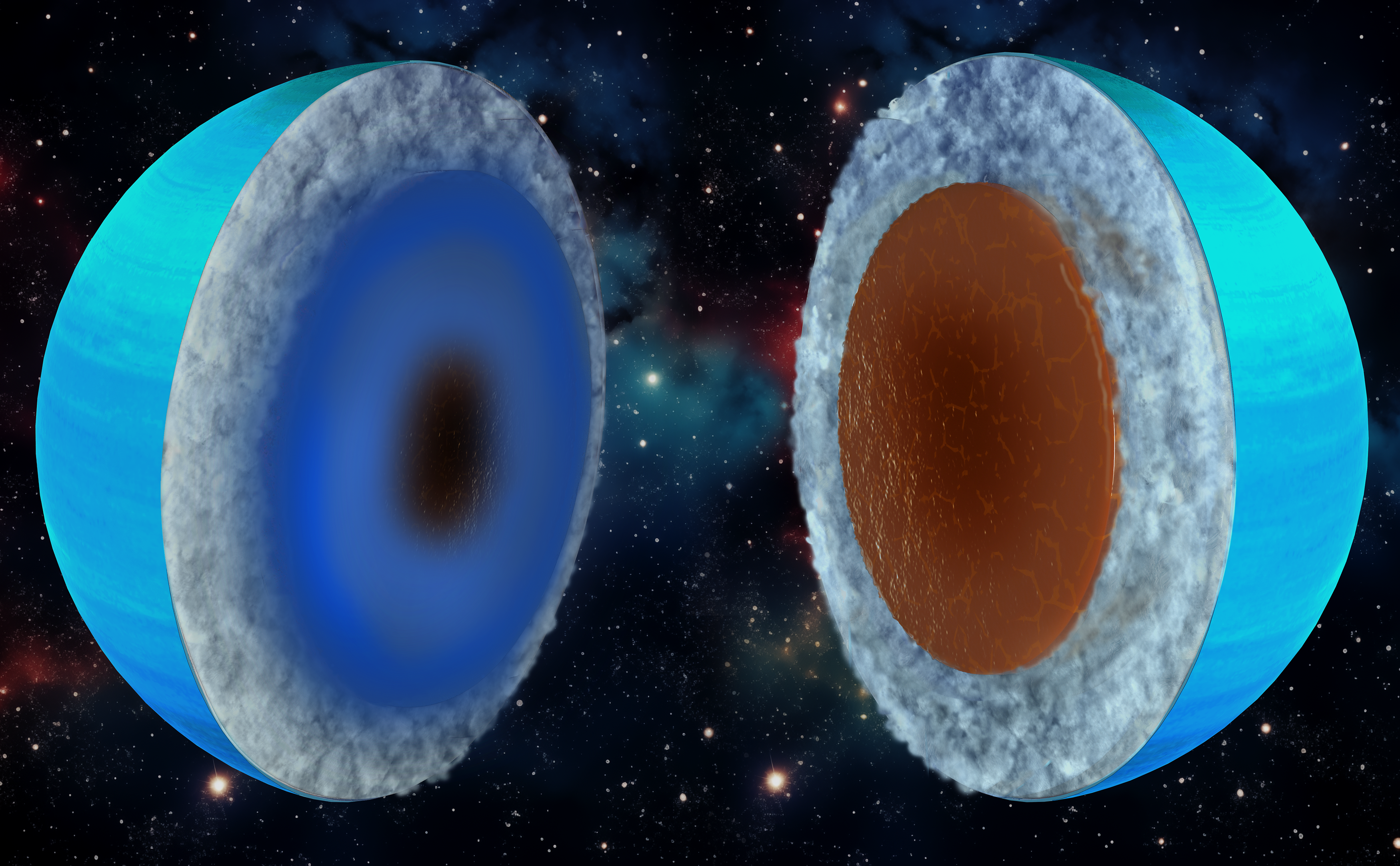}%
    \caption{Sketches of the two end-members of possible bulk compositions of Uranus: a water-dominated interior (left) and a rock-dominated interior (right). }
    \label{fig:UranusRockIce}
\end{figure*}

At the moment it is still unknown whether Uranus's bulk composition is water-dominated or rock-dominated. As we discuss below, at the moment both possibilities are valid, and further measurements are required to constrain Uranus's bulk composition.  Figure \ref{fig:UranusRockIce} illustrates the uncertainty in Uranus's bulk composition presenting the two end-members of possible solutions: a water-dominated interior and a rock-dominated interior.  
\par

To constrain the range of allowed interior structures, more precise determination of the gravity field of Uranus is important. However, these measurements alone might not be sufficient to discriminate among competing models. It might be necessary to combine determinations of the gravity field with determinations of other properties, such as the tidal response of the planet, which at present is completely unknown. The moons tug on the planet, deforming its shape and its gravitational potential slightly. The deformation is periodic and slightly dissipative, so that the direction in which the planet is deformed does not exactly align with the position of the moon. This dissipation has another important effect: in the simplest analysis, conservation of energy requires that the moons recede from the planet gradually over time.  Both the magnitude of the deformation of the gravitational potential (measured by the Love numbers $k_{nm}$), and the magnitude of the dissipation (measured by $1/Q$ where $Q$ is the quality factor of the planet), if measured accurately enough, could provide important information on Uranus' interior. It also might be necessary to better constrain the interior temperature profile and rotation rate. Refer to Appendix \ref{gravitydetails} for additional details about the science and techniques for measuring the gravity field.


\subsection{Measurements}


\subsubsection{Gravitational Moments}
\textbf{Even Zonal Harmonics}\\
For Uranus, the even gravitational spherical harmonics $J_2$ and $J_4$ have been determined from Doppler tracking of the Voyager spacecraft during its flyby in 1986 and through long-term tracking of its natural satellites and rings. The current relative uncertainties are $\sigma_{J_2}/J_2 = 2\times 10^{-4}$ and $\sigma_{J_4}/J_4=0.03$ \citep{Helled20}. However, unlike for Jupiter, the winds may have a substantial effect on Uranus's $J_4$ \citep{kaspi2013}. Constraining the wind depth through measurement of the odd harmonics is therefore important for knowing the static part of $J_2$ and $J_4$ that are used for constraining interior density distributions and thus composition. This uncertainty in the static part of $J_4$ is unique to the ice giants and comes before the uncertainties that are encountered due to the equations of state and the density distributions. 
Recently, \citep{Neuenschwander2022} predicted for the first time the higher gravitational moments $J_6$ and $J_8$ for Uranus and Neptune. It was shown that the prediction highly depends on the assumed depth of the winds (see also Fig. 6 below): without winds they predict $J_6=(46.12-59.90)\times10^{-8}$ and $-J_8= (8.42-14.39)\times10^{-9}$ for Uranus. Assuming that the winds penetrate to depths of 1100 km, the predictions shift to $J_6=(53.76-69.04)\times10^{-8}$ and $-J_8= (10.29-17.78)\times10^{-9}$. \\

It was also shown that an accurate measurement (of the order of $\sim 1\%$) of the higher gravitational harmonics $J_6$ or $J_8$ can be used to further constrain the wind depth. Similarly, it was suggested that a measurement of the moment of inertia (MoI) (to the order of $\sim 1\%$) can be used to further constrain Uranus's and/or Neptune's rotation period. \\

Finally, it was demonstrated that the solution space of possible density profiles of the planets depends on both the assumed rotation periods and wind depths: a faster rotation period and/or a higher wind depth lead to a significantly more centrally concentrated density distribution. 

\begin{figure*}[h!]
    \centering
    \includegraphics[width = 0.5\textwidth]{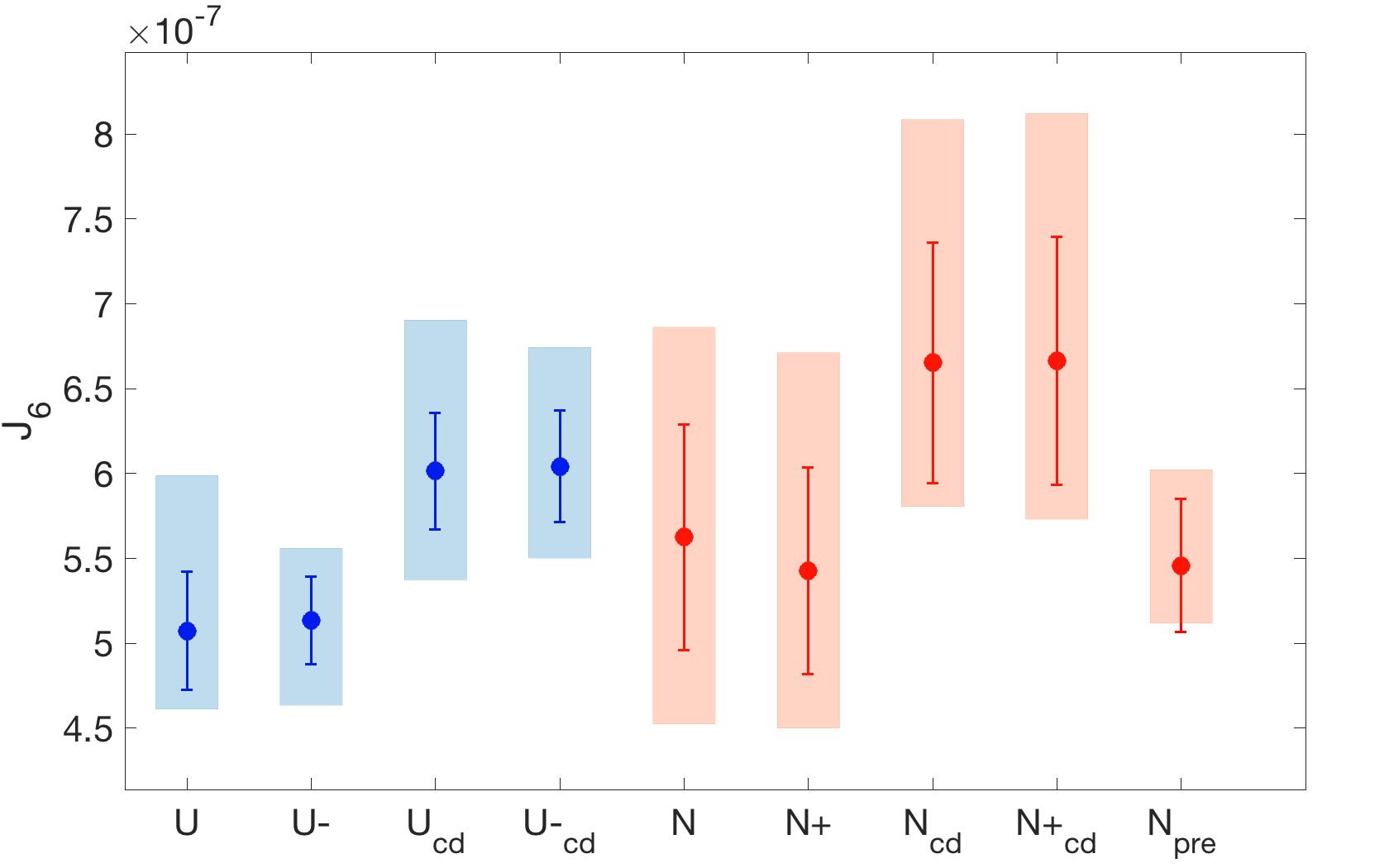}%
    \includegraphics[width = 0.5\textwidth]{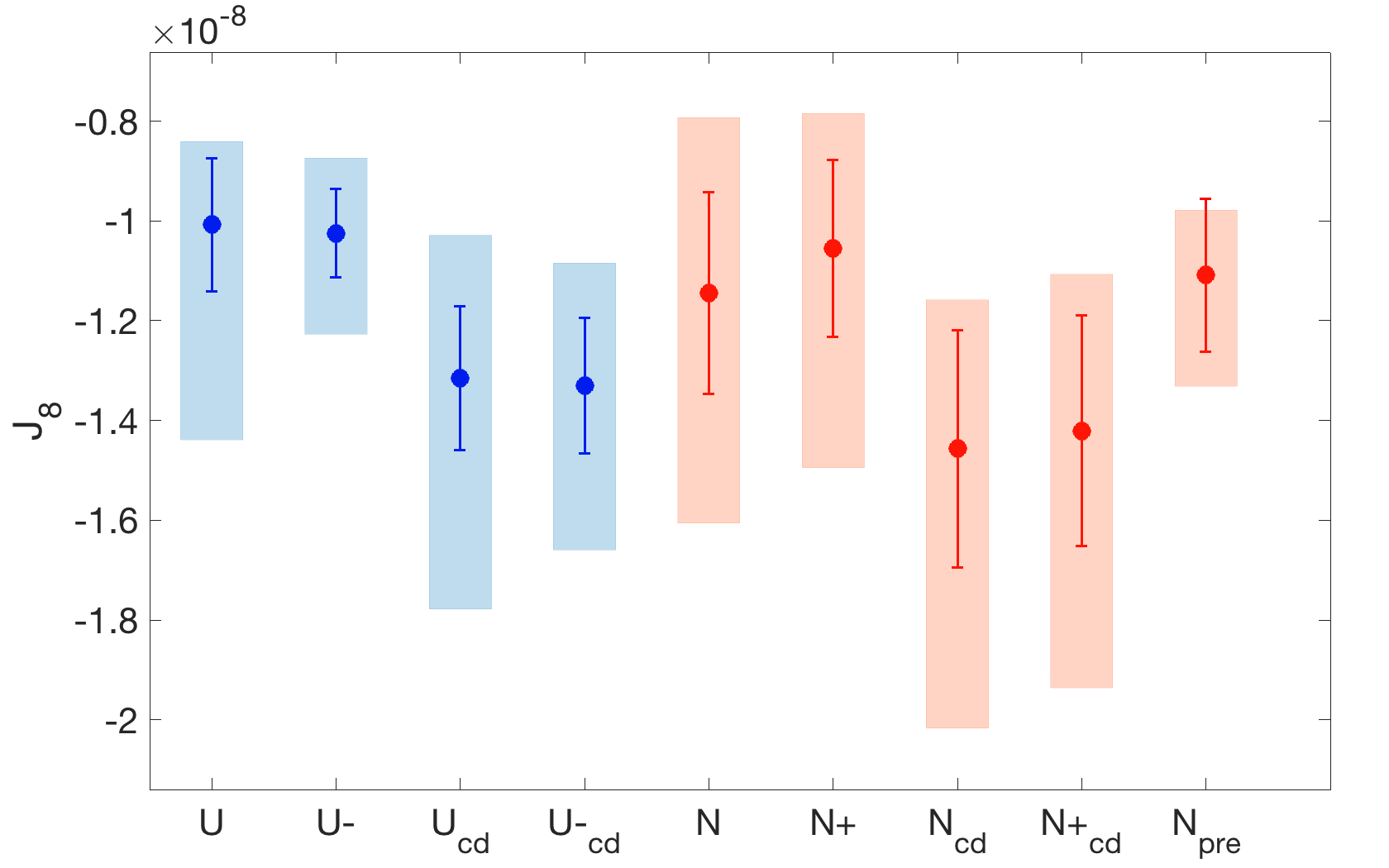}
    \caption{Predictions of $J_6$ (left panel) and $J_8$-values (right panel) for different planetary models of Uranus (blue) and Neptune (red). The planetary models are collected along the \textit{x}-axis with abbreviated names (U: Uranus, N: Neptune, cd: 1100~km deep winds). U and U$_{cd}$ (N and N$_{cd}$) have a rotation period of 17.24~h (16.11~h), U- and U-$_{cd}$ (N+ and N+$_{cd}$) have a rotation period of 16.57~h (17.46~h). The dots mark the mean values, the bars show the standard deviation, and the boxes show the full solution range. Figure from \citet{Neuenschwander2022}.}
    \label{fig:fancy_plots}
\end{figure*}

\begin{figure}
    \centering
    \includegraphics[width = 0.36\textwidth]{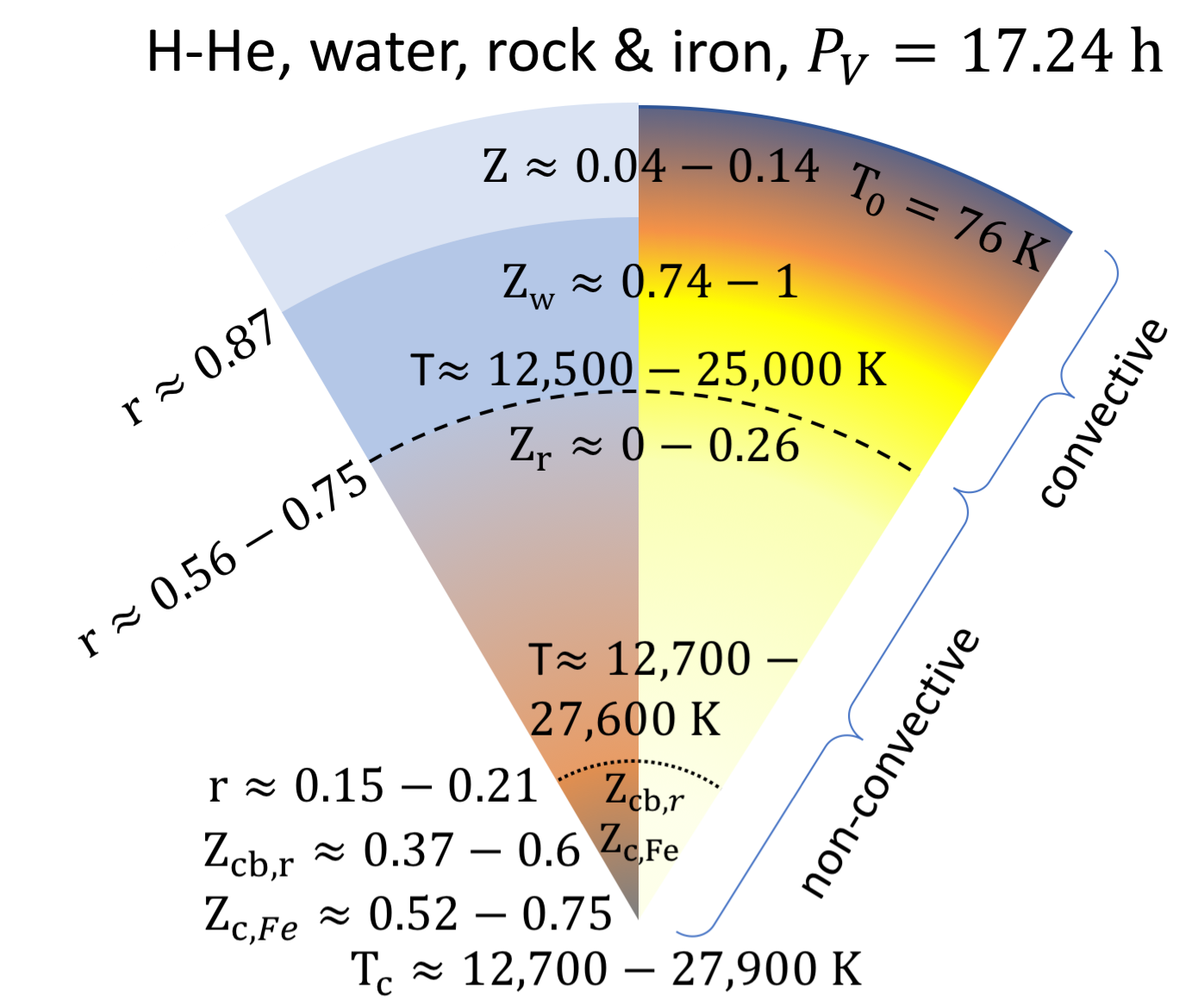}%
    \includegraphics[width = 0.36\textwidth]{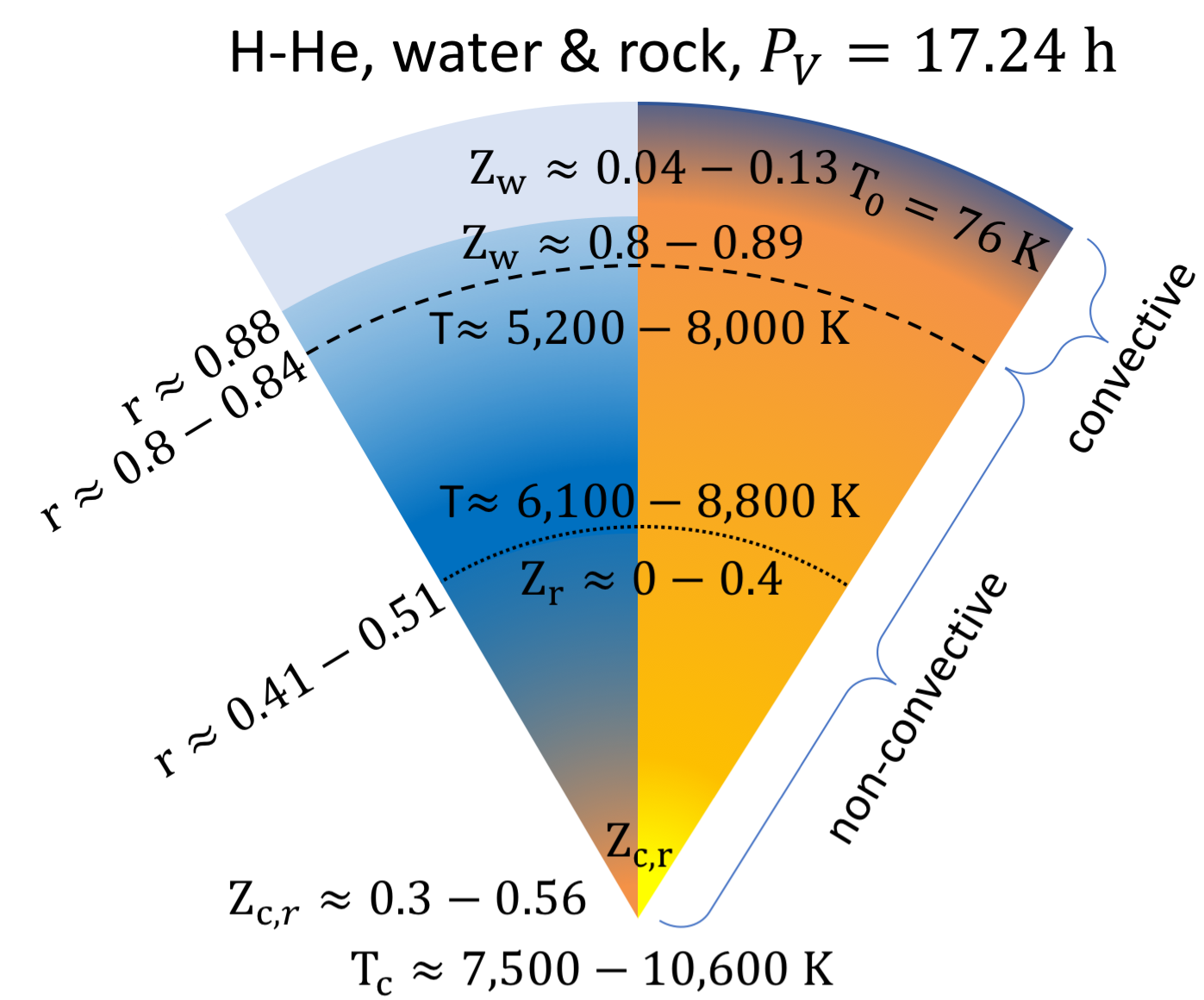}%
    \includegraphics[width = 0.36\textwidth]{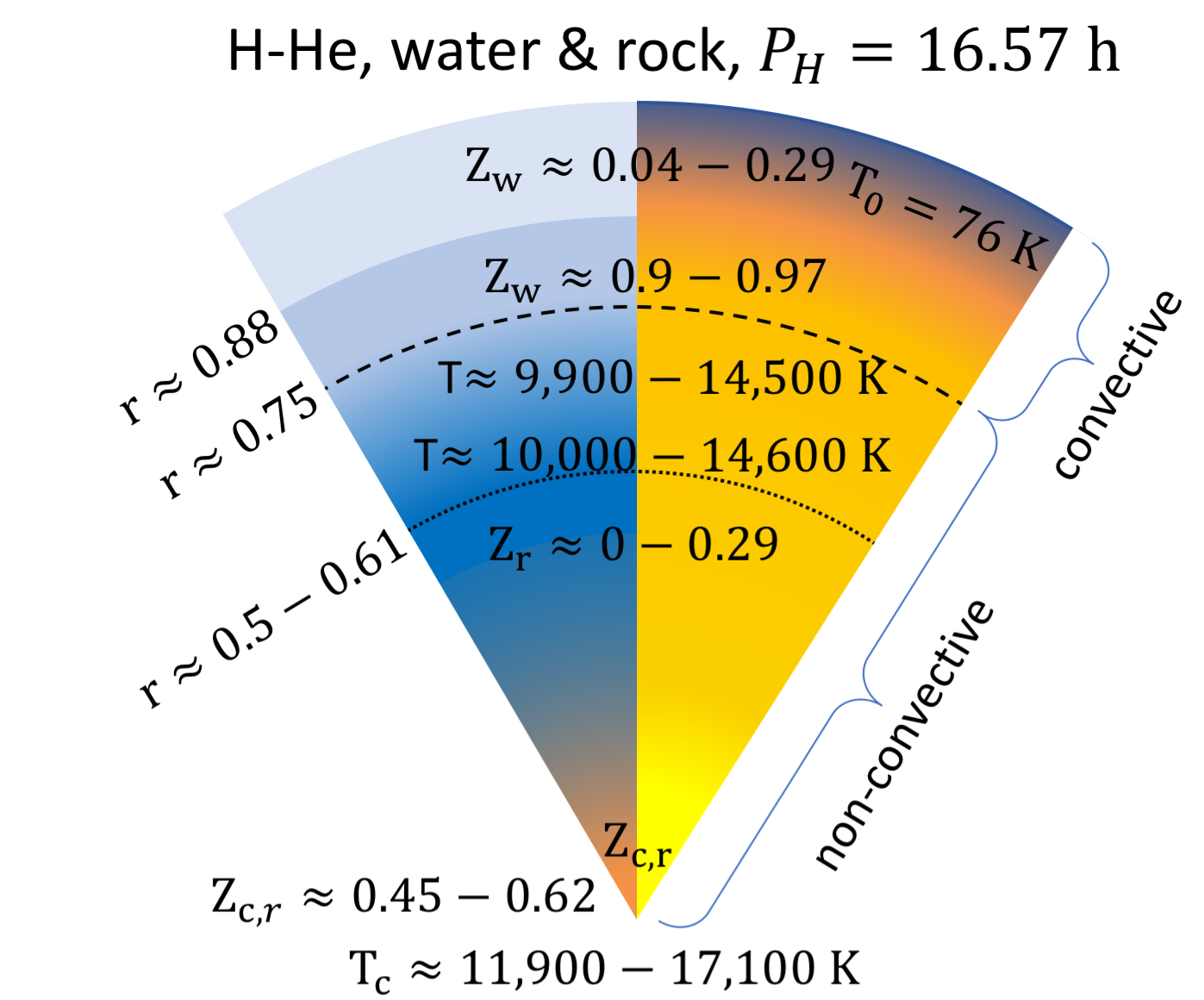}
    \label{fig:Interior_Illustration}
\end{figure}

In a recent study, \citep{Neuenschwander2024} proposed a new method to interpret empirical structure models in terms of temperature and composition and to detect nonconvective regions. They applied this method to Uranus models from \citep{Neuenschwander2022} and found that all the models consist of a convective envelope above a nonconvective region that extends down to the planet's center (see Fig. \ref{fig:Interior_Illustration}). It was shown that the existence of these nonconvective regions can prevent the heat from escaping efficiently, and have temperature gradients well in excess of the isentropic value, meaning that there can be a significant decrease in density relative to adiabatic models of the same composition due to high temperature. To obtain the same density as adiabatic models one needs intrinsically denser material (e.g., more rock relative to water). Their predicted water-to-rock ratio of 2.6--20.7 is indeed smaller than those of adiabatic models \citep[e.g., 19--35;][]{Nettelmann2013}. \\

It was also found that a faster rotation period of Uranus ($P_{rot}=16.57$~h) increases the size of the convective region (by $\sim35\%$), and leads to hotter central temperatures (higher than 10,000~K) and a higher bulk heavy element abundance (+$\sim 0.5$~M$_\oplus$).

\begin{figure}
    \centering
    \includegraphics[width = 8cm]{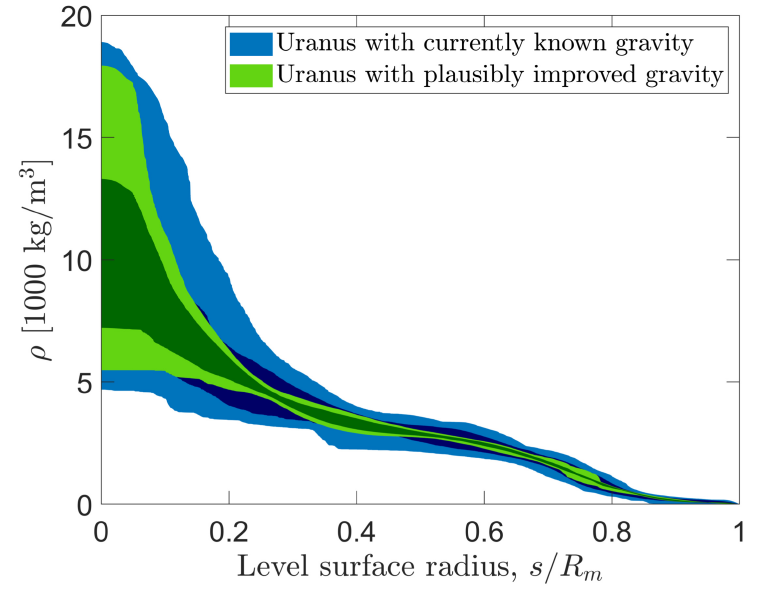}
    \caption{Uranus density profile constrained by gravity coefficients measured to current 1$\sigma$ and 3$\sigma$ accuracy (blue) and to an plausibly improved accuracy in $J_2$--$J_6$ (green) assuming a small (10 min) uncertainty in the rotation period (Fig. 14 in \cite{Movshovitz22}).}.
    \label{fig: Uranus_density_profiles}
\end{figure}

\cite{Movshovitz22} also discuss possible density distributions given current uncertainties and the effect of measuring the $J's$ more precisely (Fig. 8). The rather well-constrained density profile in the latter case suggests that the major additional uncertainty in the density profile will arise from the rotations period, which is currently unknown to perhaps 40 min. The possible composition will furthermore depend on the thermal state and the equations of state and phase diagrams of the various constituents and their mixtures (see Appendix~\ref{app:materials}).

\begin{figure}
    \centering
    \includegraphics[width = 7cm]{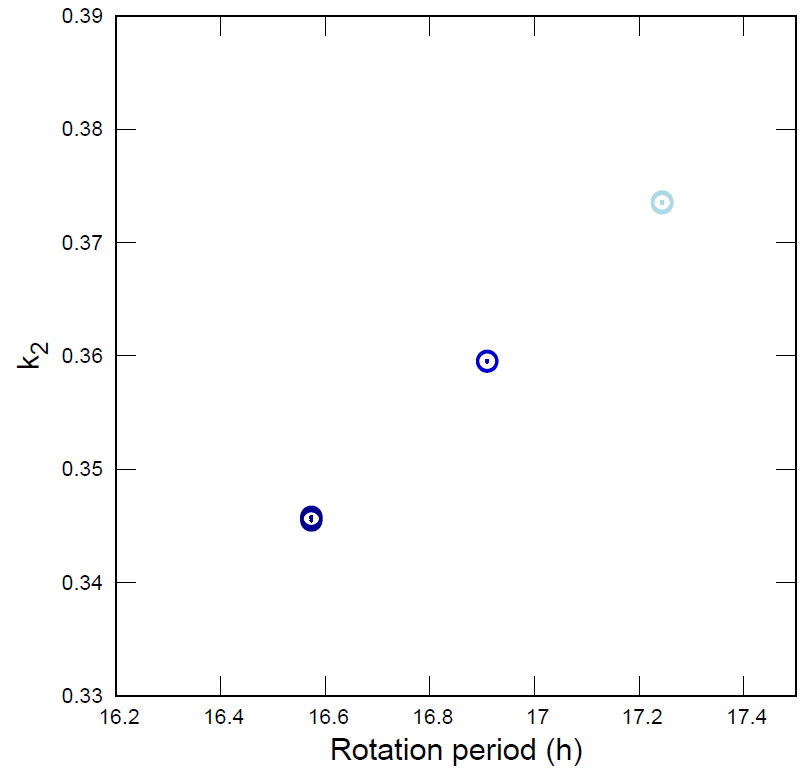}
    \caption{Static Love number $k_2$ of various Uranus models that match the observed $J_2$, $J_4$ values but differ in their assumed rotation period.
    \label{fig:U_k2_Prot}}
\end{figure}

For precisely known gravity harmonics, different spin states map onto different values of the moment of inertia and the static tidal response coefficient $k_2$. When Jupiter's $k_2$ value could be determined to within 3\%, its NMOI still lacked an observational estimate \citep{Durante20}. Therefore, a determination of Uranus's $k_2$ may seem much more feasible. For an assumed rigid body rotation period uncertainty of 40 min, its $k_2$ value is predicted to vary by 10\% in response to the tidal pull by Ariel (Figure \ref{fig:U_k2_Prot}). 

\subparagraph{Odd Zonal Harmonics}
The odd zonal ($J_{2n+1}$) gravity coefficients are dominated by the contributions from the zonal flows that circle the planet. Cloud tracking with optical imagery reveals a relatively smooth wind profile as a function of latitude \citep{sromovsky2015high} that is mostly symmetric with respect to the equatorial plane. However, small asymmetries upward to a few tens of m~s$^{-1}$ were also observed at low to mid latitudes, which could give rise to equally small asymmetries in the gravity field of Uranus. The odd zonal harmonics contain independent information about the depth of the zonal winds visible at the cloud level, since these coefficients are not influenced by uniform rotation (Kaspi et al. 2018). The structure of zonal winds at depth plays a key role in revealing the connections between the upper atmosphere and deeper interior of the planet. It follows that determining their vertical extension by precisely determining the odd harmonics is an important goal of a Uranus gravity experiment.


\subparagraph{Non-Zonal Gravity} 
 The gravity field of Uranus can have nonzonal components in several possible ways, as follows:
 \begin{itemize}
  \item Dynamic structure in the atmosphere (an example on Jupiter is the Great Red Spot [GRS]).
  \item Deep-seated convective structure in a fluid (varying on timescales longer than a mission, perhaps, and therefore effectively ``static").
  \item Deep-seated convective structure in a solid (analogous to the well-established nonzonal gravity observed for Earth and Mars and changing only on a very long timescale).
  \item Tidal distortions (time-variable in a frame of reference defined by the deep interior but with well-known temporal and spatial patterns).
  \item  Normal modes (time variable at a timescale of tens of minutes or so, comparable to the time a spacecraft is near close approach), which could allow for gravitational seismology.
\end{itemize}

Each of these possibilities for generating nonzonal components to the gravity field is discussed in more detail in Appendix \ref{gravitydetails}.

\subparagraph{Tidal Response}
Unlike the static gravity field, which is sensitive to the density distribution, tidal response is also sensitive to shear rigidity if a solid layer is present, since tidal periods may be short compared to Maxwell times for that medium. \cite{stixrude2021thermal} showed that Uranus can have a solid interior. The measurement of the tidal response may discriminate between a solid-ice versus fluid-water interior if $Re(k_2)$ is measured to sufficient accuracy. \cite{stixrude2021thermal} showed that depending on the size of the solid core, $Re(k_2)$ varies between 0.28 and 0.36. The higher value corresponds to a fluid interior. Thus, measuring $Re(k_2)$ to an accuracy of 0.02 would be sufficient to assess the state of Uranus interior. This places fundamental constraints on the thermal state of the interior, as the temperature cannot exceed the freezing point.  \citet{parisi+2024} find that representative orbits under consideration for the Uranus Flagship mission will detect Ariel's tides and can test for a large, solid core.

 The tidal response has both an amplitude and a phase, or equivalently a real and an imaginary part. The latter provides information about dissipation and the rate at which angular momentum can be transferred from the planet to the orbiting satellites. As a result, tidal dissipation sets the timescale for the dynamical evolution of the Uranian system, including the recession rates of the moons, and the possibility of resonant moon-moon encounters. A measurement of the tidal dissipation parameter $Q$ is not as easily related to interior structure as multiple dissipation mechanisms are possible, including processes that are unrelated to material properties and are fluid dynamical in nature (as appears to be the case for both Jupiter and Saturn). Viscoelastic dissipation would occur only in solid portions of the interior. Fluid dissipation related to planetary flows is also possible, as is dissipation at compositional gradients. A high value of $Q$ ($>10^9$) would point towards an entirely fluid interior. On the other hand, a low value of $Q$ ($\sim10^3$) could point towards a frozen interior, fluid dissipation mechanisms that may require certain configurations of the interior related to the thickness of distinct fluid layers, convective flow velocities, and/or the presence of compositional gradients. 

Ariel provides the largest contribution to tides. Thus, we require orbits with pericenters that sample the entire 0$^\circ$--360$^\circ$  Uranus-Ariel ecliptic plane, where Ariel is at 0$^\circ$. Multiple orbits where the pericenter is close to Ariel would disfavor an accurate $k_2$ determination. Mission studies are required to better understand the operational requirements for an accurate $k_2$ determination.

\subparagraph{Gravitational Seismology}
It is unknown whether normal modes are excited within Uranus to an amplitude that would be detected through the resulting gravity perturbations. Doppler tracking of both the Juno and Cassini spacecraft at Jupiter and Saturn, respectively, show unexpected accelerations that have been interpreted as being due to normal modes \citep{Markham2020, durante2022juno}. The large eccentricities of their orbits, however, have prevented a determination of precise mode frequencies due to the relatively short amounts of time the spacecraft are close to their planets. Measuring both the spatial and temporal frequencies of at least one mode is necessary in order to constrain the interior structure of the planet, so to constrain Uranus with gravitational seismology requires either lower eccentricity orbits than are typically flown or an improved approach to analyzing the Doppler data. 

Seismology is discussed more broadly in Section \ref{sec:seismology} and Appendix \ref{section.seismology.oscillation.modes}.

\subparagraph{Rotational Rate and State}  
The rotational motion of Uranus is related to the internal distribution of mass. An accurate rotation rate is needed for interpreting the zonal gravity coefficients, as it allows predicting the hydrostatic response of the planet to rotation. The rotation rate of a gaseous planet is currently determined using various techniques. The Voyager 2 rotation period of Uranus based on radio data is 17.24 h \citep{desch1986rotation,warwick1986voyager}. The Voyager 2 rotation period of Uranus inferred from the the magnetic field data is 17.29 h \citep{Ness1986Uranus,desch1986rotation}. Minimization of wind velocities or dynamic heights of the 1 bar isobaric surfaces and constrained by the single occultation radius and gravitational coefficients leads to solid-body rotation periods of 16.58 h for Uranus \citep{helled2010uranus}. If there is a detectable nonzonal gravity field, radio tracking data of the spacecraft can constrain the rotation rate of Uranus, which would be complementary to the magnetic field rotation. The current uncertainty in Uranus's rotation rate introduces a large uncertainty in its internal structure and bulk composition \cite{Nettelmann2013,Neuenschwander2022}. 

The Sun imposes torque on the Uranian rotational bulge and its satellites, causing the precession of the Uranus spin axis. The precession rate is inversely proportional to the polar moment of inertia of Uranus. In addition, the degree 2 gravity coefficients are related to the differences in the polar and equatorial moments of inertia. Thus, the combination of the spin axis precession and degree 2 gravity yields a full inertia tensor. Much like the gravity coefficients, the components of inertia tensor are integral properties of the mass distribution.

In principle, precession can be measured by radio tracking of an orbiting spacecraft, which in turn would constrain the moment of inertia. This has, however, proven to be difficult since precession is slow.  The fact that collecting enough precession data from a spacecraft would vastly improve our knowledge of the gravity coefficients also reduces the value of knowing the moment of inertia. This is the case for the Juno spacecraft at Jupiter, where the improvement in the gravity data overwhelms the contribution of the moment of inertia to the internal structure knowledge \citep{militzer2023relation}. Precession can also be recovered by long-term observations of Uranus's rings. 

\subparagraph{Shape from Radio Occultations} 
Knowledge of Uranus's shape can help us infer its internal structure, because structure models are set to fit the measured planetary mean radius and shape helps us constrain the zonal winds and their effect on the gravity field. At a minimum, measurements of the polar and equatorial regions are required. Measuring the radius as a function of latitude from pole to pole would provide additional information on the planetary shape and the contribution of atmosphere dynamics to the measured planetary shape.

Until now, measurements of Uranus's shape have been conducted using stellar and ring occultations, along with Voyager measurements. 
While stellar and ring occultations \citep[e.g.,][]{French1983,French1987,French1998,Hubbard1987,Elliot1984} provide accurate determinations of planetary shape, their data correspond to very low pressure levels ranging from several to thousands of microbar. Consequently, this data might not be directly applicable to the commonly used 1 bar pressure level required by interior modeling, especially since atmospheric dynamics is expected to alter the isobaric shape between the microbar and 1 bar pressure levels. Radio occultation measurements, in which the spacecraft is tracked from the Earth as it moves behind the planet (see Appendix \ref{section.radio.science}), can measure the planet's shape to pressures greater than 1 bar \citep{lindal+1987}. \citet{Lindal1992} estimates that the uncertainty in Uranus's 1 bar radius varies from 4\,km at the equator to 20\,km at the poles, although due to the uncertainty in the rotation rate, and therefore the dynamical heights, the true uncertainty could be larger.

It is crucial to emphasize that atmospheric winds and dynamics, in general, can significantly influence the planetary shape, and this effect can surpass the impact of harmonic coefficients. Consequently, the knowledge of planetary shape has a limited impact on internal structure models. Nevertheless, understanding the atmospheric shape is vital for unraveling the dynamics of planetary atmospheres. 

\begin{figure}
    \centering
    \includegraphics[width = 8cm]{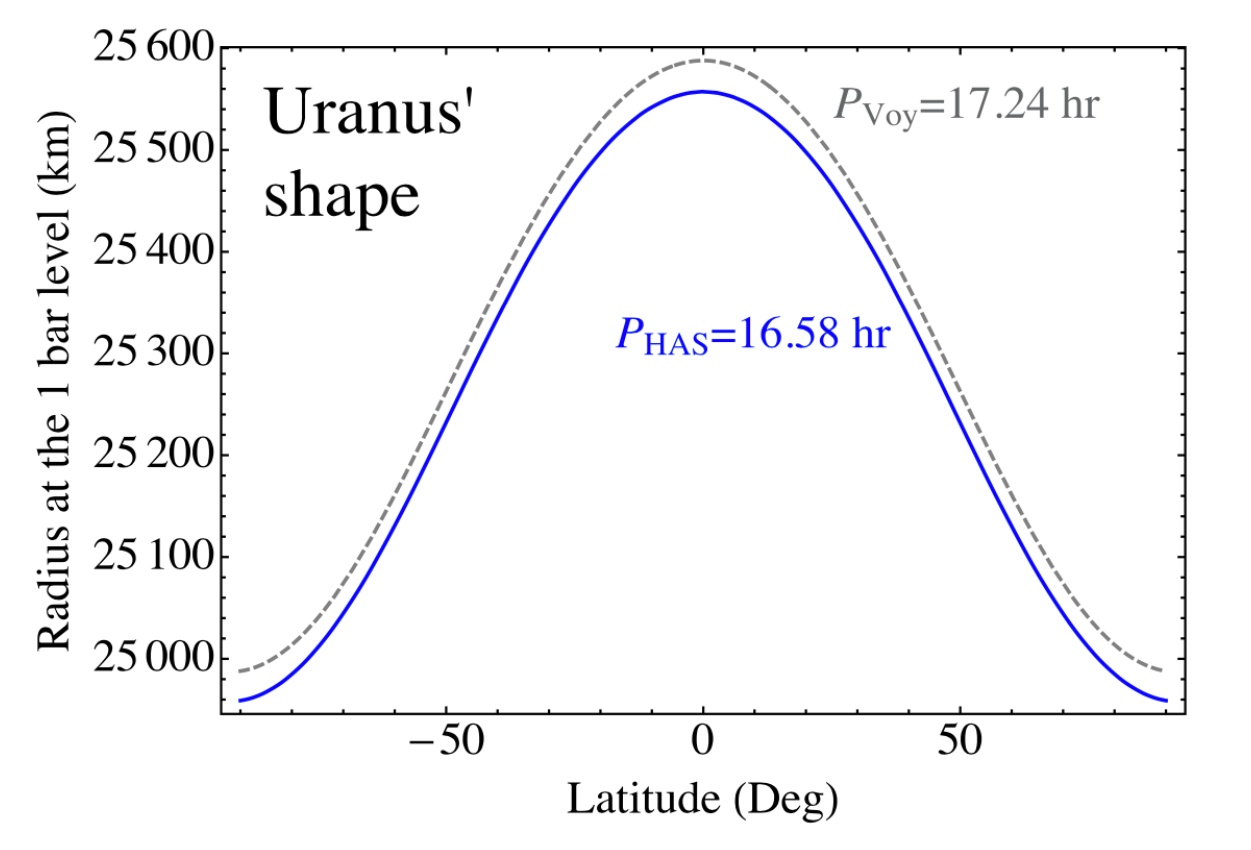}
    \caption{Uranus's shape. The dashed gray curve corresponds to the Voyager rotation period of 17.24 hr, while the solid blue curve corresponds to a rotation period of 16.58 hr, which was found to minimize Uranus’s winds and dynamical heights \citep{Helled2011}.}
    \label{fig:shape}
\end{figure}
Figure \ref{fig:shape} presents the inferred shapes of Uranus assuming the Voyager 2 and modified solid-body rotation periods for the planets as derived by \citet{helled_shape}.

\subparagraph{Zonal Winds}
\noindent Uranus exhibits fast surface zonal winds, with speeds reaching up to 200 m s$^{-1}$ relative to its assumed bulk rotation (measured by Voyager 2). Exact behavior of the wind decay is uncertain with depth, but there is evidence that winds must decay rapidly in the shallow layers of Uranus, based on zonal gravitational harmonics studies \citep{kaspi2013, soyuer2023}, where constraints on wind decay are placed through their contribution to the observed zonal gravitational harmonics. Additionally, \citet{soyuer2020} show that deep-seated fast winds are improbable due to Ohmic dissipation constraints coming from the interaction of zonal winds with the Uranian magnetic field in the region of rapidly increasing electrical conductivity. 

\begin{figure}
    \centering
 \includegraphics[width = 15cm]{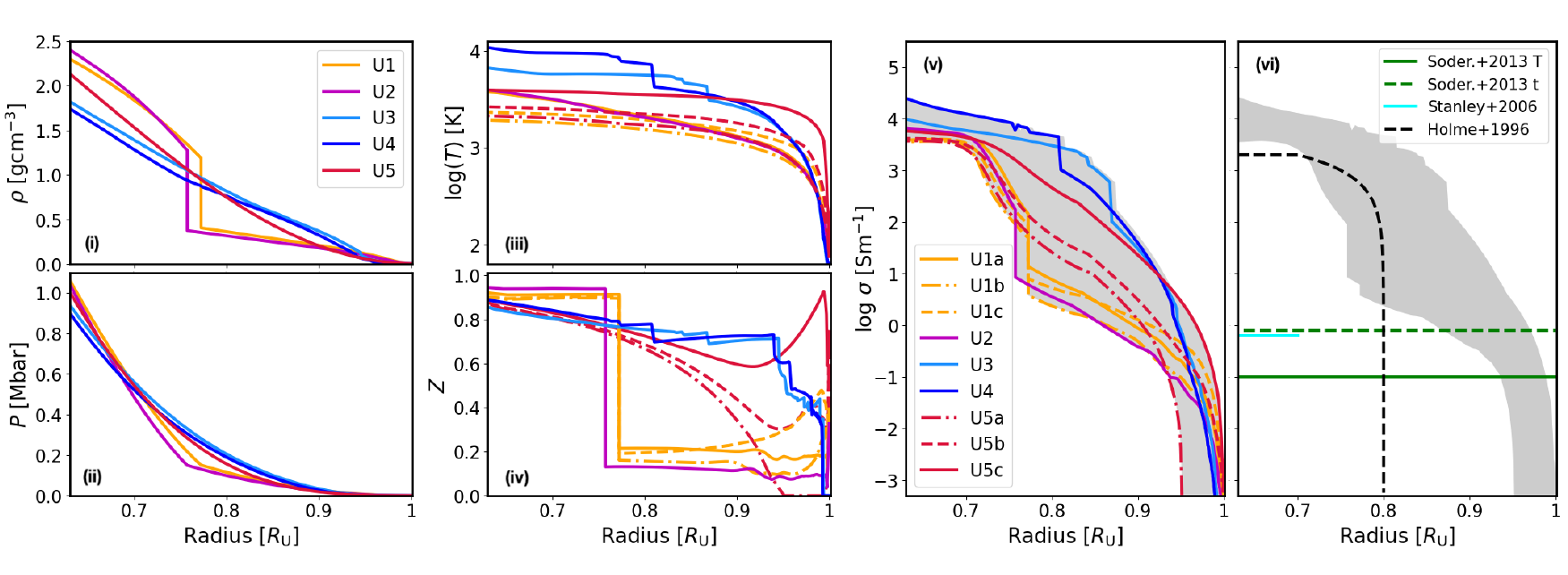}
 \caption{\footnotesize{Density \textbf{(i)}, pressure \textbf{(ii)}, temperature \textbf{(iii)} and water content \textbf{(iv)} as a function of radius of various Uranus interior structure models commonly used in the literature \citep{Nettelmann2013, Helled2011, podolak, vazan}. The steep increase and the diversity of electrical conductivity profiles can be seen in \textbf{(v)} \citep{soyuer2020, soyuer2021b}. Additionally, the assumed electrical conductivity profiles in various Uranus dynamo studies \citep{holme, stanley2006, soderlund2013} are shown in \textbf{(vi)}, where the electrical conductivity in the rapidly increasing region is usually neglected in numerical dynamo simulations \citep{stanley2006, soderlund2013}.}}
      \label{fig:winds}
\end{figure}

Understanding the zonal wind behavior with depth is crucial for exploring the zonal wind--magnetic field coupling in Uranus. There is evidence that this interaction might be creating secondary magnetic fields through the winding of primary magnetic fields above the dynamo generation region \citep{soyuer2021b}. However, most Uranian dynamo simulations or magnetic field studies \citep[e.g.,][]{stanley2004, stanley2006, soderlund2013} up to now have understandably ignored the peculiar interaction  of zonal winds and the primary magnetic field due to numerical constraints. Understanding the region of rapidly increasing electrical conductivity and the coupling of flows with magnetic field generation might be the key to understanding the non-axisymmetric and multipolar magnetic fields observed on the surface of Uranus \citep{soderlund2020underexplored}. Figure~\ref{fig:winds} illustrates the region of rapidly increasing electrical conductivity of a H-He-H$_2$O mixture in this region of Uranus.

\paragraph{Astrometry}

\subparagraph{Orbital Monitoring of the Moons}

Similar to radio science, astrometry is an integral part of space missions for navigation purposes. In particular, programming a flyby of a solar system object, or controlling the attitude of a probe to orient its instruments towards a specific target, requires a good knowledge of the ephemerides of the celestial bodies to be observed. That is why navigation teams usually include regular astrometric measurements to guarantee mission success. However, astrometric measurements can also be carried out for research purposes. In the case of the Uranus system, precise tracking of the orbital motion of the moons can reveal dynamic characteristics directly associated with some of the physical parameters already mentioned above. In the case of astrometry, the data are usually merged with older astrometric data from various surveys (ground and space), enabling the detection of long-term dynamic features, such as tidal orbital expansion.

\subparagraph{Tidal dissipation}
Astrometric measurements have been successfully used in the past to quantify the tides in Jupiter and Saturn \citep{lainey2009strong,lainey2017new}. This was possible thanks to over a century of ground-based telescopic observations, including measurements made in the late 19th century. This led to the discovery that a simple tidal dissipation model assuming a constant quality factor $Q$ cannot explain the outward motion of Saturn's moons, in particular Titan's (Lainey et al. 2020). Instead, the data favor a resonance locking scenario, implying that the moons' orbital evolution is intimately linked to the evolution of the giant planet (Fuller 2016). Since the Uranus system is much further from Earth, and its moons are still modest in size, ground-based astrometric measurements of sufficient accuracy to quantify dissipative effects began in the 1970s. As a result, the period of useful observations available to constrain tidal effects is significantly shorter, which consequently reinforces the importance of making astrometric measurements with the UOP. Since 2014, the best available measurements have been obtained from Gaia data \citep{tanga2022gaia}. These have a typical accuracy of a few tenths of a millisecond of arc, which is close to what is achievable with an orbiter. Nevertheless, only four of the main moons are observed (Miranda and the inner moons are too close to the glow of Uranus). More importantly, Gaia's measurements will come to an end in 2025. If the amount of energy dissipated by tidal friction inside Uranus matches the upper limit suggested by tidal models, quantification of the orbital expansions of a few major moons should be possible with Gaia. A combination of Voyager, Gaia, and UOP data is necessary to detect lower values of tidal dissipation.

The effect of Uranus's tidal response by means of the Love number $k_2$, on the other hand, is less obvious. In particular, the mutual influence of the tidal bulges acting on the main moons is rather weak. This is why such a measurement should be considered by studying simultaneously the inner and main system. Unfortunately, some of the inner moons are involved in a chaotic configuration, making the success of such measurements uncertain.


\subparagraph{Precession}
Astrometric data is also sensitive to precession. The long timespan of observations definitely helps here. Nevertheless, it is expected that such quantification does not compete with the precession quantification from the observation of the rings themselves.

%% file: sections/subsections/magnetic_field.tex
\section{Magnetic field}\label{sec:magnetic_field}



\subsection{Science Motivation} \label{subsec:motivation}

\begin{figure}
    \centering
    \includegraphics[width = 12cm]{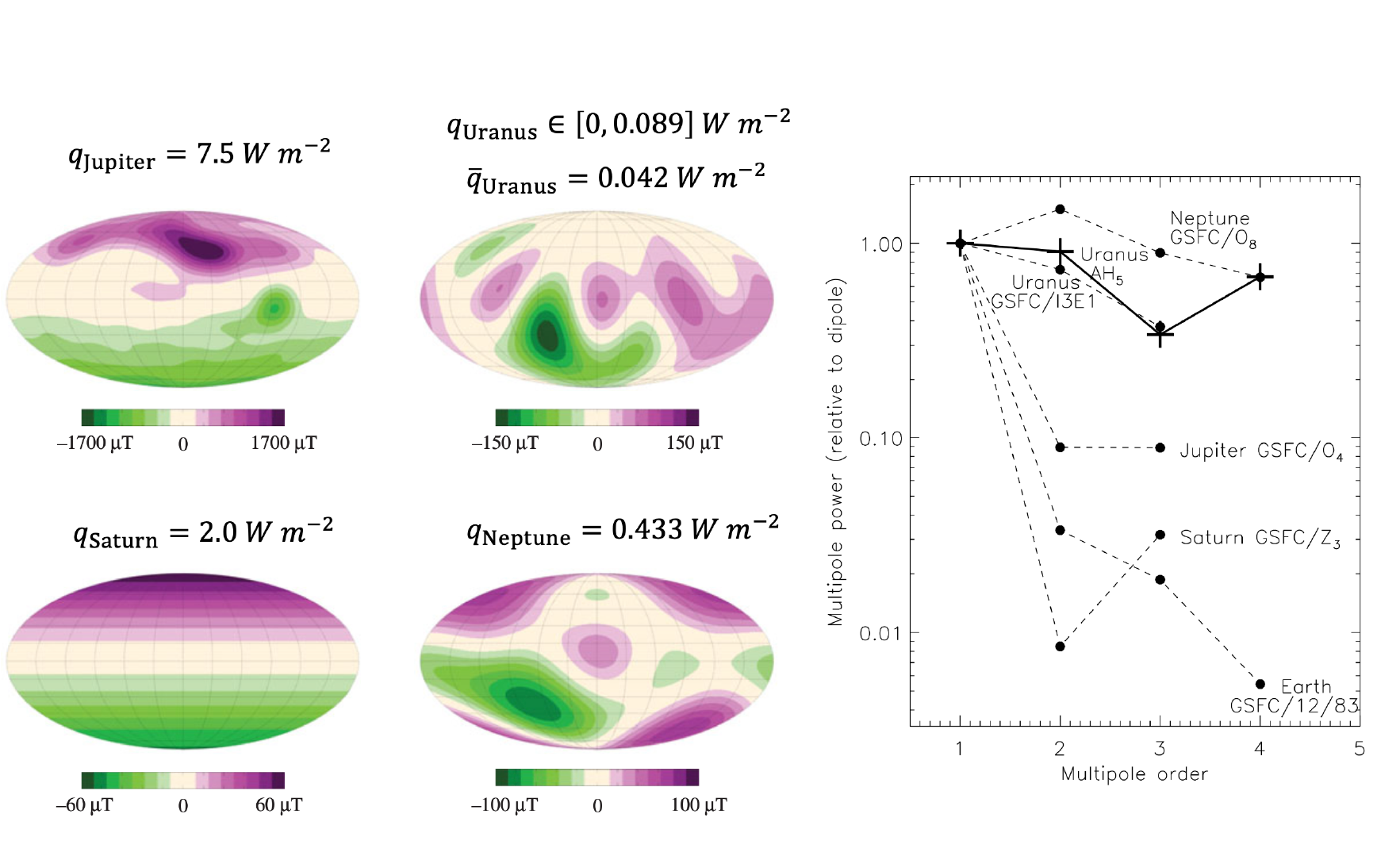}
    \caption{Current knowledge of the internal heat flux, surface magnetic field, and magnetic energy spectra of the giant planets in our solar system. The magnetic fields of Jupiter \citep[][]{Connerney2018JRM9} and Saturn \citep[][]{Cao2020} are dipole-dominant while the magnetic fields of Uranus \citep[]{Herbert2009UranusAurora} and Neptune \citep{Connerney1991Neptune} are multipolar. Internal heat flux estimates taken from \cite{Li2018N,83hanetal,pearl1} and \cite{pearl1}. Uranus's low present-day internal heat flux and the multipolar nature of its magnetic field are challenging to understand.}
    \label{fig: planetary_mag_heat_spectra}
\end{figure}

Planetary magnetic fields offer a critical window into the elusive interior of planets \citep[][]{Stevenson2003}. The maintenance of a strong magnetic field over geological timescales requires a sufficient energy source and a sizable volume of electrically conducting fluid \citep[][]{moffatt_dormy_2019}. All four giant planets in the solar system feature strong planetary-scale magnetic fields (see Fig.~\ref{fig: planetary_mag_heat_spectra}). Furthermore, the magnetic fields of solar system giant planets display a dichotomy: the magnetic fields of Jupiter \citep[][]{Connerney2018JRM9,moore2018Jupiter,Connerney2022JRM33} and Saturn \citep[][]{dougherty2018saturn, Cao2020} are axial-dipole-dominant (at the planetary surface and the expected dynamo surface), while the magnetic fields of Uranus \citep[]{Ness1986Uranus, Connerney1987Uranus, Herbert2009UranusAurora} and Neptune \citep{Ness1989Neptune,Connerney1991Neptune} are multipolar, with no apparent symmetry around any axis when evaluated at the planetary surface or below. 

At present, we do not understand the origin of this magnetic field dichotomy, although a few different hypotheses have been put forward (\citep[see][for a review]{soderlund2020underexplored}). To make progress here, the following ingredients are needed: (1) significantly improved characterization of the magnetic field of Uranus (and Neptune), in particular of their spatial and temporal spectral content given the limitations described below; (2) significantly improved knowledge of the interior structure, including composition and material properties, heat and compositional fluxes, and internal dynamics (e.g., zonal wind-magnetic field interactions); and (3) significantly improved understanding of magnetohydrodynamic (MHD) dynamo theory, in particular in the parameter regimes relevant for solar system giant planets that might be different than for the better-studied geodynamo. These three aspects are coupled: improvements in one could facilitate improvements in the others. 

The magnetic fields of Uranus and Neptune were discovered and characterized during the respective Voyager 2 flybys in 1986 \citep{Ness1986Uranus,Connerney1987Uranus} and in 1989 \citep{Ness1989Neptune, Connerney1991Neptune}. The Voyager 2 flybys remain the only in situ measurements in the Uranus system and Neptune system to date. The Voyager 2 spacecraft went significantly closer to Neptune compared to Uranus: the closest approach distances were 4.2 $R_U$ at Uranus and only 1.18 $R_N$ at Neptune. The highest magnetic field recorded by the Voyager 2 tri-axial fluxgate magnetometer \citep{behannon1977VoyagerMag} was 413 nano-Tesla (nT) during the Uranus flyby and 9923 nT during the Neptune flyby. At distances not too close to the planets (e.g., beyond about 4 planetary radii from the center), their magnetic fields can be conveniently approximated as an offset titled dipole (OTD). In the case of Uranus, that OTD would be offset from the planetary center of mass (CoM) by 0.3 $R_U$, tilted from the spin-axis by 60$^\circ$, and features a dipole moment of 0.23 Gauss $R_U^3$ \citep{Ness1986Uranus}. In the case of Neptune, that OTD would be offset from the planetary center by 0.55 $R_N$, tilted from the spin-axis by 47$^\circ$, and features a dipole moment of 0.133 Gauss $R_N^3$ \citep{Ness1989Neptune}. At distances closer to the planet, the measurements cannot be adequately described by an OTD anymore. When adopting the spherical harmonic (SH) series description centered at the planetary center, these measurements revealed that the magnetic fields of Uranus and Neptune are multipolar: the power in the resolved SH degree $n=2,3$ magnetic field components are comparable to the $n=1$ dipole power, even when evaluated at the planetary surface (Fig.~\ref{fig: planetary_mag_heat_spectra}, rightmost panel). 

Even with the simple, approximate OTD description of their magnetic fields, an interesting dilemma arises: the magnetic field of Uranus appears to be stronger than that of Neptune. This is unexpected because the present-day Uranus seems to feature a much weaker internal heat flux (see section \ref{sec:atmospheric_energy_balance} for more discussion of the observations and interpretation of the heat flux and atmospheric energy balance at Uranus and Neptune). Within the framework of rapidly rotating convective dynamos, \citet{Christensen2010} proposed an energy-based magnetic field scaling law where the dynamo-generated magnetic field strength is proportional to the cubic root of the available energy flux
\begin{equation}
\frac{B^2}{2\mu_0}=c f_{ohm} \overline{\rho}^{1/3} (Fq_o)^{2/3},
\end{equation}
where $B$ is the magnetic field strength in the active dynamo region and $q_o$ is the heat flux at the dynamo surface. Thus, a factor of 2 larger magnetic field strength requires an eight-fold increase in heat flux, all other parameters being held fixed.

At present, we have very limited observational constraints on the magnetic field strength (and morphology) in the active dynamo region inside Uranus and Neptune. It is much more difficult to study their internal dynamos than those of other planetary dynamos for a number of reasons.

\begin{enumerate}
  \item The behavior of the $n>4$ components of their magnetic fields are highly unconstrained at present: they could remain high compared to that of the dipole until a very high $n$ before rapidly dropping off, or they could drop off soon after $n=4$. When combined with the fact that higher SH degree components increase faster when downward extrapolated to the dynamo surface, this could lead to significantly stronger magnetic field strength at the dynamo surface of Uranus and Neptune than current estimates. 
  \item We do not have a good understanding of the depth and thickness of the dynamo layer inside Uranus and Neptune. To estimate the magnetic field strength and morphology at the dynamo surface, we can downward extrapolate the surface field model to the dynamo surface as long as there is no significant electrical currents in between. At Uranus and Neptune, however, we do not know where the dynamo surfaces are. This depends on the bulk composition (e.g., rock- versus ice-giants) and temperature as well as the mixing and separation of different constituents. Furthermore, we do not know whether the dynamos of Uranus (and Neptune) can operate in the double-diffusive convection regime. Very little is known about the efficiency of double-diffusive dynamos and the characteristics of the resultant magnetic fields. 
  \item We have not resolved any time variations in the internal magnetic fields of Uranus and Neptune. Downward extrapolation from externally measured potential fields only works above the dynamo surface, where no significant electrical currents flow. To probe the bulk magnetic field strength within the dynamo layer, detection of MHD dynamo waves such as torsional oscillations (TO) is required. Detection of time variation of the internal magnetic field would also enable an estimation of the flow speed and pattern at the dynamo surface \citep[e.g.,][]{Bloxham2022}, which would constrain the internal energetics and dynamo mechanism. 
  \item The present and future magnetic field and related measurements need to be placed and interpreted in our imperfect, but evolving, theoretical framework of MHD dynamo theory. 
\end{enumerate}


It is therefore clear that future in situ magnetic field measurements with adequate spatial and temporal coverage are imperative for understanding the origin of the unique magnetic field of Uranus (and Neptune) and the closely connected interior state (e.g., rock vs. ice giants, convective vs. stably stratified or double diffusive). A well-resolved spatial power spectrum of the magnetic field (e.g., to SH degree 12 or much higher) could lead to an independent estimate of the depth of the dynamo, as has been done at the Earth and Jupiter. Note that this estimation method relies on dynamo theory for its validity, the case for which is less established for multipolar magnetic fields. Resolving time variation in the internal magnetic field of Uranus could lead to an estimation of the flow speed and pattern inside Uranus near the dynamo surface \citep[e.g., see the study with Juno measurements at Jupiter by][]{Bloxham2022} and further diagnosis of interior properties with MHD dynamo waves. 

Furthermore, the magnetic fields of Uranus and Neptune offer a unique opportunity to study natural multipolar dynamos in greater detail with combinations of in situ and remote observations. These investigations would serve as a benchmark for studying the origin and consequences (e.g., on atmospheric loss) of natural multipolar dynamos outside the solar system, such as those in low-mass stars \citep{reiners2012coolstarmag} and exoplanets \citep{cauley2019magneticHotJupiter,Turner2021ExoplanetRadioEmission,ben2022signatures}. 
Further information on the theoretical calculations relevant for Uranus's magnetic field such as dynamo theory, interior structural models, and thermal evolution models is given in Appendix \ref{subsec:mag_theory}. 




\subsection{Synergies with other spacecraft measurements} \label{subsec:ties}

In order to understand Uranus's interior including the origin of its magnetic field, an iterative and integrative process taking into account the measurements of gravity, atmospheric composition, magnetic field, heat flow, and our improving knowledge of planetary material properties and dynamo processes, is needed. Below, we outline the magnetic field's connection to other spacecraft measurements. 

 \begin{figure}
        \centering
        \includegraphics[width = 11cm]{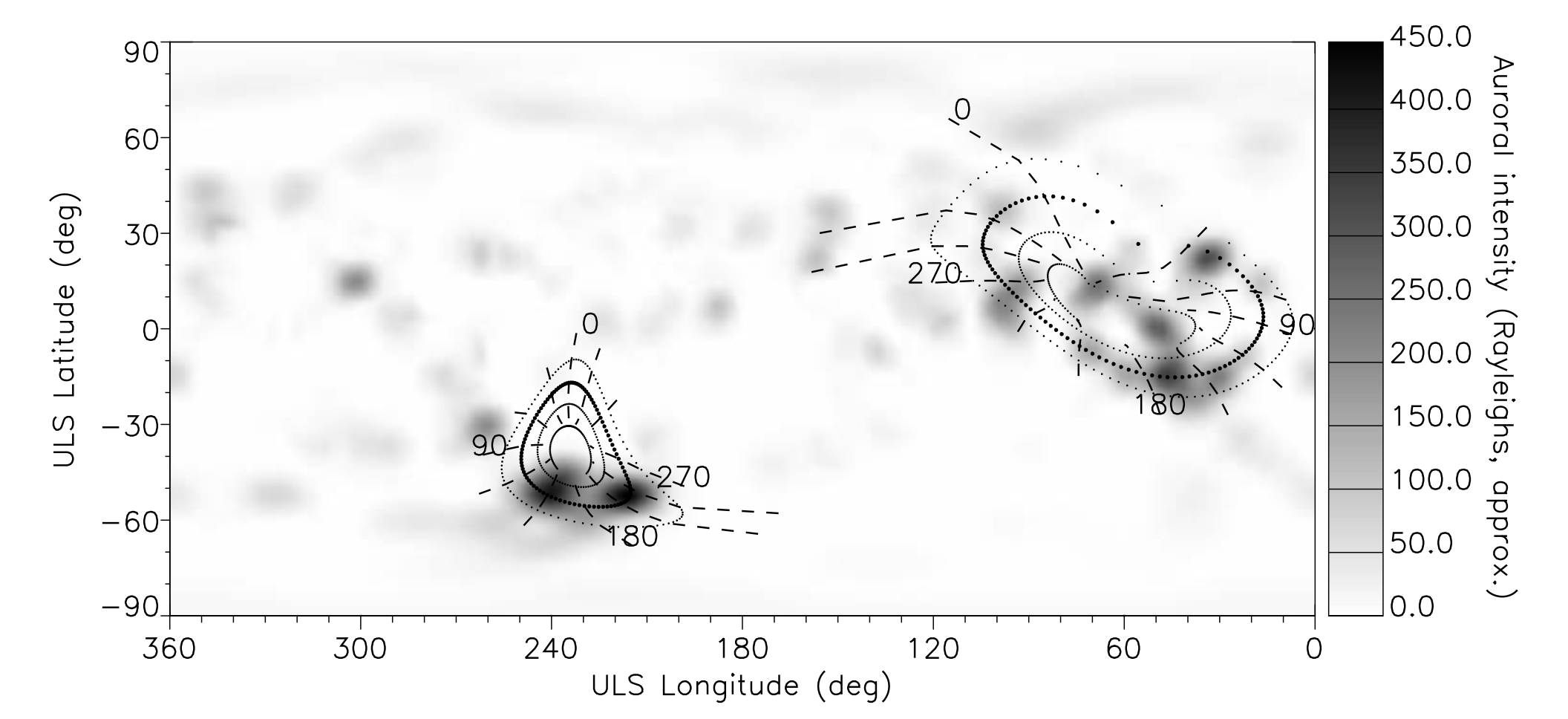}
        \caption{Uranian UV aurora image from Voyager 2 UVS measurements (halftone image) and mapping of magnetic fields from the magnetosphere to the surface of Uranus (dotted and dashed lines) \citep{Herbert2009UranusAurora}. The dotted lines correspond to $L$=3, 5 (large dots), 10, and 20 $R_U$. $L$ is defined as the maximum distance (in $R_U$) along the field line from the center of the offset, tilted dipole (OTD) of Uranus.}
        \label{fig: Uranus_Aurora}
    \end{figure}

\begin{itemize}

    \item {\bf Aurora.} 
    In terms of characterizing the spatial and temporal structure of Uranus's internal magnetic field, imaging of its auroral emission (e.g., the Uranian UV aurora image from Voyager 2 UVS measurements shown in Fig. \ref{fig: Uranus_Aurora}) including the footprints of various satellites in the Uranian system could augment the in situ magnetic field measurements as those auroral emissions trace the geometry of the magnetic field lines. This technique has been applied to the Uranian system with the Voyager 2 measurements \citep{Herbert2009UranusAurora}. Satellite footprints along the planetary magnetic field lines are particularly useful since the locations of the natural satellites are known to high precision. The footprints have been detected at the Jovian system \citep{connerney1998vip4} and Saturnian system \citep{pryor2011EnceladusFootprint} but are yet to be detected at the Uranian system. Complementary long-baseline, Earth-based measurements (e.g., Keck and JWST) to monitor long-term variations in morphology of Uranus's aurora emission could, potentially, place constraints on the secular variation of the dynamo if seasonal effects in the magnetosphere could be distinguished from changes in Uranus's internal magnetic field.
   
    \item {\bf Gravity.} Both the static gravity and ``dynamical" gravity can constrain 1) the physical setup of the planetary dynamo process, 2) the importance of deep atmosphere-dynamo coupling \citep[e.g.,][]{liu2008, CS2017Icarus}, and 3) the pattern and amplitude of flows in the dynamo region \citep[e.g.,][]{Kulowski2020}. As discussed in section \ref{sec:gravity_and_winds}, the interpretation of the static gravity field at Uranus can be highly nonunique. Due to the relatively smooth nature of the surface zonal flows in the latitudinal direction, the separation of the zonal-wind-induced gravity and the solid-body-rotation-induced gravity can be more difficult than for Jupiter and Saturn \citep[see Fig.~5 in][]{wisdom2016winds}. For the nonaxisymmetric ``convective" flows in the dynamo layer, they could in principle lead to correlated signals in the nonaxisymmetric magnetic field and the nonaxisymmetric gravity field. However, more theoretical work is needed to demonstrate this correlation and reveal its quantitative form. 
    
    \item {\bf Seismology.} Seismology at Uranus could provide important observational constraints to a critical uncertainty regarding its interior: the extent and characteristics of stable stratification throughout the planet. Stable stratification can fundamentally alter the heat transfer, material mixing, and magnetic field generation inside Uranus. The prospect of narrowing down this possibility and its bulk property (e.g., the B-V frequency distribution as a function of radius) with seismology is certainly very exciting. 
    
    \item {\bf Energy balance (heat flow).} Heat flow at the dynamo surface is a critical factor controlling the operation of the dynamo inside Uranus. This applies to thermal convection, compositional convection, and double-diffusive convection possibilities.  Long-term monitoring of infrared emission and atmospheric dynamics at Uranus with ever more powerful Earth-based observatories, such as JWST, could help disentangle short-term atmospheric events, seasonal effects, and long-term secular cooling of the bulk interior.
    
    
    
\end{itemize}

\subsubsection{Possible Instrument}
\label{subsec:mag_instruments}

High-precision in situ magnetic field measurements and adequate sampling of the near-surface environment are the two pillars to improve our knowledge of Uranus's magnetic field.  
The two common types of space-based magnetometers are (1) vector fluxgate magnetometers and (2) absolute scalar magnetometers with or without a vector mode. Generally speaking, the fluxgate magnetometer systems are more mature and have more heritage in outer planet explorations. For example, Voyager 1 and 2, Galileo, Cassini, and Juno carried fluxgate magnetometers that functioned well on decadal timescales.  

In addition, the intrinsic noise level of the state-of-the-art fluxgate or absolute scalar magnetometers tends to be lower than required to achieve the objective to significantly improve the characterization of the magnetic field of Uranus. The limiting factors on the engineering side tend to be the magnetic cleanliness of the spacecraft, the rigidity or stability of the boom, and our knowledge of the orientation of the magnetometer axes. For characterizing Uranus's internal magnetic field, the limiting factor on the science side is the presence of complex and time-varying magnetospheric-ionospheric magnetic fields. 

Relevant details about the two common type types of space-based magnetometers include:  

\begin{itemize}
    \item Vector fluxgate magnetometers. Vector fluxgate magnetometers are the most common type of in situ magnetic field measurement instrument for space applications. Nearly all space missions to the outer planetary system carried fluxgate magnetometers with them. For example, Voyager 1 \& 2 \citep{behannon1977VoyagerMag}, Galileo \citep{kivelson1992GalileoMag}, Cassini \citep{Dougherty2004SSR}, Juno \citep{connerney2017junomag}, the upcoming Europa Clipper mission \citep{kivelson2023ECM}, and the recently launched ESA JUICE mission. Fluxgate magnetometers can be designed to measure an extremely wide range of magnetic field amplitudes, from $<$ 0.1 nT to over 10 Gauss (1 G=100, 000 nT). Here at Earth, vector fluxgate magnetometers are also the main instrument for high-precision geomagnetic field measurements onboard Low-Earth-Orbit (LEO) survey satellites such as the ESA SWARM mission \citep{merayo2008swarm}. 
    \item Absolute scalar magnetometers with or without a vector mode. There are benefits to adding an absolute scalar magnetometer to the payload, in addition to the vector fluxgate magnetometers, because of their ability to make extremely accurate magnetic field determination to a precision much better than 0.1 nT. These absolute scalar magnetometers are typically optically pumped and make use of the Zeeman effect. Examples in outer planet explorations include Ulysses, the S/VHM on board Cassini \citep{Dougherty2004SSR}, which operated only in the first few years, and the SCA onboard the ESA JUICE mission. For near-Earth space environments, absolute scalar magnetometers have been used to calibrate the vector helium magnetometer onboard the ESA SWARM mission \citep{merayo2008swarm, fratter2016SwarmAsm} and the Chinese Seismo-Electromagnetic Satellite \citep{pollinger2018magnetometer}.
\end{itemize}

%% file: sections/subsections/atmospheric_energy_balance.tex
\section{Atmospheric Energy Balance}\label{sec:atmospheric_energy_balance}



\subsection{Science Motivation}

%
%


The emitted power from a planet can be from a wide range of sources including heat left over from formation and differentiation, latent heat due to condensation of constituents, radioactive decay, and insolation.
For example, gravitational energy release, and to a lesser extent latent heat release from helium condensation, could be a heat source in the interiors of Jupiter and Saturn, and the work done to transport heavier material upward could be an energy sink. Helium differentiation is not expected to be important in Uranus and Neptune, but there might be other phase transitions present or an even stronger effect arising from redistribution of heavy material. From the first law of thermodynamics, a hotter interior of Uranus today (e.g., arising from the difficulty of removing heat by convection in the presence of compositional gradients) could mean a lower heat flow. The primordial thermal and compositional states are not known, and both can be important for the current energy balance. Planet formation models predict very hot internal temperatures, but the exact profile is clearly model-dependent. In addition, giant impacts could also play a role in changing the thermal state of the planets shortly after their formation \citep[e.g.,][]{Reinhardt2020,Valletta2022}.  

In any case, the emitted power contains important and unique information about the internal composition, heat transfer, layering, and structure.
Often this is simplified into the ratio of emitted flux to received insolation, referred to as the ``energy balance,'' where a ratio larger than unity indicates an excess emission and implies an internal heat source.

For Uranus, the emitted power has been estimated using Voyager 2 IRIS measurement at 0.69$\pm$0.01~Wm$^{-2}$ corresponding to an effective temperature of 59.1$\pm$0.3~K \citep{pearl1}.
However, the IRIS observations cover only a limited spectral region (180--400~cm$^{-1}$) that does not fully cover the peak emission from Uranus, such that 75\% of the emitted energy must be estimated from a spectral model \citep{pearl1}.
The spatial and emission angle coverage was also incomplete, leading to further uncertainty.

The Voyager 2 IRIS instrument also included a broadband spectrometer covering 0.3--1.8~$\mu$m that could be used to estimate Uranus's reflected solar competent and a Bond Albedo of 0.30$\pm$0.05 \citep{pearl1}.
This suggests a radiative equilibrium temperature of 58.2$\pm$1.0~K.
However, model fitting was also required to derive the albedo, because phase angle and geographical coverage were incomplete.
Therefore, Uranus's implied energy balance of 1.06$\pm$0.08 is only marginally above unity and has significant error.
The excess emission is thus a modest 40$\pm$55~mWm$^{-2}$.

For Neptune, the Voyager 2 measurements suggest a similar emitted flux to Uranus at 0.70$\pm$0.01~Wm$^{-2}$ with an effective temperature of 59.3$\pm$0.8~K \citep{91peacon}.
The albedo of Neptune is also similar within error at 0.29$\pm$0.07, but due to Neptune's great distance, the radiative equilibrium temperature is only 46.6$\pm$1.1~K and the energy balance ratio is 2.61$\pm$0.28, suggesting an excess emission of 430$\pm$75~mWm$^{-2}$. At first glance Neptune seems to have a much more significant internal heat source than Uranus. 

We are then left with the rather puzzling result that there are two broadly similar ice giant planets, with essentially the same emitted flux and albedo, but Uranus is close to being in balance with solar input, whereas Neptune has significant excess emission.
There are a number of possible explanations for this.

First, Voyager 2 might have witnessed Uranus during a quiescent period (or conversely, observed Neptune during an  active period).
Long-term comparisons of Saturn from Voyager \citep{83hanetal} and Cassini \citep{10lietal}, [Wang et al. 2024, submitted to Nature Com.] indicate that the emitted flux can vary significantly on seasonal scales. Similar variations have also been observed for Jupiter \citep{12lietal}.
Convective storms can also alter the energy balance by significant amounts. On Saturn, storms with an approximately 20 year cadence occur that can have global influence \citep{15lietal} (+Wang et al. 2024). Uranus's $98^\circ$ tilt yields the most dramatic seasonal effects of any planet in the solar system, with likely consequences on the planet's global output. 

Second, Uranus's low excess heat flux could be due to nonadiabatic stably stratified internal layering (e.g., Figure~\ref{fig:ev_models}) that prevents internal heat escaping through convection. However, if the dynamo is thermally driven, there would still need to be a nonzero heat flux across the field-generating layer, although this could be quite small. Compositionally driven dynamos would not have such a requirement. Double-diffusive convection, where thermal and compositional gradients oppose each other and diffuse at different rates, is a particularly important aspect to consider in this context (see also section~\ref{sec:magnetic_field} on magnetic fields). 

Third, a giant impact sometime in Uranus's past could have created a shock wave that propagated through the planet and stirred up the interior, thus allowing the bulk of any residual heat of formation to escape.
Such an impact could also explain Uranus's high obliquity \citep[e.g.,][]{Reinhardt2020}.

Another possibility is that the available albedo measurement is incorrect. 
It was recently suggested that the albedo used for Jupiter and Saturn for decades is inappropriate \citep{Li2018N}(+Wang et al. 2024). This might also be the case for Uranus, which in turn influences Uranus's expected energy budget and internal heat. It is therefore important to have reliable measurements of Uranus's albedo. 



Finally, other process can also affect Uranus's energy budget. These include different amounts of radiogenic heating due to different rock contents and latent heat release associated with the formation of superionic ice. 
It is important to identify all the factors and processes that can influence our assessment of Uranus's energy budget today and therefore its cooling history. 

A key question in all of this is how close Uranus's excess emission is to to 0~Wm$^{-2}$.
Voyager 2 measurements are limited in time, wavelength, spatial coverage, emission angle coverage, and phase angle coverage, which makes the estimated emitted and absorbed fluxes uncertain.
A neutral energy balance cannot be ruled out with these current observations. New, more accurate constraints are an important goal of any new mission.

\subsection{Observations required / remote sensing}

The critical measurements for determining accurate energy balance are the emitted flux and the Bond albedo. These measurements are not trivial, as they have stringent instrument and orbital requirements.

The emitted flux requires integrating the infrared flux over all wavelengths and emission angles.
For a non-spatially-uniform planet with extreme seasonal variations such as Uranus, it also requires coverage at all latitudes and over a very long time.
It is especially important to cover the peak in thermal emission fully,
and ideally the measurement should cover 10--1000~$\mu m$ in the mid- and far-IR ranges.
This requires measurements much further into the far-IR than was possible with Voyager 2's IRIS spectrometer and requires an extended range such as Cassini's CIRS instrument \citep{17jenetal} or a simpler a wide-band bolometer or net flux radiometer. Achieving the coverage at all latitudes is possible only from an orbiter. The coverage of seasonal variations can be achieved through regular observations from the ground or from a space telescope in the inner solar system. 

The Bond albedo is defined as the fraction of solar radiation reflected over all wavelengths and directions.
This again requires measurements covering a wide range of emission angles, phase angles, and latitudes at all wavelengths over the peak of the solar radiation (0.2--5~$\mu m$).
Voyager 2's IRIS broadband spectrometer covers most of this range, but is limited in phase angle coverage, which limits the accuracy possible.

Both emitted flux and albedo measurements require an orbit that has polar coverage and covers day and night sites.
Long-term monitoring is also beneficial to determine seasonal dependence and to measure the effect of transient clouds or storms.

In the context of atmospheric dynamics, mapping with optical and Doppler imagers is beneficial to recognize updrafts/downdrafts and better understand possible seasonal variations in heat flux and energy transport.



Understanding the difference between the energy balance on Uranus and Neptune requires knowledge of both planets' internal structure and composition, as well as the stability of any internal layers. 
This requires detailed knowledge of the heat transport mechanisms in planetary interiors, especially in the presence of composition gradients. From studies of Jupiter and Saturn, we know that they are likely not fully convective today \citep[e.g.,][]{wahl2017comparing,nettelmann2021theory,IdiniSteve:2022,miguel2022jupiter,MankovichFuller2021,militzer2022juno,howard2023jupiter}. However, formation-evolution models of Jupiter suggest that it is difficult to sustain composition gradients over billions of years against large-scale convection \citep[e.g.,][]{2020A&A...638A.121M}. One potential solution is that energy transport in inhomogeneous regions could be through oscillatory double-diffusive convection (semi-convection), which was suggested to provide stability due to the more efficient heat transport \citep{LC}. Numerical simulations, however, suggest that semiconvective layers are transient, since chemical transport is efficient and tends to erase composition gradients \citep[e.g.,][]{2013ApJ...768..157W}, or are difficult to form at low Prandtl numbers \citep{2022PhRvF...7l4501F}. In general, semi-convection is still poorly understood for planetary conditions \citep[e.g.,][]{2018AnRFM..50..275G}, and therefore further studies are required to constrain the heat transport and mixing processes in Uranus and Neptune. These mixing processes will additionally impact dynamo processes, which also require further study. 


The material properties (e.g., equation of state, thermal and electrical conductivity, transport properties) also influence the transport of heat from the deep interior in the planet, which in turn is determined by the balance between convection and conduction and the evolution of the planet over a timescale of 10$^9$ years \citep{VazanHelle:2020,scheibe2019thermal}. This is discussed in more detail in Appendix \ref{section.seismology.oscillation.modes}. 








\subsection{Synergies with other measurements}

Energy balance depends on many competing factors, so there are many synergistic measurements.

First, we need internal structure to interpret measurements properly, especially regarding internal layering. Seismology would be most helpful, as this would provide powerful information about layering, but it is currently unclear whether the measurements are practical (see Section~\ref{sec:seismology}). 

Second, measuring Uranus's magnetic field will also provide constraints on structure and on the planetary thermal state. For example, the stability and presence of fluid layers and modes of convection affects both the field intensity and the spectral distribution of its magnetic energy (see Section~\ref{sec:magnetic_field}). 

Third, measuring the gravity field will provide some interior constraints as described in Section~\ref{sec:gravity_and_winds}, but not much information on layering due to the width of its sensitivity kernels. J$_2$ and J$_4$ are estimated from Voyager reanalysis, and they are the ones that probe the deepest. Nevertheless, more accurate determination of the low-order harmonics is needed. In addition, although higher order gravity moments do not constrain the deep  interior directly, they are important for determining the depth of the winds in Uranus as well as its rotation rate, and therefore eventually also the deep interior \citep{Neuenschwander2022}. 
Furthermore, if the magnetic field is generated rather shallow, the higher moments could provide valuable constraints to Uranus's interior. 

Fourth, as discussed in section~\ref{sec:noble_gasses_and_isotopes}, the detection of noble gases, for example $^{40}$Ar, would indicate that there is some direct material communication with the deep interior. When that communication occurred and how complete it was, however, might not be obvious. Future studies should investigate how to address those aspects of the communication. 

Fifth, temperature profile measurements in the upper atmosphere at microwave and far-IR wavelengths can provide information on heat transport and mechanisms for emitting energy to space (Section~\ref{sec:atmospheric_pressure_temperature_profile}). Doppler wind circulation cell measurements are also relevant to this, as they can provide a more direct measure of flow and advection of heat from the interior than can measurements of the temperature profile.

Finally, the study of exoplanets can provide wider insight. If a statistical study of ice-giant-sized exoplanets is performed to determine how many have anomalously low emissions, perhaps with the James Webb Space Telescope (JWST), we will learn how common it is for ice giants to have some property that limits the radiation of internal energy to space.

%
%
%
%
%




%% file: sections/subsections/seismology.tex
\section{Seismology}\label{sec:seismology}

Seismology is a powerful tool for probing the interiors of giant planets, providing information complementary to the traditional approach of gravity science. Background information on the technique can be found in  Appendix \ref{section.seismology.oscillation.modes}.  Detecting seismic oscillations, however, can be difficult because their amplitudes can be very small, and since we cannot easily place a seismograph on gaseous planets, remote sensing techniques are necessary. One possibility is to measure the radial velocity of the planet's surface using high-resolution spectrographs (Doppler imaging [DI]; see Appendix \ref{section.seismology.detection.doppler}). This is how the Sun's oscillations have been detected, which produce radial velocity variations of $\sim \! 1 \, {\rm m/s}$ on timescales of $\sim$5 minutes. There have been several efforts to use this technique to detect Jovian oscillation modes from the Earth  \citep{Schmider1991,Gaulme2011,Shaw2022,gulledge_phd} with conflicting results. Uranus is too faint to attempt an Earth-based detection with a DI, even with a 10 meter telescope, but one could consider flying a DI on the UOP spacecraft, since at close range relatively low amplitude oscillations could be detected.  
Despite the potential of a DI for seismology, justifying a DI at the expense of alternative payload options would require additional research. Either confident constraints must be placed on Uranian seismic amplitudes, or a sufficiently compelling science case must be made for additional uses for a DI, for example three dimensional wind mapping \citep[see Section~\ref{sec:atmospheric_energy_balance} for further discussion and][for an example of how a DI has been applied in this way to Jupiter]{schmider++2023}. The theory of seismic excitation on giant planets is poorly understood \citep[e.g.,][]{markham-stevenson2018}, and is unlikely to become predictive without unambiguous observations. If it is possible to place useful constraints on Uranian seismic amplitudes using Earth-based observations, such a possibility has not yet been studied in detail. We itemize specific recommendations to address these knowledge gaps in Section~\ref{sec:recommendations}. 

Another possibility is to indirectly detect global oscillation modes through their affect on planetary rings. The gravity variations produced by planetary oscillation modes excite density and bending waves at Lindblad resonances with orbiting ring particles, which can be detected in images of the ring or through stellar occultation measurement \citep{Marley1991,MarleyPorco1993}. This technique has  been successful at Saturn, with a series of Saturnian oscillation mode detections \citep{Hedman2013,Hedman-Nicholson-2014,French2016,French2019KronoIII,Hedman2019,2021Icar..37014660F}, which have allowed for precise measurements of the oscillation mode frequencies and azimuthal pattern numbers $m$ of $>$30 of Saturn's oscillation modes. Seismic modeling has utilized this information to place novel constraints on Saturn's interior \citep{Fuller2014}, finding the planet must have a ``diffuse" core characterized by a gradual transition from a gaseous envelope to an icy or rocky core \citep{MankovichFuller2021}. \citet{2022PSJ.....3..194A} provide a recent assessment of ring seismology at Uranus, and \citet{FRENCH+2024} report detection of a density wave in Uranus's gamma-ring that does not appear to be due to satellite forcing. Given that the study of waves in the rings has proven value for probing the rings themselves and satellite densities \citep{chancia+2017}, and that detection of such waves can be achieved without specialized instrumentation, it is likely that the UOP mission will be able to search for normal modes within the rings. The prospects of this technique for Uranus are discussed further in Appendix~\ref{section.seismology.detection.rings}. 

A third possibility is to detect the gravity variations produced by global scale planetary oscillation modes, for example, through their affect on the trajectory of an orbiting spacecraft. This equates to measuring time variations of the gravity harmonics $C_{\ell m}$, and is discussed in Appendix~\ref{sec:gravity_and_winds}. These affects appear to have been detected at Jupiter \citep{durante2022juno} and Saturn \citep{Markham2020} through residuals in Doppler tracking data of the \textit{Juno} and \textit{Cassini} satellites, but the very eccentric orbits of those spacecraft have, so far, prevented identification of individual mode frequencies. Without that frequency information, very little can be learned about the interior structure. More circular orbits might be suitable for identifying normal modes \citep{friedson2020ice}, but the amount of fuel required to get a spacecraft into such an orbit is prohibitive unless we know in advance that detectable oscillations are present.

The central issue with all seismological techniques is detectability at Uranus. There is clear evidence of detection at Saturn using the rings and gravity, and less but likely evidence for detectability in gravity at Jupiter. The current lack of understanding for their excitation prevents any confident estimate of what amplitudes might be expected at Uranus. The various detection techniques on a Uranus mission including their strengths and weaknesses are summarized in 
Table \ref{tab:seismology-procon-table}. Clearly, each of these methods is sensitive to different parts of the planetary spectrum. Additional background information on seismology can be found in Appendix~\ref{section.seismology.detection.rings}.


\begin{sidewaystable}
\hspace{5cm}
\caption{Comparison of various seismic detection techniques on a Uranus mission.}
\begin{center}
\begin{tabular}{ |p{.15\textwidth}|p{.4\textwidth}|p{.4\textwidth}| }
\hline
 Technique & Mission considerations & Science impact \\ \hline
\vspace{-10pt}
\begin{flushleft}
\hyphenpenalty=10000
\exhyphenpenalty=10000
Doppler Imager
\end{flushleft}  & 
\begin{minipage}{.4\textwidth}
\textbf{Pros:}
\vspace{-5pt}
    \begin{itemize}
        \item May be integrated with a standard imaging system.
        \vspace{-4pt}
        \item Unique opportunity for 3D wind mapping, relevant to heat transport and atmospheric dynamics.
    \end{itemize}
\vspace{-5pt}
\textbf{Cons:}
\vspace{-5pt}
    \begin{itemize}
        \item Significant additional payload.
        \vspace{-4pt}
        \item Risk of non-detection if seismic amplitudes are insufficient.
    \end{itemize}
\vspace{3pt}
\end{minipage} & 
\begin{minipage}{.4\textwidth}
\textbf{Pros:}
\vspace{-5pt}
    \begin{itemize}
        \item Largest range of spectral coverage.
        \vspace{-4pt}
        \item Tight constraints on eigenmode frequencies.
        \vspace{-4pt}
        \item Potential to construct \'{E}chelle diagrams (Fig.~\ref{fig.uranus_interface_pmodes}).
    \end{itemize}
\vspace{-5pt}
\textbf{Cons:}
\vspace{-5pt}
    \begin{itemize}
        \item Unknown a priori seismic amplitudes.
        \vspace{-4pt}
        \item Less sensitive to lower frequency (f and g) modes.
    \end{itemize}
\end{minipage}
\\
\hline
\vspace{-10pt}
\begin{flushleft}
\hyphenpenalty=10000
\exhyphenpenalty=10000
Ring Seismology
\end{flushleft} &
\begin{minipage}{.4\textwidth}
\vspace{3pt}
\textbf{Pros:}
\vspace{-5pt}
    \begin{itemize}
        \item Requires no specialized instruments beyond an optical camera.
        \vspace{-4pt}
        \item Synergy with other ring science.
    \end{itemize}
\vspace{-5pt}
\textbf{Cons:}
\vspace{-5pt}
    \begin{itemize}
        \item Risk of non-detection if mode resonances happen not to fall near narrowly confined rings.
    \end{itemize}
\vspace{3pt}
\end{minipage} & 
\begin{minipage}{.4\textwidth}
\textbf{Pros:}
\vspace{-5pt}
    \begin{itemize}
        \item Tight constraints on eigenmode frequencies.
        \vspace{-3pt}
        \item Potentially sensitive to g modes and hence composition gradients.
    \end{itemize}
\vspace{-5pt}
\textbf{Cons:}
\vspace{-5pt}
    \begin{itemize}
        \item No sensitivity to p modes (too high frequency).
        \vspace{-4pt}
        \item Lack of continuous spectral coverage.
    \end{itemize}
\end{minipage}
\\
\hline
\vspace{-10pt}
\begin{flushleft}
\hyphenpenalty=10000
\exhyphenpenalty=10000
Gravitational Seismology
\end{flushleft} &
\begin{minipage}{.4\textwidth}
\vspace{3pt}
\textbf{Pros:}
\vspace{-5pt}
    \begin{itemize}
        \item Utilizes the spacecraft telecom system without additional payload.
    \end{itemize}
\vspace{-5pt}
\textbf{Cons:}
\vspace{-5pt}
    \begin{itemize}
        \item Ideal orbital configuration is likely expensive in fuel.
        \vspace{-4pt}
        \item Risk of non-detection if seismic amplitudes are insufficient.
    \end{itemize}
\vspace{3pt}
\end{minipage} & 
\begin{minipage}{.4\textwidth}
\textbf{Pros:}
\vspace{-5pt}
    \begin{itemize}
        \item Can be sensitive to a wide range of frequencies, from g and f modes to p modes.
    \end{itemize}
\vspace{-5pt}
\textbf{Cons:}
\vspace{-5pt}
    \begin{itemize}
        \item Deconvolution of gravitational signal into a high-confidence power spectrum is not straightforward.
        \vspace{-4pt}
        \item Difficult to yield precise eigenfrequencies with a high eccentricity spacecraft orbit.
    \end{itemize}
\end{minipage}
\\
\hline
\end{tabular}
\end{center}
\label{tab:seismology-procon-table}

\end{sidewaystable}

%% file: sections/recommendations.tex
\newpage
\section{Recommendations}\label{sec:recommendations}

Given how little we know about Uranus's interior, a broad set of measurements from within the Uranus system as well as Earth-based work is needed to resolve the open questions about its interior structure and composition.  The key relevant questions presented in the OWL Decadal Survey and its UOP mission study are:
\begin{itemize}
\item When, where, and how did Uranus form and evolve, both thermally and spatially?
\item What is the bulk composition of Uranus and its depth dependence?
\item Does Uranus have discrete layers or a fuzzy core?
\item What is the true rotation rate of Uranus, does it rotate uniformly, and how deep are the winds?
\item What dynamo process produces Uranus's complex magnetic field?
\item What mechanisms are transporting heat / energy in the planet today?
\end{itemize}
Collectively, these questions are getting at the higher-level goal of understanding how our solar system and exoplanetary systems form and evolve.  In this section we summarize the recommendations and requirements that were presented in Sections 2 through 6.  We divide them into three categories:  measurement requirements and mission requirements (Section 7.1), and knowledge gaps to be filled (Section 7.2).

\subsection{Measurement Requirements and Associated Mission Requirements}
For each measurement, we have attempted to provide some guidance on how accurately the measurement must be made or the spatial and temporal sampling required.  Those values are only guides and should not be treated as hard limits.  Our lack of knowledge about Uranus's interior is so profound that doing the best that can be practically achieved will in almost all cases be sufficient to revolutionize our understanding of Uranus. The following bullets summarize the critical measurements to make as well as the suggested mission requirement (MR).\\  

* The abundance relative to hydrogen of all noble gases.  Also, all noble gas stable isotopic ratios.  In each case, abundances should be measured with sufficient accuracy to distinguish variations of one solar abundance (i.e., distinguish a solar abundance from two solar abundances).\\
{\it MR: An atmospheric entry probe carrying a mass spectrometer to pressures of at least 2 bars, and ideally to 10 bars}\\


*  The isotopic ratios D/H, $^{15}$N/$^{14}$N, $^{13}$C/$^{12}$C, $^{18}$O/$^{16}$O, $^{17}$O/$^{16}$O, each to 10\%, at pressures $\geq$~2~bars, but ideally to 10 bars. Note that measuring the nitrogen and oxygen isotopes may be problematic due to the relatively low abundances of species carrying those elements in the upper troposphere; distinguishing the bulk Uranus ratios from that of possible exogenic sources may require information on the ratio vertical profiles. \\
\textit{MR:  An atmospheric entry probe carrying a mass spectrometer or tunable laser spectrometer to pressures of at least 2 bars but ideally to 10 bars.}\\

*   The spherical harmonics of the gravity field to J8, measured to 1 part in 10$^8$. \\
\textit{MR: Polar orbiting spacecraft with $\sim$10 periapse passes within 2,500 km of the 1 bar level while tracking the spacecraft around periapse in two-way coherent mode. Periapses near the equator are highest priority, but it is desired to sample as wide a range of latitudes as possible.}\\
Supporting measurements not identified elsewhere are measuring Uranus's moment of inertia, spin axis precession, and zonal wind speeds and their depth.\\

*    The interior rotation rate of the planet to within $\sim$~10 minutes.\\
\textit{MR:  Determine the shape of the planet via dual-band radio occultations that sample $\sim$~5 latitudes; Determine periodic variations in the planetary magnetic field measured separately in the Northern and Southern Hemispheres; Measure periodic variations in radio and plasma waves tied to interior processes.}\\

*    Tidal love number of Uranus, k2, raised in response to Ariel's orbit to 5\%.\\
\textit{MR:  At least $\sim$~5 near-equatorial periapse passes within $\sim$~2,500~km of the 1-bar level that sample a range of angles from 0 to 360 degrees from the Uranus-Ariel line.}\\
Supporting measurements not mentioned elsewhere are to search for tidal evolution of Ariel's and other satellite orbits via astrometric measurements.\\

*    Magnetic field to spherical harmonic degree~12.\\
\textit{MR:  Polar orbits with peripase within 2,500~km of the 1-bar level, sampling as large as possible of a range of rotational and magnetic latitudes and longitudes.}\\
Supporting measurements not mentioned elsewhere are observations of auroral footprints via UV, visible, and IR imaging.\\

*   Atmospheric energy balance\\
This measurement has two components:
\begin{itemize}
\item Total outgoing IR flux to $\sim$ 3\% (preferably with some spectral resolution) at wavelengths from 10 to 1000 $\mu$m, sampling all emission angles, solar phase angles, and latitudes with $\sim$ 15$^\circ$ resolution in latitude/longitude.
\item Visible reflectance to $\sim$3\% (preferably with some spectral resolution) at wavelengths from 0.2 to 5 $\mu$m, sampling emission angles, solar phase angles, and latitudes with $\sim$ 15$^\circ$ latitude resolution.
\end{itemize}
\textit{MR: Polar orbits with ground tracks spaced $\sim$1 hour apart in local time of day.}\\
Supporting measurements not mentioned elsewhere are to map atmospheric up and down drafts via the abundance of trace species, the hydrogen ortho-para ratio, or cloud morphology, and to make atmospheric energy balance measurements over decadal time scales.\\

*   Atmospheric temperature-pressure profile.\\
Temperature to ±1 K, Pressure to $\sim$ 5\%, both measured from the tropopause to pressures $\geq$~5 bars.\\
\textit{MR:  Atmospheric entry probe with an atmospheric structure instrument to provide ground-truth at a single location representative of conditions within 30$^\circ$ of the equator and sampling pressures from $\sim$~0.5 to $\geq$~5 bars; An IR spectrometer to provide global coverage (including the probe entry location) of T-P profiles at pressures $\leq \sim$1~bar;  Dual-band radio occultation experiment sampling as many latitudes as possible; Microwave radiometer measurements at wavelengths from $\sim$1~cm to $\sim$1~meter, providing limb-darkening measurements for global temperature retrievals (including the probe entry location) at pressures $\geq$~5 bars with $\sim$5~degree resolution in latitude and longitude (but note that validation of this technique also appears in the knowledge gaps section).}\\
Supporting measurements not mentioned elsewhere are:  Earth-based visible and near-IR spectroscopy and microwave radiometry to provide seasonal context for the spacecraft measurements; H$_2$O, CH$_4$, NH$_3$, H$_2$S, CO, CO$_2$, PH$_3$, C$_2$H$_6$, N$_2$, and GeH$_4$ abundances and the hydrogen ortho-para ratio with $\sim$5$^\circ$ longitude and latitude spatial resolution.\\

*   Search for density and bending waves in the rings indicative of normal mode oscillations in the planet.\\
\textit{MR:  Stellar occultations with fast sampling at VIS, IR, and/or UV wavelengths across all the rings;  Radio occultations across the rings if the spacecraft/Earth geometry allows.}\\
(See Knowledge Gaps regarding alternative approaches for detecting normal modes.)\\

\subsection{Knowledge Gaps}
In this section we identify knowledge gaps which, if addressed, would likely enable critical science discoveries at Uranus from the above measurements.  We focus on efforts that are achievable with Earth-based studies in the near term.\\

\noindent Regarding interpreting measurements of noble gases, the following knowledge gaps were identified:
\begin{itemize}
\item Solubility and mixing properties of $^{40}$Ar under Uranus interior conditions.
\item Partitioning of K in a potential iron core.
\item The statistics of Na abundance in the envelopes of ice giant-sized exoplanets (used for determining how commonly envelopes are polluted with heavy elements during accretion).
\item Theoretical predictions of Uranus envelope pollution during accretion.
\end{itemize}

\noindent To advance our ability to interpret gravity field measurements, knowledge gaps include:
\begin{itemize}
\item Melting curves and solubility of superionic ice and its possible alloys and of the O-H system.
\item Excitation mechanisms for normal-mode oscillations in giant planets.
\item Analysis of which normal modes are most likely to complement gravity measurements when trying to determine interior structure.
\item A better theoretical understanding of magnetic coupling between the atmosphere and dynamo region.
\end{itemize}

\noindent Our ability to interpret magnetic field measurements would be enhanced by:
\begin{itemize}
\item A better understanding of MHD theory, in particular in modeling the strength, morphology, and secular variations of multi-polar fields, double-diffusive dynamos, and dynamo waves.  
\item Improved knowledge of the electrical conductivity, viscosity, and thermal/compositional diffusivities of possible interior compositions.
\end{itemize}

\noindent To better understand the implications for planetary evolution of the current atmospheric energy balance, it is desired to know:
\begin{itemize}
\item The thermal conductivity of possible interior materials.
\item The likelihood of internal layering or stable regions within an ice giant. This can be determined from the statistics of exoplanets emitting very low amounts of internal energy.
\end{itemize}

\noindent To better determine the atmospheric temperature-pressure profile, Earth-based efforts could include:
\begin{itemize}
\item Determine the microwave opacity of H$_2$S at high temperatures ($\sim$ 500~K) and pressures (100's of bars).  This knowledge improves our ability to interpret microwave radiometer measurements.
\item Assess the ability of microwave limb-darkening to accurately estimate temperature profiles at depth.
\end{itemize}

\noindent Finally, the field of normal-mode seismology is currently a high-risk, high-reward area. To reduce the risk, Earth-based studies in the near term could include:
\begin{itemize}
\item Quantify the value of detecting \textit{p}-, \textit{f}-, and \textit{g}-modes given various levels of knowledge of the gravity field.
\item Quantify the ability of techniques other than ring seismology to detect normal modes (e.g., from gravitational perturbations to spacecraft or natural satellites, or a Doppler Imager instrument).
\item Advance our understanding of normal-mode excitation mechanisms on giant planets.
\item Detect and characterize normal-modes on Jupiter or Saturn with Earth-based instruments.
\end{itemize}

%% file: sections/summary.tex
\newpage
\section{Summary}\label{sec:summary}
Albert Einstein once said {\it ``the important thing is not to stop questioning"}. In this report 
we summarize the current knowledge of Uranus's internal structure including the many key open questions. A better understanding of Uranus's internal structure requires a dedicated Uranus mission performing a range of complementary measurements as well as progress in numerical simulations, lab experiments, and ground-based observations. 
The key points from this study can be summarized as follows:
\subsection*{Uranus's atmosphere}
\subsubsection*{(i) Noble gases}
\begin{itemize}
    \item Radiogenic $^{40}$Ar can be used as a proxy for the abundance of rocky material within Uranus and the extent of interior-atmosphere communication, but must be combined with other indicators to eliminate ambiguities between the two factors. 
    \item All the noble gas elemental ratios should be measured as they provide distinct types of information; the lightest (He, Ne) on deep envelope processes, the others on the composition of the solids that delivered the heavy elements to Uranus. 
\end{itemize}

\subsubsection*{(ii) T-P Profiles \& Condensible and Disequilibrium Species}
    \begin{itemize}
    \item A direct in situ measurement of pressure and temperature down to $\sim$5 bars in a representative region of Uranus will serve as a reference for all future studies.
    \item Remote sensing measurements of the probe entry region with radio occultations and microwave radiometry will be crucial to lift degeneracies in remote sensing retrievals and to provide context for the probe measurements. This has never been done in any planet. 
    \item Mapping of the atmosphere in latitude and longitude with visible and IR spectroscopy and with microwave radiometry will constrain atmospheric properties (e.g., the abundance of condensible and disequilibrium species, temperature) from the stratosphere down to deep tropospheric levels.  
 \end{itemize}

  
\subsection*{Gravity and winds}
\begin{itemize} 
\item Measuring Uranus's gravitational field (including J$_6$ and J$_8$) is crucial for constraining its composition and internal structure. Measuring the zonal harmonics are also important to study atmospheric dynamics and constrain its rotation period, shape, and depth of the winds. 
 \item An accurate measurement of $k_2$ (within $\sim$ 5\%) can be used to further constrain Uranus's interior. 
\end{itemize}

\subsection*{Magnetic field}
\begin{itemize}
 \item Measuring Uranus's magnetic field spatially and temporally (to $\sim$ degree 12) will provide key and complementary constraints on its interior structure and dynamics. 
 
\end{itemize}

\subsection*{Energy balance} 
\begin{itemize}
    \item It is required to determine Uranus's energy balance and therefore heat flux (and its time variation) using a wide range of viewing and solar geometries, and to repeat the measurements over as long a time span as possible. This is important for constraining Uranus's internal structure, temperature profile, and long-term thermal evolution. 
\end{itemize}

\subsection*{Seismology}
\begin{itemize}
\item 
Seismology, in particular ring seismology, could be used to  determine whether Uranus's internal structure involves compositional gradients, stable regions, and interfaces.
\end{itemize}

\subsection*{Recommendations and Knowledge Gaps}
As was highlighted in the recent OWL Decadal Survey and other mission studies going back decades, in order to dramatically increase our knowledge of the interior structure and composition of Uranus, and to better understand planetary formation and evolution in general, it is necessary to fly a long-lived, Flagship-class orbiter in the Uranus system. The crucial measurements for that mission regarding Uranus's interior are (see Section 7.1 for details):
\begin{itemize}
    \item Radio science measurements, specifically two-way Doppler tracking of the spacecraft to map out the planet's gravity field in detail and (ideally dual-band) radio occultation studies of the atmosphere.
    \item Magnetic field measurements. 
    \item Visible, IR, and microwave remote sensing of the atmosphere.
    \item Occultation studies of the rings at visible, IR, and/or UV wavelengths.
    \item In situ measurements of temperature and composition from an atmospheric entry probe in the 0.5 to 5 bar pressure region.
    \item The magnetic field and Doppler tracking measurements must be made as close to the planet as possible, and sample as wide a range of latitudes as possible.
\end{itemize}

Earth-based work is also critical for understanding Uranus's.  Laboratory and theoretical studies as well as Earth-based observations of Uranus enable us to properly interpret spacecraft measurements and guide what measurements the spacecraft should make and where they should be made. Earth-based observations spanning many decades also are needed to provide seasonal context for the conditions encountered by the spacecraft (each season on Uranus lasting 21 years). The knowledge gaps to address are discussed in more detail in Section 7.2, but can be summarized as:
\begin{itemize}
    \item Determination of material phase boundaries, solubility and mixing properties, viscosity, diffusivity, and the thermal and electrical conductivity of likely components of Uranus under interior conditions.  
    \item Theoretical studies of magneto-hydrodynamics and dynamo generation processes relevant to Uranus, including magnetic coupling between the atmosphere and the dynamo region.
    \item The microwave opacity of H$_2$S under deep-tropospheric conditions and an assessment of the ability of microwave limb-darkening to accurately estimate temperature profiles at depth.
    \item Theoretical studies to understand mechanisms for exciting normal-mode oscillations in giant planets and to quantify the utility of various methods for detecting such oscillations.
    \item Ground-based observations to detect or place upper bounds on the amplitude of seismic oscillations in the giant planets.
    \item Theoretical studies and exoplanet surveys to better understand the statistics of envelope pollution and internal heat release in candidate ice giant planets.
\end{itemize} 

%% file: sections/appendix.tex
\newpage
\appendix
\section{Appendices}
\subsection{Radio Science}
\label{section.radio.science}
Planetary radio science employs radio frequency links between interplanetary spacecraft and ground-based antennas to investigate the interior structures and atmospheres of planets and their satellites. The success of these experiments relies on precise measurements of the Doppler shift in the radio carrier frequencies. The space segment of radio science instrumentation typically consists of deep-space transponders, capable of establishing coherent and highly stable links with the ground, via the large antennas of NASA's Deep Space Network (DSN). Modern radio science experiments operate within specific frequency bands: X-band (7.2--8.4 GHz) and Ka-band (32.0--34.0 GHz), as documented by \cite{asmar2017juno} and \cite{deTiberis2011} for NASA's Juno and ESA's BepiColombo experiments, respectively. 

In the context of gravity science, these Doppler observables provide a means to measure the line-of-sight velocity of the spacecraft. As the spacecraft's trajectory is affected by the gravitational potential of the celestial body it orbits, analyzing the velocity profile enables an accurate determination of the body's gravitational parameters, including spherical harmonic coefficients, pole position and tidal Love numbers. Radio science experiments have different configurations, contingent on the number of links between the spacecraft and the ground segment's antennas. The main configuration employed for gravity science is the two-way mode. Within this mode, a radio signal is transmitted from the deep-space station, received by the spacecraft, and then re-transmitted back coherently to the same ground station. When we refer to coherent modes, it implies that the downlink signal is referenced to the frequency standard of the uplink signal, provided by the ground station. Therefore, the spacecraft is freed from the necessity of providing a highly stable frequency standard, as the ground station serves as the reference for both legs of the experiment.

During radio occultation experiments, the spacecraft vanishes (ingress) and reappears (egress) from behind the planet while maintaining radio communication with Earth's deep-space antennas. Although the signal is temporarily lost while the spacecraft is occulted, the links are maintained as the signal path traverses the upper layers of the planetary ionospheres and atmospheres. In this scenario, the Doppler measurement is linked to the bending angles of the electromagnetic wave as it traverses the planetary medium. When the refractive index differs from unity, the ray's trajectory bends compared to the straight path it would have followed if it were traveling through vacuum. This behavior allows us to retrieve the refractivity profiles as a function of ray path altitude above the body's surface, which in turn can be converted into vertical profiles of pressure and temperature. The retrieval of isobaric surfaces is crucial for determining the shape of the planet as a function of latitude and longitudes (depending on the location of the occultation points). The optimal configuration for radio occultations typically involves a one-way mode, wherein the spacecraft transmits the radio signal which is then received at the ground. In this scenario, the spacecraft needs to be equipped with an Ultra-Stable Oscillator (USO) to provide a frequency standard for the one-way downlink signal. The preference for the one-way, as opposed to the two-way configuration utilized in gravity science, stems from the fact that the signal crosses the atmosphere only once, simplifying the analysis of radio occultation data. However, radio occultations can also be conducted in a two-way mode \citep{buccino2022ganymede,parisi2023europa}. In this case it is important to consider that, during the analysis, one must account for the radio signal traversing the medium on both legs of the link. In cases where both an ionosphere and atmosphere are present, it is preferred to conduct radio occultations in dual-frequency mode (e.g., X- and Ka-band). This approach allows for the decoupling of the effects of charged particles and neutral atmosphere by combining the two radio links. In fact, electrically charged layers exhibit a dispersive behavior, where the phase delay on radio signals depends on the carrier frequency, enabling a more refined analysis. 

Instrumentation noise (space and ground segments) is rarely the dominant source of data fluctuations for two-way experiments, at least for integration times considered in gravity science (30-60 s). They include disturbances linked to the transponders' frequency standard and antenna mechanical noise, with typical Allan deviations of less than 5$\cdot 10^{-15}$ at 1000 s \citep[see Table 1 of][]{asmar2005spacecraft}. The finiteness of the signal-to-noise ratio (SNR) is also a source of thermal noise, however, at 60-sec integration time its effect is limited provided the SNR remains high enough \citep[e.g., for Juno this is around 40 dB-Hz;][]{buccino2023precision}.

The primary noise sources in two-way experiments are fluctuations in the medium. The use of dual-frequency links is not only useful for performing radio occultation experiments (as highlighted in section \ref{section.radio.science}), but also for calibrating other sources of dispersive noise, such as interplanetary plasma and Earth's and Uranus's ionosphere \citep{mariotti2013experimental}. Employing an X uplink, X/Ka downlink configuration allows calibration of up to 50\% of plasma noise. This percentage increases with a shift to X/Ka uplink and downlink (75\%) or a triple link (100\%). When the Sun-Earth-Probe angle is adequately high ($>$120$^{\circ}$) and the spacecraft is far from solar conjunction, the presence of dual- or triple-links, combined with Ka-band's inherent robustness to solar plasma, ensures resistance to this noise source. Often, the main noise source for two-way experiments is tropospheric noise, arising from phase delay as the radio signal traverses the dry and wet components of Earth's neutral atmosphere. \cite{asmar2005spacecraft} indicate that these fluctuations can be as high as 3$\cdot 10^{-14}$ at 1000 s if uncalibrated. 

When conducting radio occultation experiments, additional factors must be considered. Firstly, integration times for data analysis are typically set around 1 second, in order to preserve resolution of the vertical profiles (the ray path velocities are usually high, upward to tens of km~s$^{-1}$). Within these integration periods, thermal fluctuations due to the finite value of the SNR become non-negligible compared to other sources of noise. to effectively mitigate this noise source, maintaining a high SNR is essential, often achieved by employing the spacecraft's High Gain Antenna (HGA) and using large dishes on the ground (e.g., DSN's 70 m antenna). 

Furthermore, since these experiments are often done one-way, as discussed in Section \ref{section.radio.science}, the radio science instrumentation must include an Ultra Stable Oscillator (USO). USOs are highly stable clocks serving as the reference for onboard transponders when the signal is originated onboard. Modern examples of USOs for interplanetary radio science are NASA's Cassini \citep{kliore2004cassini} and ESA's 3GM experiment onboard the JUICE spacecraft \citep{shapira2016ultra}.

Instrument requirements for a Uranus radio science experiment follow from these considerations. The spacecraft's transponder stability must be $< 2\cdot 10^{-15}$ at 1000 s for X- and Ka-band \citep{asmar2005spacecraft}, in order to ensure low instrumental noise. Dual-frequency links (if not triple link) that enable plasma noise cancelation must be available for gravity and occultation investigations. Radio science experiments must be conducted with the HGA pointed towards the Earth. Desired levels of SNR are around 40-42 dB-Hz \citep{buccino2023precision}. However, transmitting high power levels at typical Earth-Uranus distances (19.2 AU on average) poses a challenge, requiring substantial onboard energy generation for transmission. Hence, meticulous budgeting of power consumption during these observations is crucial. 

Earth-based instrumentation is equally important for the success of radio science experiments. Troposphere delays represent the bulk of the noise on Doppler observations. Therefore the use of Advanced Water Vapor Radiometers \citep[AWVRs;][]{buccino2021performance} or Embedded Water Vapor Radiometers \citep[EWVRs;][]{tanner2021embedding} in the proximity of NASA's DSN antennas is crucial for future experiments at Uranus. Currently, two AWVR units are available only at the Goldstone DSN complex, next to DSS-25 (34 m antenna). They have been successfully used by the Juno gravity science experiment. In fact, these calibrations are especially useful to mitigate the effects of the wet, less predictable, component of Earth's troposphere. 

Radio occultations' noise level strictly depends on the frequency stability of the onboard USO. The instrument requirement for this component is for an Allan deviation $< 2\cdot 10^{-13}$ at 1000~s. These values are adapted from the Cassini and JUICE USOs \citep{asmar2005spacecraft,shapira2016ultra}.

Considering these instrument requirements, expected measurement accuracies for two-way Doppler measurements are around 1.2 mHz  at 60 sec for Ka-band, and around 0.3 mHz at 60 sec for X-band, which translate to about 5 $\mathrm{\mu m~s^{-1}}$ for the same integration time in range-rate errors. This is the typical noise level for measurements that have been calibrated with AWVR data \citep{buccino2021performance}. If these are not available at the time of Uranus observations, the noise level would increase by about a factor of 2. For one-way Doppler measurements, the expected accuracy depends mostly on the USO stability, with expected levels around 7 mHz for Ka-band, and 1.7 mHz for X-band at 1 sec integration time.

Astrometric measurements do not require any particular constraints on the camera system. In practice, almost any visible camera with a field of view wide enough to guarantee the presence of stars in the background will do. A field of view of 0.2 by 0.2 degrees up to a few square degrees is typical for this type of measurement. The option of a clear filter is a plus, as astrometry looks for a short exposure time, usually less than 1 second, to avoid stars trailing in the background (especially with a small field of view). The image-taking procedure, meanwhile, requires a few precautions. Fields with a low star population should be avoided. And, if possible, performing a few images in a row for each astrometry session can help to tease out the presence of cosmics in the image close to the expected star positions during field calibration.

\subsubsection{Mission Requirements for Radio Science}

Gravity science measurements generally require proximity of the spacecraft to the body that is being investigated (e.g., Juno pericenters routinely occur at $\sim$1.05--1.07 Jupiter radii). However, each measurement discussed above comes with its own mission requirements, essential for effectively determine key parameters, and often these requirements can clash with each other. 

The accurate resolution of the zonal gravity field requires the spacecraft to sample a wide range of latitudes at close range, in order to capture the axisymmetric signal from even and odd zonal harmonics. A realistic scenario for a Uranus flagship mission is that of Juno-like or Cassini Grand Finale-like orbits (high-eccentricity, high-inclination), where the spacecraft gets very close to the planet (at high velocities) for a limited amount of time and over a limited latitudinal band and spends the majority of the time far away from the planet at apoapsis. In this case, in order to reach measurement requirements for the zonal harmonics, it is necessary to have a sufficient number of closest approaches to the planet with periapsis latitudes as varied as possible. In order to achieve the latter condition, the spacecraft would either have to spend a $\Delta V \sim$100 m/s to change the periapsis latitude by $\sim$1$^{\circ}$, or let the orbit naturally precede due to Uranus's $J_2$. In either case, the periapses are most likely confined to one hemisphere. These requirements work for the estimation of pole position (right ascension and declination) as well. Measuring the rotation rate with Doppler is challenging, unless a static nonzonal gravity field (e.g., from the solid core) is present. Precession can be recovered if the time frame of Doppler observations is long enough (years or decades).

The study of nonzonal features (e.g., atmospheric phenomena or topography of the solid interior) requires the spacecraft to fly as close as possible to the anomaly. In this case, it might be necessary to not only target specific latitudes (as for measuring zonal harmonics), but also specific longitudes \citep[e.g., the case of Jupiter's GRS;][]{parisi2021depth}. This also comes with a required $\Delta V$, which is however smaller than the one required to change periapsis latitude. 

In order to measure Uranus's tidal response to Ariel's perturbations, it is necessary to sample the gravity field of Uranus at close range (i.e., during periapsis) at different spacecraft-Ariel phase angles. In other words, it is necessary to sample the tidal bulge from different viewpoints. This in general requires that the spacecraft orbit is not in resonance with Ariel's orbit around Uranus. 

For gravitational seismology, \cite{friedson2020ice} has demonstrated that the most favorable orbit for detecting normal modes with gravity science is an equatorial, circular orbit. However, it is very challenging from the $\Delta V$ standpoint to achieve such an orbit, especially if the orbital altitude is very low (e.g., 3000~km over Uranus's surface). A possibility could be to extend the radius of the spacecraft orbit to outside of the rings (in the gap at $\sim$ 55,000 km), if normal modes' amplitudes are large, we might be able to detect their effect on the spacecraft orbit also at such distances. Juno and Cassini-like orbits have proved to be not ideal for normal mode detection \citep{Markham2020,durante2022juno} as the period of time that the spacecraft spends in proximity to the planet is comparable to the periods of some of the normal modes, therefore the frequency and azimuthal order cannot be uniquely resolved. 

A key element for all Doppler observations is that the spacecraft orbital plane is as perpendicular as possible to the line-of-sight direction with the DSN stations. This configurations is called edge-on and is favorable because the changes in the line-of-sight velocity are the largest. Also the spacecraft must be in visibility (no occultations) around periapsis. 

Numerical simulations of a Uranus orbital phase that envisions 8--10 periapsis at close range ($\sim$1.04 Uranus's radii) were run. The orbit is nearly polar and the periapsis latitudes are close to the equatorial plane (this requires the spacecraft to fly between the rings and the upper troposphere, which comes with its own risks). We found that $J_2$ and $J_4$ may be resolved with a relative accuracy of a few parts in a million, $J_6$ with a relative accuracy of 0.1\% and $J_8$ of 10\%. $k_2$ can be resolved to an accuracy of $\sim10^{-3}$.

\subsection{Gravity Science}
\label{gravitydetails}
We do not know what Uranus is made of. Based on its mass and radius, it must contain a hydrogen-helium envelope comprising $\sim10 \%$ of its mass, and a core made of some combination of ice and rock \citep{Teanby2020}. Our current knowledge of Uranus's gravity field confirms this basic picture \citep{Neuenschwander2022,Soyeur2023}, but it cannot distinguish between an ice or rock interior, liquid or solid phases of ice, and whether the heavy elements are mixed with or separated from the gaseous surface layers. It also remains unknown whether Uranus's interior consist of gradual compositional transitions with extended stably stratified regions or there are sharp compositional boundaries as expected from terrestrial planets. 
\par 

Interior models of Uranus that relax the assumption that the interior is adiabatic (i.e., fully convective) showed that  Uranus and could have an internal structure without distinct layers, suggesting the existence of composition gradients \citep[e.g.,][]{Helled2011}, and that Uranus may not contain large fractions of water. This implies that Uranus might not be as water rich as typically thought. The possibility that Uranus could be a “rock giant” (instead of an “ice giant”) and that its interior  could consist of large amounts of silicates is now more accepted \citep{Teanby2020,Helled2020b}, and it is clear that determining Uranus's bulk compositions is highly desirable.
\par

The current gravity data allows for extreme end-member models. The available data are insufficient to constrain Uranus's composition: the data can be fitted with Uranus being  entirely ice-free (rock-hydrogen planet) or rock-free \citep[ice-hydrogen planet;][]{Helled2011}. It is also possible that the deep interior consists predominantly of pure ices \citep[icy Uranus;][]{Bethkenhagen2017}. 
It is crucial to address the uncertainty in composition to effectively utilize Uranus as a test bed for competing views on planetary formation: is Uranus more like Pluto, with a 2:1 rock:ice ratio \citep[e.g.,][]{2017Icar..287....2M}, or is its composition similar to a protosolar composition, with a rock:ice ratio of 1:2 \citep[e.g.,][]{Lodde:2003}?


On top of the wide range of possibilities described above that are available for adiabatic models, two additional major factors also contribute to the ambiguity: the internal temperatures and the rotation rate are also very uncertain. The warmer the deep interior, the higher the heavy-element content must be in order to provide sufficient mass density. When the ice mass fraction has reached 100\% and the density is still too low to fit the gravity data, the rock mass fraction must be increased. This may result in an ice-rock interior. In the absence of gas, the maximum rock mass fraction as constrained by current gravity data is about 20\% even if the central temperatures are assumed to remain at typical hot-start values of 20,000 K \citep{nettelmann2016uranus,scheibe2019thermal}. Higher rock mass fractions require the presence of the light elements H and/or He in the deep interior.


The nominal value of Uranus's rotation period is obtained from a period in the radio emission measured at the Voyager flyby \citep{warwick1987radio}. The same measurement at Saturn, however, is off by several minutes from the rotation periods inferred from several other independent methods: pattern speeds from \textit{f}-modes excited vertical resonances in Saturn's rings \citep{2019ApJ...871....1M}, shape data from the Cassini mission \citep{2015Natur.520..202H}, and wind energy and dynamical height minimization \citep{Helled2011}. The latter method applied to Uranus yields a 40 min faster rotation of 16 h 34 m \citep{helled2010uranus}. Such a substantially different rotation rate reflects on the inferred interior mass distribution. Therefore, determining the rotation period of Uranus through different methods including gravity is important for understanding its internal structure. 

\textbf{Gravity Zonal Harmonics}\\
The even zonal gravity coefficients $J_{2n}$ are dominated by the response of the planetary density distribution to rotation. The shape of a fluid planet in hydrostatic equilibrium is symmetric with respect to the equatorial plane and with respect to the rotation axis. Due to these symmetries only even zonal harmonics are nonzero (tesseral harmonics $C_{nm}$, $S_{nm}$ and odd zonal harmonics disappear, if hydrostaitc equilibrium applies). The value of $J_{2n}$ depends primarily on $q^{2n}$ where $q$ is the ratio of centrifugal to direct gravity, assuming rigid rotation. Since $q$ is smaller for Uranus and Neptune than for Jupiter and Saturn, the gravity harmonics are smaller and the determination of interior structure is accordingly more difficult. In addition, the strong winds of these bodies means that the effect of differential rotation can be more important and arise at lower $n$ than for Jupiter and Saturn, complicating the interpretation, although, as discussed below, the winds depth in Uranus and Neptune are expected to be $\sim$ 1000 km. 

In the ice giants Uranus and Neptune, $J_4$ is substantially affected by zonal winds \citep{kaspi2013,Neuenschwander2022}. Tides and normal modes are small effects and mainly affect non-zonal harmonics, as we discuss below. 


The gravitational fields of Jupiter and Saturn were primarily established through Juno and Cassini missions respectively. These posed some similar and some different challenges for interpreting gravity relative to Uranus and Neptune. The similarity is the universal ambiguity of gravity: It cannot give you a density profile but rather limits on the range of acceptable radial profiles. Unlike Uranus and Neptune, where three components (rock, ice, and gas) exist and increase the uncertain interpretation, Jupiter and Saturn benefit from the dominance of hydrogen and helium. Even there, ambiguity remains in Jupiter, for example, because of the persisting uncertainties in the H/He equation of state and the H/He phase diagram. As a result, a wide range of models for Jupiter is proposed \citep{militzer2022juno,miguel2022jupiter,howard2023jupiter}. On the other hand, only a small subset of the Jupiter models satisfies the constraints within their uncertainties \citep{militzer2022juno}, and constructing Jupiter models that allow for enhanced atmospheric abundances is still an open challenge \citep{nettelmann2021theory,howard2023jupiter}.
\par 

\textbf{Non-Zonal Gravity}\\
 As mentioned in Section \ref{sec:gravity_and_winds}, the gravity field of Uranus can have nonzonal gravity in several possible ways: (1) Dynamic structure in the atmosphere \cite[an example on Jupiter is the Great Red Spot [GRS][]{parisi2021depth}, (2) Deep-seated convective structure in a fluid (varying on timescale longer than a mission, perhaps, and therefore effectively ``static"), (3) Deep-seated convective structure in a solid (analogous to the well-established non-zonal gravity observed for Earth and Mars and changing only on a very long timescale), (4) Tidal distortions (time-variable in a frame of reference defined by the deep interior but with well known temporal and spatial patterns), and (5) Normal modes, time variable at a timescale of tens of minutes or so, potentially comparable to the time a spacecraft is near close approach but with unknown phase relationship or spatial pattern from encounter to encounter for an orbital geometry like that used for Juno or Cassini. This could allow for gravitational seismology. Each of these possibilities for generating nonzonal components to the gravity field is discussed below. 
 
 We do not expect near-surface features to dominate the nonzonal gravity field. This is because absolute density variations (the cause of gravity variations) due to dynamics (e.g., convection, differential rotation, or vorticity) are likely to be smaller in the atmosphere than those deeper down. Moreover, they are expected to be confined to a thin layer relative to the radius of the planet. Nonetheless, the Juno mission successfully detected the GRS gravity field, and should Uranus have a similar structure, it would be detectable as well. Juno constrained the depth of the GRS by combining gravity science and microwave radiometer observations during 3 close flybys of the vortex, between 2017 and 2019 \citep{parisi2021depth,bolton2021microwave}.

 \begin{figure}
    \centering
    \includegraphics[width = 14cm]{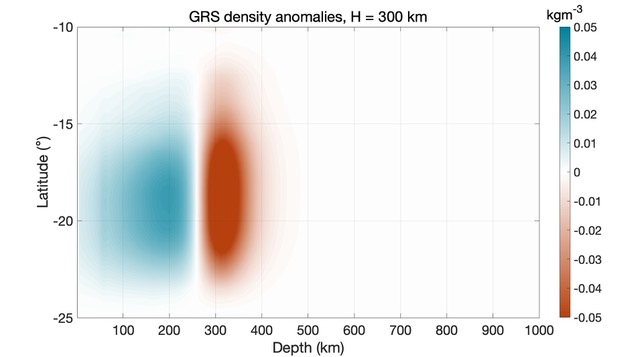}
    \caption{Expected density anomalies produced by the turbulent flow of Jupiter’s Great Red Spot. The scale height for the vortex is 300 km. Figure modified from  \cite{parisi2021depth}.}
    \label{fig:parisi_grs}
\end{figure}

 On Uranus, the presence of long-lasting features such as Jupiter's Great Red Spot is much less certain. Southern polar circulation was observed by the Voyager 2 mission \citep{karkoschka2015uranus}, while the Hubble Space Telescope discovered the first Uranus Dark Spot in 2006 \citep{hammel2009dark}, followed by possible further sightings in Keck images \citep{sromovsky2009uranus}. Furthermore, recent VLA observations have put forward evidence for a polar cyclone in the northern hemisphere as well, that extends down to at least a few tens of bars \citep{akins2023evidence}. Uranus's winds are considerably more equatorially symmetric than Jupiter's, a condition that might explain Uranus's more subdued atmospheric dynamics when comparing the two planets. However, the asymmetries increase toward the poles, as shown by \cite{sromovsky2015high} and \cite{karkoschka2015uranus}, which might favor the presence of polar cyclones. 
 
 The mechanisms for forming polar cyclones at all four giant planets have been investigated by \cite{brueshaber2019dynamical}, who identified the Burger number, related to the ratio between the Rossby deformation radius and the mean planetary radius, as a key factor in the formation of different dynamical regimes for the giant planets. Whether these features could be probed with gravity measurements depends on their depth and mission constraints. Assuming that the cyclonic (or anticyclonic) winds decay at some depth (the wind decay could be influenced by the surrounding zonal winds, or by interaction with the conductive layer), they would most likely be characterized by a vertical dipole of concentrated masses, with the decay of the winds' strength and increase in the hydrostatic density balancing each other out. Because there is a balance between the positive and negative density anomalies, the resulting gravity signal is rather small even for relatively deep features \citep[as shown by][]{parisi2021depth}. 

It is unknown whether Uranus has a deep, solid region. It is more likely if there is an adiabatic structure 
in comparison to a structure that has compositional 
gradients, since the temperature gradient can be much higher in that case. Candidates for a solid region in Uranus include superionic water (which has a solid lattice of oxygen atoms with interstitial, migrating protons) and rock. A similar situation to terrestrial planets (albeit at greater depth relative to the atmosphere) cannot be excluded. In terrestrial planets, where the mantles are solid and have high effective viscosities, the density anomalies associated with convection can be orders of magnitude larger than those in a fluid because one may have to carry a comparable heat flow but at a much lower vertical velocity, and heat flow is proportional to the correlation of vertical velocity with temperature (hence density) anomalies. It is not possible to predict with confidence these anomalies, even if we knew there is a solid region, since the viscosity is highly uncertain and might be quite low in the case of dirty superionic water, for example. 

 If this source of gravity is present, it would be at low harmonic degree because of the radial attenuation of high harmonics. With sufficient data of many nonzonal harmonics, one could separate atmospheric from deep-seated anomalies by considering the power spectrum. Cassini and Juno benefited from having apparently little power in these harmonics since that meant one could get a good measure of the zonal harmonics without a large number of periapses.


\subsection{Seismology}

\subsubsection*{Types of planetary oscillation modes}
\label{section.seismology.oscillation.modes}

\begin{figure}
\centering
  \includegraphics[width=.8\linewidth]{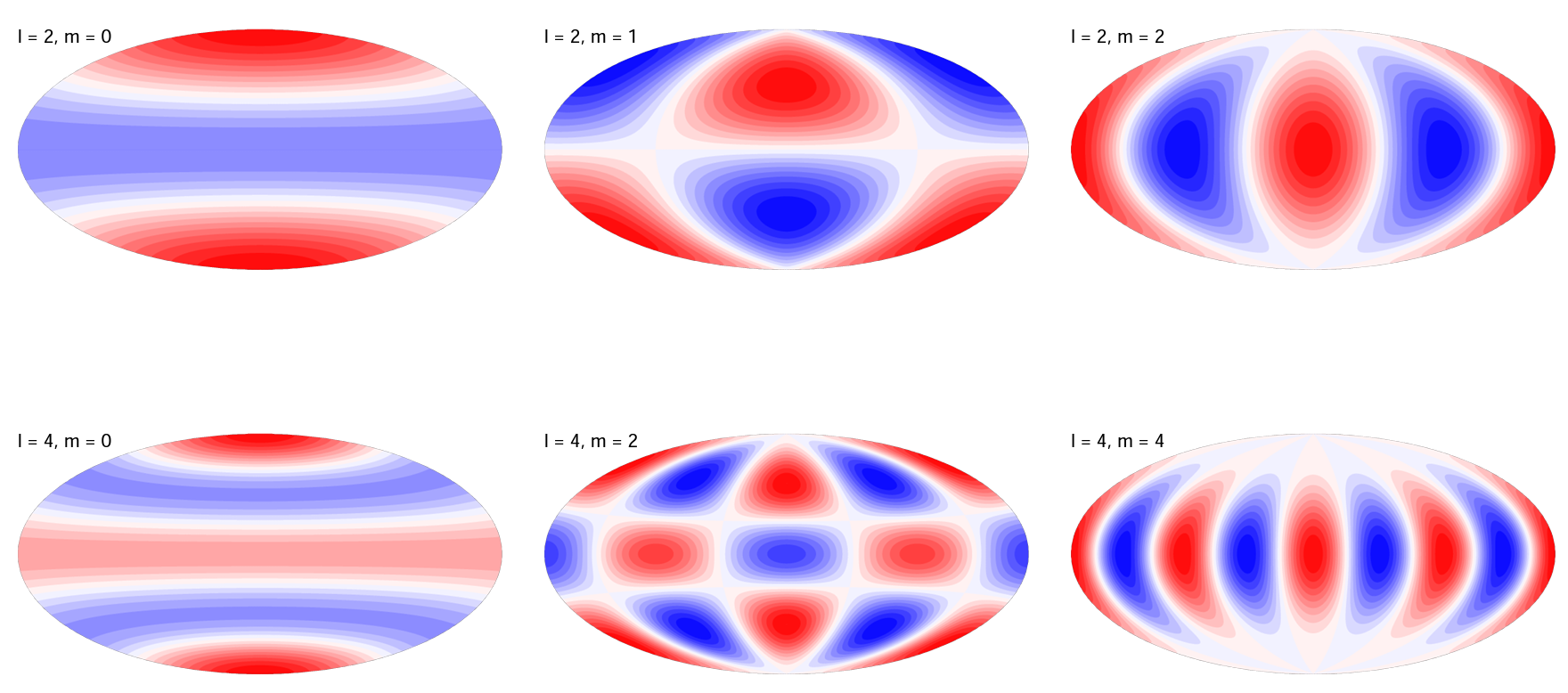}
  \caption{Angular patterns of global oscillation modes of angular degree $\ell=2$ (top row) and $\ell=4$ (bottom row). Fluctuations in fluid displacement, radial velocity, and gravity produced by oscillation modes are all characterized by these types of angular patterns.}
  \label{fig:Ylm}
\end{figure}

The detection techniques mentioned above are sensitive to normal modes, that is, global scale planetary oscillation modes that have a discrete set of temporal frequencies and spatial structures. Each normal mode has a unique frequency $\omega$ whose value depends on the planetary structure. For a spherical planet, each mode has an integer angular pattern number $\ell$ and azimuthal pattern number $m$ corresponding to a spherical harmonic $Y_{\ell m}$ pattern of the distortion produced by the oscillation mode (see Figure \ref{fig:Ylm}). 

\begin{figure}
\centering
  \includegraphics[width=.7\linewidth]{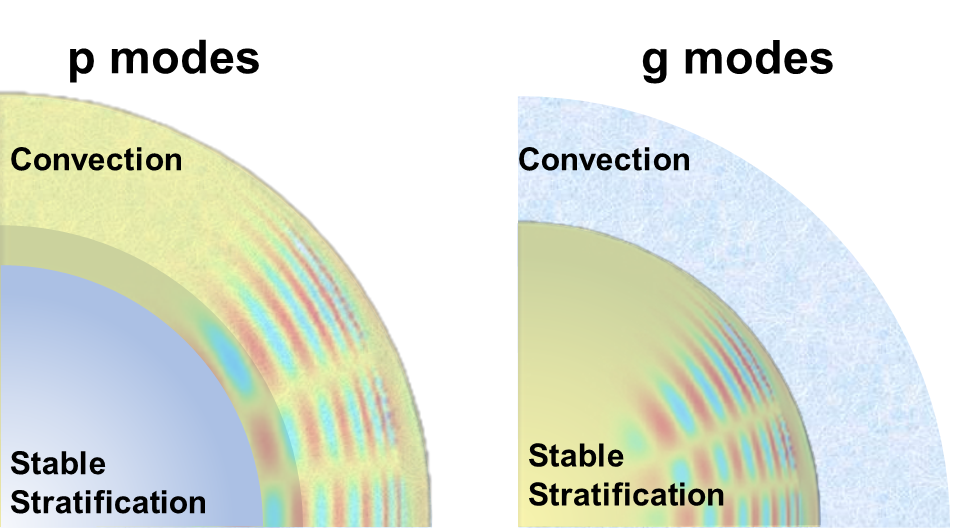}
  \caption{A cartoon showing a cross section of a planet with a convective shell overlying a stably stratified (non-adiabatic) core. The planet's \textit{g} modes are confined to the stably stratified layers. The \textit{p} modes are confined to the surface layers, but can penetrate into stably stratified regions.}
  \label{fig:Pulsation}
\end{figure}

Normal modes are often labeled as fundamental modes (f modes), pressure mode (\textit{p} modes), or gravity modes (\textit{g} modes). The \textit{f} modes have frequencies comparable the planet's dynamical frequency, $\sqrt{G M/R^3}$, and are primarily sensitive to the planet's density structure. The \textit{p} modes are essentially high-frequency standing sound waves that propagate through the planet, and are primarily sensitive to the planet's sound speed profile. The g modes are low-frequency waves restored by buoyancy forces and are restricted to stably stratified (i.e., nonconvective) fluid regions of the planets, and they are sensitive to the buoyancy frequency (Brunt-V\"ais\"al\"a frequency) within these regions (see Fig. \ref{fig:Pulsation}). Another type of modes are interface modes which propagate at discrete transition layers within the planet (e.g., the Earth's core-mantle boundary), and they are sensitive to the density and sound speed change at these boundaries. Finally, solid regions of a planet support shear modes (\textit{s} modes) that are sensitive to the shear modulus of the solid material. Each of these modes can be affected by planetary rotation, which also introduces other classes of modes such as inertial modes or Rossby modes.

Each detection technique discussed above is only sensitive to some of the planet's oscillation modes. Since \textit{p} modes typically produce large fluid velocities and are more confined to surface layers of the planet, they are most accessible to Doppler imaging techniques. The \textit{f} modes (followed by low-frequency \textit{p} modes or high-frequency \textit{g} modes) produce the largest gravity perturbations and are most accessible to satellite tracking measurements. Ring seismology is most sensitive to \textit{f} modes or low-frequency \textit{g} modes because only these modes have the right frequencies to cause Lindblad resonances with orbiting ring particles.

Seismology typically exploits patterns in the spacing between the mode frequencies that translate to specific aspects of the planetary interior, for example, its density and sound speed profiles. For instance, successive \textit{p} modes of the same $\ell$ are labeled by a radial number $n$, and they are spaced in frequency by $\Delta \omega = \omega_{n+1} - \omega_n \approx \sqrt{G M/R^3}$. Hence measuring $\Delta \omega$ translates to a measure of the bulk planetary density. However, the actual spacing between \textit{p} modes has variations about this mean value, and these deviations carry information about the sound speed profile in the planet. The spacing in oscillation mode period between successive \textit{g} modes of the same $\ell$ is given by $\Delta P = P_{n+1} - P_{n} \approx (2 \pi^2/\sqrt{\ell(\ell+1)}) [\int dr N/r]^{-1},$ where $N$ is the Brunt-V\"ais\"al\"a frequency, and hence this quantity provides information about the size and stratification of non-convective regions of the planet.

Seismology is most powerful if the geometry of each oscillation mode can be measured. For Doppler imaging techniques, the radial velocity variations can be measured on different parts of the observable planetary hemisphere, allowing for mode identification. For ring seismology, the angular pattern number $m$ can be measured, as well as the mode parity (i.e., whether $|\ell-m|$ is even or odd), but the value of $\ell$ must be inferred from modeling or \textit{g} mode period spacings. For detection using gravity fluctuations, mode identification and frequency measurements is more difficult, but could potentially be achieved with enough data and properly designed spacecraft orbits.

\subsubsection{Detection methodologies}
\label{section.seismology.detection}
\subsubsection*{Doppler imaging seismology}
\label{section.seismology.detection.doppler}
A Doppler imager is an instrument that measures minute shifts in the frequency of spectral lines to infer a line-of-sight velocity map of a resolved object. 
\begin{figure}
\centering
  \includegraphics[width=.85\linewidth]{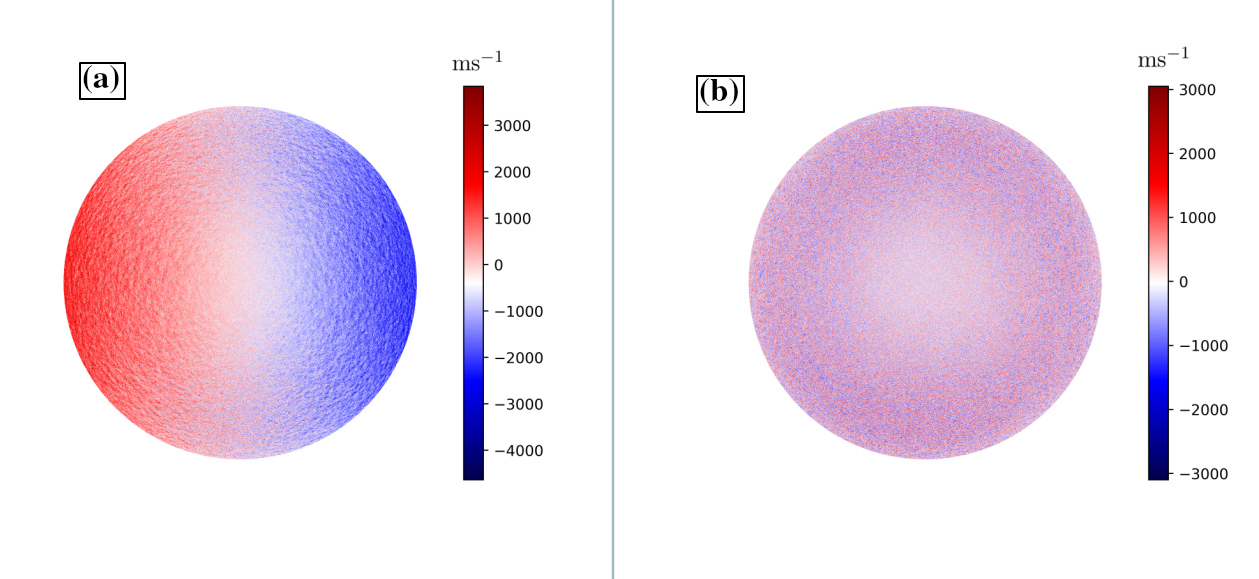}
  \caption{An example image of the velocimetry measurements of the sun by the helioseismic and magnetic imager (HMI) from the Solar Dynamics Observatory.  Figure adapted from \citep{kashyap+hanasoge2021} before (a) and after (b) correcting for rotation and meridional circulation.}
  \label{fig:doppler}
\end{figure}
Various instrumental techniques can be used to achieve the same basic effect. 
This technique has been used with tremendous success to detect \textit{p} modes on the sun. 
The sensitivity of this technology to \textit{p} modes has the potential to constrain the interior properties of Uranus. 
Since \textit{p} modes are sensitive to sound speed profiles, the reflection and transmission of sound waves at compositional interfaces and/or rapid compositional variations would make \textit{p} modes an ideal diagnostic of such features that are crucial to constraining the formation and evolution conditions of Uranus. 
Furthermore, the additional constraint of sound speed as complement to density profiles obtained using gravity field measurements or \textit{f} modes allows for the possibility to break the degeneracy between a continuous subspace of mixtures with identical densities. \\

For a slowly rotating body, global oscillations due to \textit{p} modes can be decomposed according to \citep[e.g.,][]{aerts++2010}
\begin{equation}
\mathbf{\xi} = \sum_{nlm} a_{nlm} \mathbf{\xi}_{nlm}(\mathbf{r}) e^{- i (\omega_{nlm} t + \beta_{nlm})}
\end{equation}
where $\beta_{nlm}$ is the temporal phase of the normal mode and $\mathbf{\xi}_{nlm}(\mathbf{r})$ is the displacement eigenfunction given by 
\begin{equation}
\mathbf{\xi}_{nlm}(r, \theta, \phi) = \left[ \xi_r(r) \mathbf{\hat{r}} + \xi_h(r) \mathbf{\hat{\theta}} \frac{\partial}{\partial \theta} + \frac{\xi_h(r)}{\sin \theta} \mathbf{\hat{\phi}} \frac{\partial}{\partial \phi} \right] P_l^m(\cos \theta) e^{i m \phi}
\end{equation}
where $\xi_r$ and $\xi_h$ are scalar eigenfunctions for radial and horizontal displacement respectively, and $P_l^m$ are Legendre polynomials. 
Near the surface for sufficiently low $l$ modes, one can demonstrate that $\xi_h(R) / \xi_r(R) \ll 1$. In other words, near the surface the displacement due to \textit{p} modes is predominantly radial. 
To an observer in the body's equatorial plane, the projection of the surface velocity from a set of oscillating modes at a given surface coordinate $\theta, \phi$ is therefore \citep{christensen2002}
\begin{equation}
    v_\perp(\theta, \phi, t) = \sin \theta \cos \phi \sum_{nlm} a_{nlm} \omega_{nlm} P_l^m(\cos\theta) \sin[m\phi - \omega_{nlm} t - \beta_{nlm}]
\end{equation}
If one can isolate the contribution to the signal from a single spherical harmonic, one can reduce the signal into a scalar time series $v_{lm}(t)$. 
This can be done by introducing a weighting function for a given spherical harmonic, which when integrated over the full visible disk, quantifies the contribution from that spherical harmonic. 
Although spherical harmonics and normal mode eigenfunctions are formally orthogonal, in practice the data only samples the projection of the surface displacement along the line of sight, and there is not full simultaneous coverage of the sphere. 
Further discussion can be found in \cite{christensen2002}. 
Nevertheless, one can reduce the data to a scalar time series, upon which Fourier analysis can be performed. 
By doing this, one can obtain a detailed power spectrum with narrow peaks corresponding to the eigenfrequency $\omega_{nlm}$ of a given mode. 
As discussed in the previous section, these eigenfrequencies are diagnostic of the interior sound speed profile, which can place powerful constraints on interior models. 
Such is the methodology and purpose of Doppler imaging seismology. 
\\

\begin{figure*}
\centering
  \includegraphics[width=12cm]{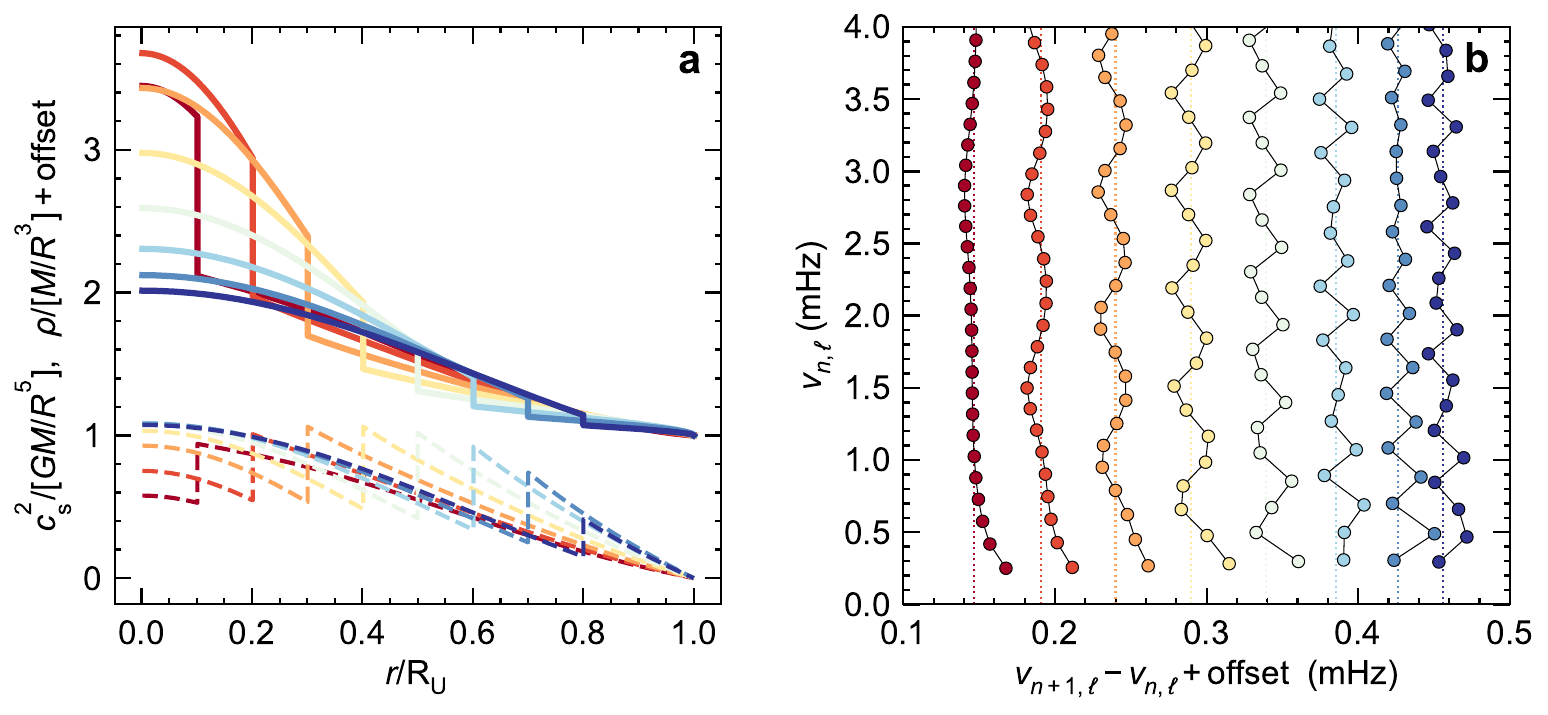}
  \caption{Interior sound speed and density profiles (a) and $\ell=1$, $m=0$ p mode spectra (b) for piecewise polytropic models matching Uranus's mass, radius, and $J_2$. 
  \'Echelle diagrams like panel (b) are diagnostic of abrupt features in Uranus's internal structure. 
  Vertical lines show the asymptotic frequency spacing $\Delta\nu$ (Equation~\ref{eq:delta_nu}) for each model. 
  All models but the red model with an interface at $0.1\,{\rm R}_{\rm U}$ are shifted horizontally by a constant for clarity.
  These and higher order frequency differences have been used in helioseismology to locate the base of the solar convection zone and the second helium ionization zone \citep{christensen2002}.
  }
  \label{fig.uranus_interface_pmodes}
\end{figure*}

As an example, Figure~\ref{fig.uranus_interface_pmodes} shows a family of simple Uranus models with a single internal interface in the density and sound speed. These models assume an inner polytropic index $n_1=1$ and enforce a density enhancement factor of 2 at a chosen radius coordinate; the outer polytropic index $n_2$ is optimized to match $J_2=3,510.68\times10^{-6}$ \citep{2014AJ....148...76J}. In each layer the squared sound speed $c_s^2=\Gamma_1P/\rho$ is chosen such that $\Gamma_1=1+1/n$, representing an adiabatic stratification. The normal mode frequencies $\nu_{n,\ell}=\omega_{n,\ell}/2\pi$ are calculated using GYRE \citep{2013MNRAS.435.3406T} neglecting the effects of rotation, and including internal jump conditions to preserve continuity of the radial displacement and the Lagrangian pressure perturbation. The presence of such an interface can be deduced from an \'echelle diagram (Figure~\ref{fig.uranus_interface_pmodes}b), where the frequency differences $\nu_{n+1,\ell}-\nu_{n,\ell}$ deviate from the asymptotic large frequency separation 
\begin{equation}\label{eq:delta_nu}
    \Delta\nu=\left(2\int_0^Rc_{\rm s}^{-1}dr\right)^{-1}
\end{equation}
with a periodic modulation whose period corresponds to the acoustic radius of the interface (e.g., \citealt{2001MNRAS.322...85R}).\\

Various instrumental techniques have been used to obtain the velocity maps needed to conduct the analysis described above. 
A Fourier tachometer \citep{beckers+brown2013} uses interferometry centered on a narrow spectral line to detect small phase changes to infer minute shifts in light frequency to exquisite precision. 
These frequency shifts then relate linearly to velocity at the non-relativistic speeds relevant to seismology. 
Two examples of interferometry techniques include a modified Michelson interferometer \citep{scherrer++1995} and a Solid Polarizing Interferometer \citep{evans1981}, both of which have been used extensively for observations of the sun. 
Attempts have also been made to conduct seismology experiments on Jupiter using Doppler imaging techniques. 
Claimed observations of Jovian oscillations on the order of 50 cm/s by the SYMPA instrument \citep{Gaulme2011}, which used a modified Michelson interferometer technique. 
The design of SYMPA has informed the design of next-generation instruments JIVE \citep{soulat++2014} and the JOVIAL \citep{gonccalves++2016} global network, which has since been used to reproduce Jupiter's zonal wind profile \citep{gonccalves++2019}. 
More recently this network has been used to detect vertical and meridional winds \citep{schmider++2024}. 
Another Doppler imaging technique using a megneto-optical filter called PMODE has likewise had success in reproducing Jupiter's zonal wind profile \citep{Shaw2022}, but did not reproduce the \cite{Gaulme2011} seismic spectrum. 
The application of Doppler imaging technology to Jupiter and Saturn is currently impeded by under-investment. 
Greater investment into instrumentation, telescope time, and funded analysis of retrieved data is necessary to clarify the seismic activity of Jupiter and Saturn, and to constrain excitation models \citep[see][]{markham-stevenson2018, wu-lithwick2019} necessary to motivate and calibrate a Doppler seismology experiment on Uranus. \\

A Doppler imager on board an orbiter has been studied in the context of the JUICE mission. That instrument, named Echoes, was ultimately not selected \citep{soulat++2011}. The mass of Echoes was estimated to be about 14 kg.

Preliminary studies of instrumental capabilities and requirements at Uranus have also been carried out \citep{schmider++2023}, finding plausible photon noise levels as low as 1--2 mm/s for realistic mission architectures assuming continuous observations of Uranus near apoapse. 
This analysis was conducted using an instrument design similar to Echoes for JUICE with a 3.2 cm telescope. 
Further reductions in mass may be possible if the instrument can be combined with other instruments, for example a narrow angle camera.

\subsubsection*{Ring seismology}
\label{section.seismology.detection.rings}
Planetary rings serve as systems of test particles whose orbits are sensitive to gravitational forcing. At locations where periodic forcing is commensurate with some characteristic frequency of the ring orbits themselves, the resonance leads to collective motion that we can hope to observe. This is the underpinning of ring seismology, wherein waves excited at Lindblad and vertical resonances are used as a means of measuring oscillation frequencies $\sigma_{\ell m n}$ and azimuthal pattern numbers $m$ for the planetary normal modes that are responsible.

Nonradial ($\ell>0$) planetary oscillation modes can be detected through observations of spiral bending waves excited near Lindblad resonances, where the frequency of azimuthal forcing felt by orbiting ring material is similar to the local epicyclic frequency: 
\begin{equation}\label{eq.lindblad_resonance_condition}
|m|(\Omega-\Omega_{\rm p})=\pm\kappa.
\end{equation}
Here $r$ is radial location in the ring plane, $\Omega$ is the ring mean motion, the ``pattern speed'' $\Omega_{\rm p}=\sigma_{\ell m n}/m$ is the rotation rate of the $m$-fold symmetric perturbing pattern in an inertial frame, $\kappa$ is the epicyclic frequency, and the plus (minus) corresponds to the condition for an outer (inner) Lindblad resonance. A different class of wave known as a bending wave can meanwhile be excited near vertical resonances, where the frequency of vertical forcing felt by a ring orbit is similar to the local vertical frequency. These resonances obey a condition identical to Equation~\ref{eq.lindblad_resonance_condition} but with the ring vertical frequency $\mu=\sqrt{2\Omega^2-\kappa^2}$ in place of $\kappa$. The frequencies $\kappa$ and $\mu$ are functions of position and the planet's gravity field and their difference from $\Omega$ is of order $J_2$. Symmetry properties of the spherical harmonics imply that only perturbations proportional to $Y_\ell^m$ with even $\ell-|m|$ excite density waves (Lindblad resonances) because they generate purely horizontal forcing, and likewise harmonics with odd $\ell-|m|$ excite only bending waves (vertical resonances) thanks to their purely out-of-plane forcing.

Classification of an observed normal mode in the rings is helped by the nature of the observation: imaging or occultations reveal not only the location of the wave in the ring plane, but also the number of spiral arms in the resulting wave pattern, corresponding simply to the $m$ value of the perturbing planet mode. Furthermore, the resonance locations naturally separate as a function of $m$ and $\ell-|m|$ thanks to rotational splitting of the planet mode frequency (e.g., \citealt{aerts++2010}).

Saturn ring seismology has proven a valuable means of accessing new information about Saturn's interior structure, thanks to wave detections using stellar occultations observed by Cassini's Visual and Infrared Mapping Spectrometer. Spurred on by wave patterns seen in Voyager radio occultations \citep{1991Icar...93...25R} and later in Cassini UVIS \citep{2011Icar..216..292B}, as well as specific predictions by \cite{Marley1991} and \cite{MarleyPorco1993} that \textit{f} mode resonances should be observable in the C ring, Hedman and colleagues \citep{Hedman2013,Hedman-Nicholson-2014,French2016,French2019KronoIII,Hedman2019,2021Icar..37014660F} have measured more than 30 density and bending waves in the C ring apparently excited by oscillation modes in Saturn. Of these waves, 28 are at outer Lindblad resonances (OLRs), 6 at outer vertical resonances (OVRs), and 2 at inner Lindblad resonances (ILRs). (This is not counting the additional $m=-3$ waves with pattern speeds close to Saturn's rotation, some of which seem to change frequency over time \citep{2022PSJ.....3...61H} and some of which have been interpreted as low-frequency inertial modes of Saturn \citep{2023Icar..40515711F}.) Each of the two detected ILRs is a direct counterpart to a known OLR with the same pattern speed \citep{2021Icar..37014660F}, meaning in each of these cases a single Saturn mode is responsible for two distinct spiral density waves, observed thousands of kilometers apart.

The vast majority of the detections correspond to prograde ($m<0$) f modes in Saturn, but additional modes (not present in the \textit{f} mode spectra anticipated by Marley and Porco) were observed at $m=-2$ and $m=-3$ \citep{Hedman2013}. \cite{Fuller2014} proposed that these modes arise naturally if Saturn's outer core region is stably stratified, yielding a spectrum of g modes supported by buoyancy. Further work interpreting these and newer low $m$ constraints found that the $m=-2$ spectrum suggests a stably stratification supported by a rock/ice gradient extending to at least $0.6\, R_{\rm S}$ \citep{MankovichFuller2021}. The full set of modes detected in the rings also provide new constraints on Saturn's bulk and differential rotation \citep{2019ApJ...871....1M,2021PSJ.....2..198D,2023PSJ.....4...59M}.

Uranus's highly structured ring system is fundamentally different from the extensive, relatively uniform rings of Saturn. Rather than an extended sheet of orbiting ring material that is sensitive to a continuum of forcing frequencies, Uranus is encircled by a series of very narrow rings, their widths typically of order 10 km or less. Hence, a methodology exactly in the vein of Saturn ring seismology is not feasible, barring a fortuitous detection of a wave embedded within one of these narrow rings.

However, the normal modes might play a role in sculpting the rings themselves, which in the absence of external forcing would tend to spread out as a consequence of collisions between ring particles. Their sharply defined edges might result in some cases from proximity to Lindblad resonances with nonradial Uranian normal modes \citep{Marley1988UranusAbstract}, a process closely analogous to shepherding of rings by satellites \citep{1979Natur.277...97G}. While shepherd satellites are well documented in the Saturn system, at Uranus only the $\epsilon$ ring is known to be confined by this mechanism, in that case by a pair of resonances with the nearby satellites Cordelia and Ophelia \citep{1987AJ.....93..724P}. Putting aside the broad, dusty, and low density $\zeta$, $\lambda$, $\nu$, and $\mu$ rings, the process responsible for confining the narrow 6, 5, 4, $\alpha$, $\beta$, $\eta$, $\gamma$, and $\delta$ rings remains unknown. 

\begin{figure*}
\centering
  \includegraphics[width=10cm]{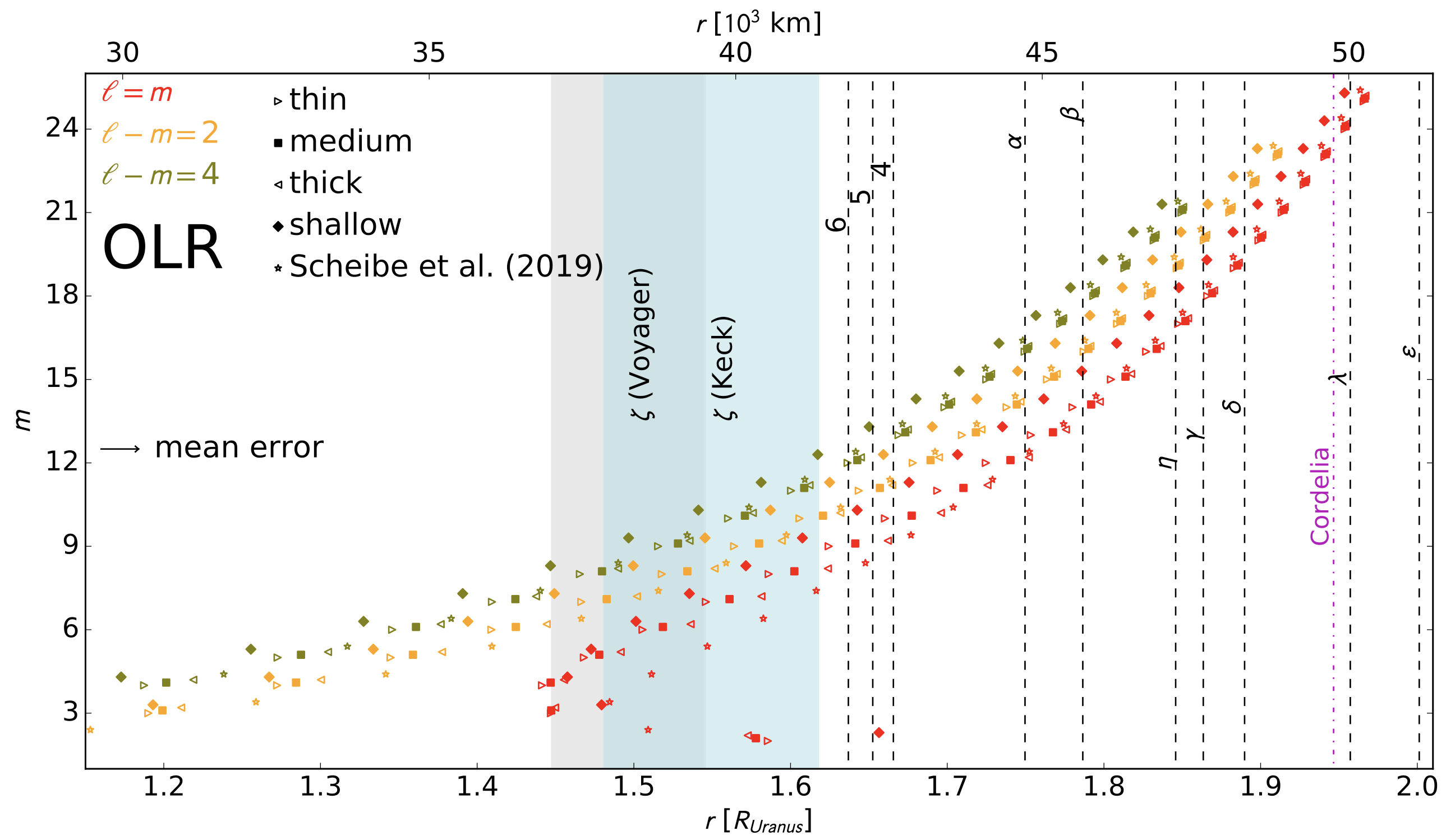}
  \caption{Predictions for outer Lindblad resonances with prograde Uranian \textit{f} modes compared with orbits of the rings. Different point styles correspond to different interior models constrained by Uranus's radius and gravity field. Point colors correspond to different values of $\ell-|m|$ as given in the legend. Note that the sign convention for \textit{m} in the figure is opposite that of the text, so \(m=3\) in the figure corresponds to \(m=-3\) in the text. Reproduced from \cite{2022PSJ.....3..194A}. 
  }
  \label{fig.uranus_ring_olrs}
\end{figure*}

\cite{Marley1988UranusAbstract} suggested that these rings could be confined by Uranian normal modes, showing that the $\ell=2$, $m=-2$ Uranian f mode would have a Lindblad resonance close to the $\delta$ ring, which itself shows evidence of an $m=-2$ perturbation \citep{1979Natur.277...97G}. \cite{2022PSJ.....3..194A} broadly pursued this same premise of Uranian ring seismology, applying forward modeling with modern interior models compatible with Uranus's gravity field to find that many resonances between Uranian normal modes and rings orbits are possible. Figure~\ref{fig.uranus_ring_olrs} shows the their predictions for outer Lindblad resonances among the rings, including f modes spanning $\ell-|m|=0,2,4$. While all models produce an $\ell=2$, $m=-2$ f mode frequency higher than in \cite{Marley1988UranusAbstract}, locating its outer Lindblad resonance well interior of the $\delta$ ring, that resonance does typically fall among the 6, 5, and 4 rings. Similarly, $-13\leq m\leq -7$ f modes of various $\ell$ values produce outer Lindblad resonances throughout the 6, 5, and 4 rings, and $-18\leq m\leq -11$ Uranian f modes fell among the $\alpha$ and $\beta$ rings. The more distant $\eta$, $\gamma$, and $\delta$ rings could resonate with Uranus f modes of still higher azimuthal order $m\lesssim-17$, although A'Hearn et al. argued that those rings' dynamics may be dominated by the influence of Cordelia and Ophelia and evidence of seismicity might be less clear. 

The range of possible resonances in Figure~\ref{fig.uranus_ring_olrs} reflects the huge range of interior structures permitted by current knowledge of the gravity field, as well as Uranus's uncertain rotation. This highlights the diagnostic potential of one or more detections of Uranian modes in the rings. These detections would be most easily made in repeated stellar occultations by the rings, where repeated cuts as a function of orbital phase can reveal the telltale $m$-fold symmetric perturbations to ring edges that result from forcing by an order $m$ planet mode. As an example, if repeated occultations show a clear $m=-2$ perturbation on the outer edge of the 5 ring, and an $m=-10$ perturbation on the outer edge of the 6 ring, the \textit{f} mode spectra from \cite{2022PSJ.....3..194A} would immediately rule out all but one of their interior models. In this case the locations (orbital frequencies) of the outer edges of the 5 ring and 6 ring would provide two new seismic constraints on Uranus's interior structure, precious information complementary to the zonal gravity harmonics expected from the radio science (see Section~\ref{section.radio.science}). 

Note that each individual resonance might be responsible for confining only a single edge of a narrow ring, not both edges. Nonetheless, the edge to edge difference is minor in terms of planet mode frequency: in the limit of a spherical planet, $\kappa\to\Omega$ such that Equation~\ref{eq.lindblad_resonance_condition} reduces to the corotation condition $\Omega_{\rm p}\propto\Omega=\sqrt{GM/r^3}$. Hence, over the typical 10 km scale of a narrow ring, the pattern speed (and hence the planet mode frequency $\sigma_{\ell m n}$) only varies by a fractional amount
\begin{equation}
    \frac{\delta\Omega_{\rm p}}{\Omega_{\rm p}}
    \approx(10\,{\rm km})\left|\frac{d\Omega/dr}{\Omega}\right|
    =(10\ {\rm km})\left(\frac32r^{-1}\right)
    \approx 3\times10^{-4}
\end{equation}
at a representative radius $40\times10^3$ km from Uranus's center (see Figure~\ref{fig.uranus_ring_olrs}). As a point of comparison, the relative contribution of rotation to Uranian mode frequencies is of order $\Omega_{\rm rot}/\sqrt{GM/R^3}\approx0.2$. Useful information about Uranus's rotation and internal structure could therefore be extracted from \emph{any} indication of $m$-fold forcing on a given ring. If occultations can be used to deduce the location of a Lindblad resonance even more precisely, then the frequency constraints available for seismology would be proportionally more precise. 
For comparison, Cassini VIMS stellar occultations were able to pinpoint the locations of Lindblad resonance locations with Saturnian modes to within hundreds of meters \citep[e.g.,][]{2021Icar..37014660F}. 
This technique is described in greater detail in Section~\ref{sec:gravity_and_winds}.

\newpage
\subsection{Material Properties (Equation of State and Transport Properties)}
\label{app:materials}

Our understanding of planet interiors has always involved three essential ingredients: The planetary data (e.g., gravity and magnetic fields), the theoretical framework  (e.g., the equation of hydrostatic equilibrium) and the material properties of the likely constituents (e.g., the pressure-density relationship determined by laboratory measurement, aided by condensed matter theory). For example, we infer the presence of elements other than iron in Earth's core by determining the extent to which the data-driven determination of the density at a known pressure (provided by hydrostatic equilibrium) differs from the laboratory and theoretical data we have for pure iron. 

For Uranus, as discussed above, we would like to know its compositional and thermal structure, the way heat escapes, and the location and means whereby the magnetic field is created. Even if we had a density profile (from gravity and other measurements), we would not know the composition in general, even if we knew the equation of state (EoS) of relevant constituents and thermal structure. This is different from the terrestrial and giant planet experience. The main reason is ambiguity: In the case of a planet such as Uranus, you have three components of roughly comparable importance. You can get a specified density by a range of mixtures of these components; for example a mixture of rock and gas can have similar density (and rather similar compressibility) to pure ice. The problem is less central to an understanding of Jupiter and Saturn because they are dominated by a hydrogen-helium component, and the problem does not exist for terrestrial planets where one has two immiscible components (the mantle and core) each with limited compositional variability. The problem is further compounded by the uncertainty in temperature structure: A large increase in temperature can give the same density as a decrease in ice relative to rock for example. The temperature ambiguity is largely absent in terrestrial planets because of the formation of nearly homogeneous layers, each of which may be almost isentropic.

This ambiguity diminishes the value of knowing precise EoSs (the function $P(\rho,T)$ where $P$ is pressure, $\rho$ is density and $T$ is temperature) but enhances the importance of knowing phase relationships (especially the properties of mixtures),  melting curves and non-EoS issues such as the formation and evolution. The ambiguity can only be diminished or eliminated by consideration of other issues: How was the planet built? How can we create a convecting, electrically conducting fluid layer that is needed to allow generation of magnetic field? How do we explain the heat flow? How can we reconcile interior models with the observable atmospheric composition? This section concentrates on a narrower issue: To what extent does it help to have better knowledge of the EoS of relevant constituents, the phase diagrams and the microscopic transport properties (electrical conductivity, viscosity, and diffusivity)? These considerations point to a knowledge gap that should be at least partly filled through laboratory and theoretical work, preferably prior to a planetary mission.

\subsubsection{The EoS}

A standard approximation is \textit{volume additivity}. This provides an estimate of the density for a known mixture of pure constituents, each of which has a known equation of state. It is desirable to have experiments and theoretical tests of this assumption, although experience to date suggests that it is a good enough approximation in most cases.\footnote{It is easy to think of important possible counterexamples, however. They arise when the end member state is a different phase from the mixture state. An example would be where ``rock" is present as a vapor (or supercritical fluid) when mixed with hydrogen, say, whereas the pure rock phase at the same P and T is a dense fluid or solid.} 

One way to appreciate the experimental or theoretical accuracy required is to ask the following: By how much would I get the wrong mass ratio of rock to ice in a mixture if I got the density of an end member (e.g., ice) wrong by x$\%$?. Since rock is typically two to three times more dense than ice, the answer is also +x$\%$  within a factor of two, assuming volume additivity and comparable amounts of ice and rock. (The plus sign means that an increase in the estimated density of ice would require an increase in the amount of ice in the mixture.) Typically, we are interested in this issue at the level of 10$\%$ (with the caveat that we don't know exactly what ``ice" means given the uncertainty in the C/O ratio and the speciation of carbon). This suggests that the current understanding of end member densities at a given pressure (typically several percent or less) may be adequate. 

\subsubsection{Other Thermodynamic Variables}
 
The EoS is actually a derivative property of the more fundamental Gibbs or Helmholtz free energy. Fortunately, Maxwell provided interrelationships among the relevant variables, reducing the range of possibilities. An important parameter is $\gamma \equiv (\partial \ell nT/\partial \ell n\rho)_S$, often called the Gruneisen parameter. Specific heat is also important though generally less uncertain (simply by counting the available degrees of freedom), though very different for hydrogen than for ice or rock. Compressibility comes from the EoS and the remaining parameter $\alpha$, the coefficient of thermal expansion, can then be obtained from the Maxwell relations (or the appropriate derivatives of the free energy). All of these must be known in order to compute a model of Uranus thermal structure and evolution.  The most challenging (and often most difficult) is $\gamma$, both for end members and for mixtures. It is needed to assess the temperature gradient and the possibility of convection. The need to have these parameters as well as EoS points to the desirability of a free energy model as the starting point, guided by experiment and theory. Of course, the form (phase or speciation) the material takes becomes most important and is considered next.

\subsubsection{The Phase Diagrams}

There are two main issues with the relevant phase diagrams: What is the melting curve (more precisely, the liquidus and solidus) for some assumed mean composition and what are the solubilities of constituents (the extent to which constituents can mix down to the atomic level)? This also requires an understanding of speciation; that is, what form does carbon take, for example. (Candidates include methane, heavier hydrocarbons, carbon atoms in a soup dominated by hydrogen, diamond, oxidized as in CO, or some as yet undiscovered alloy). These two issues of partial solidification and mixing are likely to be interconnected because solid phases are usually lower entropy phases with specific stoichiometries (integer ratios of relevant atoms, as in H$_2$O). By the Gibbs phase rule, a coexisting liquid is likely to have a different composition and may be entropically favorable because it lacks a required stoichiometry. (An example is the iron -sulfur system, which has a low solidus at low pressure because the coexisitng solid lacks the benefit of the mixing entropy arising from sulfur in the coexisting liquid; water-ammonia is conceptually similar.) Solid phases in Uranus may also be less likely because the temperature gradient may be higher than in adiabatic models, but still merit some attention because of our current ignorance.

It is not known whether the three components (gas, ice and rock) mix in all proportions under Uranus conditions. It is likely at deep interior conditions, at sufficiently high temperature, because then the entropic benefit of mixing outweighs any energy benefit from having pure phases. Certainly mixing is not present at sufficiently low temperature and pressure, near or within the atmosphere. In the standard picture, water clouds form below the observable atmosphere but the base of those clouds at still greater depths\footnote{ defined as the place where the vapor pressure of water at that temperature divided by the total pressure equals the mixing ratio of water to hydrogen, roughly} is not known because we don't know the mixing ratio of water to hydrogen (it is likely not the same as the ratio of water to hydrogen averaged over the entire planet nor is it necessarily similar to the atmospheric ratio of methane to hydrogen, despite the similar solar system abundances of C and O).

With these points in mind, below is a partial list of the likely important systems to consider in experiments and theoretical studies, needed to support a Uranus mission:

\textit{1. Superionic ice and its possible alloys or analogues.}

Pure superionic ice is of interest because its high entropy implies (by Clausius-Clapeyron) that it have a large increase of freezing point with pressure, increasing the likelihood that it will be encountered within Uranus. Since the material from which it might form is unlikely to be pure H$_2$O, it is of importance to understand how other elements are incorporated or excluded from this compound, and the extent to which other superionic materials might be present. Its transport properties are also of interest and are considered below. 

\textit{2. Water-hydrogen (more generally H-O)}

Water is slightly ionic at low pressure and increasingly so as pressure increases. Hydrogen remains almost insulating until pressures approaching a Megabar, even for temperatures of a few thousand degrees. Not surprisingly, these two constituents may unmix at quite high temperatures by Uranus standard (as much as a few thousand degrees), provided the pressure is sufficiently high. Some lab data at lower pressure suggest this though more data are needed at multi GPa pressures. As hydrogen becomes more able to give up electrons (not necessarily metallize), a mixture may be preferred, so it is also of importance to understand the likely solubility at very high (Megabar) pressures since we cannot exclude the presence of hydrogen deep down (even though many Uranus models assume so). Different ratios of H to O must also be considered as alternatives to ice and molecular hydrogen.

\textit{3. Carbon and its Compounds}

Carbon is only modestly less abundant than oxygen by mass. Moreover, it is usually very different in its chemical and bonding behavior. At low pressure but high hydrogen partial pressure, methane is the expected main carrier of carbon within the outer regions of Uranus and is present in the atmosphere at a level far higher than the C/H primordial solar value. However, methane has limited solubility in water at depth, raising the possibility that it has partitioned between atmosphere and deep interior. Equally important, methane is known to decompose at high pressure and temperature, forming heavier hydrocarbons and eventually even diamond. The latter phase cannot be excluded as relevant for Uranus at present, but entropy considerations (the benefit of mixing) suggest many other possibilities.

\textit{4. Nitrogen and its Compounds}

Nitrogen is less abundant than oxygen by about a factor of 6, assuming primordial solar system abundances. For this reason it may be less important. Even so, it is of value to understand its disposition. It may also diagnose phase transitions within the planet. For example, nitrogen (as ammonia) is far more soluble than methane in water. If water has a high-pressure (somewhat ionic) phase that separates at depth, then it might partition ammonia downward. This of interest because ammonia is observed (though apparently depleted) in the Uranus atmosphere.

\textit{5. Rock and Ice (e.g., Water and Silicates)}

This is possibly the most important ``binary" system to consider at high pressure. Current evidence suggests that it is indeed an intimate mixture at the relevant T and P deep within Uranus and fails to separate into two components (as it undoubtedly does in Ganymede, for example, where the pressure and temperature are far lower). This system needs to be better understood: What is the liquidus and solidus?

\subsubsection{Microscopic transport Processes}

These are diffusive processes (thermal diffusion, atomic or molecular interdiffusion, viscosity, magnetic diffusivity or equivalently electrical conductivity). They range enormously, though in most cases the required accuracy is far less than for thermodynamic variables. Three specific issues are highlighted here:

\textit{1. Magnetic Diffusivity}

This is likely to be one of the most important since it is needed to understand the magnetic field observations, specifically the ability to operate a dynamo. The requirement of an adequate magnetic Reynolds number typically can be met at quite modest electrical conductivity (i.e., you don't need a metal), especially for fast flow speeds and/or large dynamo domains. Traditionally, Uranus's dynamo has been attributed to a material that is mostly water and has protonic conduction. However, this is not known and a small ``free" electron contribution could easily be of importance if the temperature is high enough. Guided by the phase diagram considerations above, both theoretical and experimental work is needed on the electrical conductivity of candidate Uranus materials.

\textit{2. Thermal Diffusivity}

The measured heat flow from Uranus's interior is sufficiently low that the possibility of conductive heat transport cannot be excluded. As an analogy, the heat flux from Earth's core (which may not be more than a factor of a few smaller than the heat flux deep within Uranus) is thought to be comparable to what is carried by electronic conduction along an adiabatic temperature gradient. Thermal diffusivity is also potentially relevant in any model that involves a large number of boundary layers (e.g., double diffusive convection) since the properties of those layers depends on microscopic heat transport.

\textit{3. Viscosity and Interdiffusion}

Boundary layers (as in double diffusive convection, where there may be many staircase ``steps" or layers) also depends on the value of viscosity and diffusivity of atoms or molecules. 

Viscosity (more precisely, rheology) is also of interest for any solid phase since it affects the tidal response and the ability to convect heat.


\subsection{Magnetic field} 
\label{subsec:mag_theory}

\subsubsection{Dynamo models}

\begin{figure}
    \centering
    \includegraphics[width = 15cm]{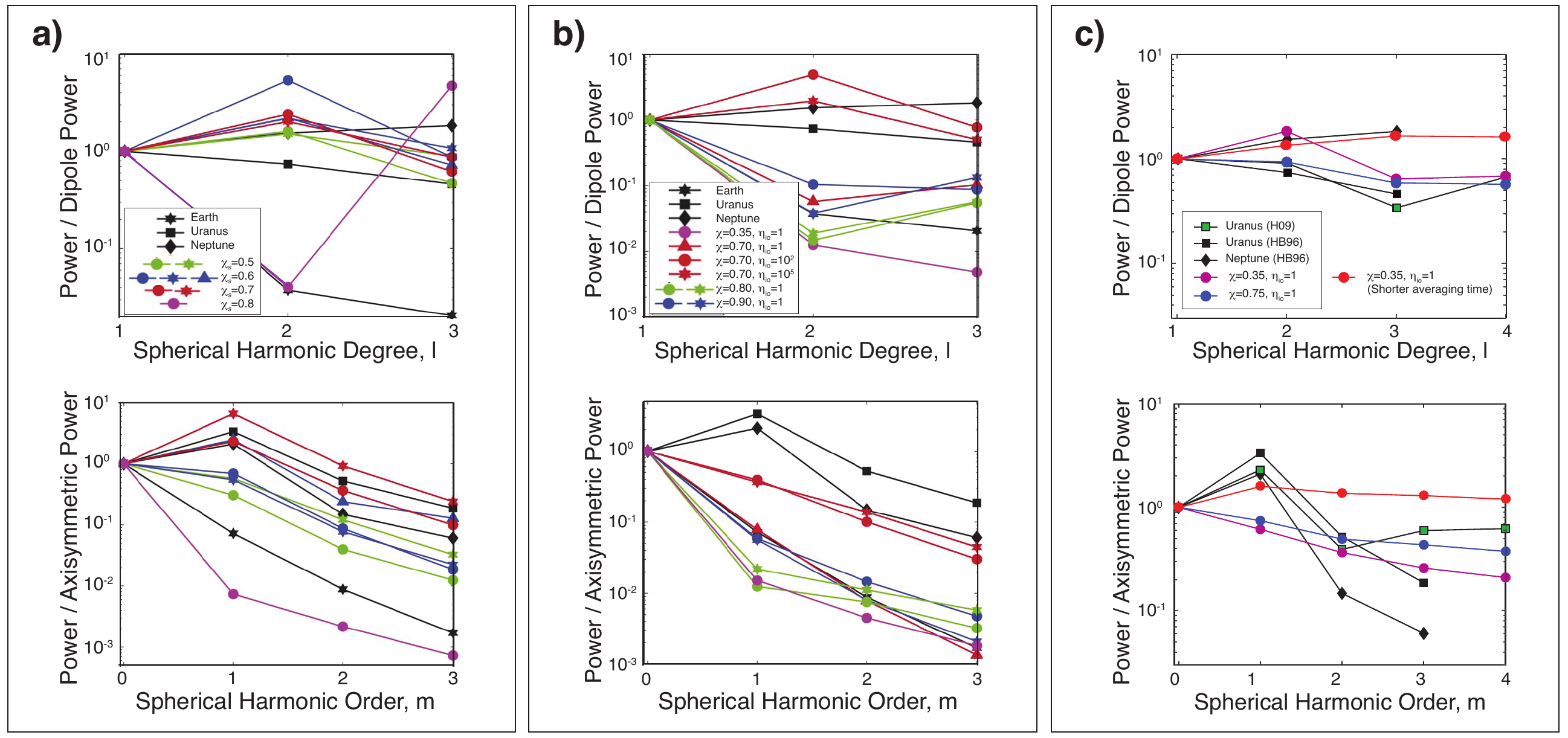}
    \caption{Magnetic power spectra for published dynamo models focused on the magnetic fields of Uranus and Neptune \citep{soderlund2020underexplored}. Dynamo models with solid inner cores of a) different radius ratios $\chi$ and b) different inner/outer core magnetic diffusivity ratios $\eta_{io}$ \citep[adapted from][]{stanley2006}. (c) Dynamo models with strongly forced convection that is moderately constrained by rotation with thick and thin shell geometries \citep[after][]{soderlund2013}. Spectra in (a-b) are averaged in time and upward continued to the surface assuming the top of the dynamo region is located at 0.7 planetary radii. Spectra in (c) are time-averaged and taken at the outer boundary; H09 refers to \cite{Herbert2009UranusAurora}, HB96 refers to \cite{holme}.}
    \label{fig: Uranus_dynamo_models}
\end{figure}

\begin{figure}
    \centering
   \includegraphics[width = 14cm]{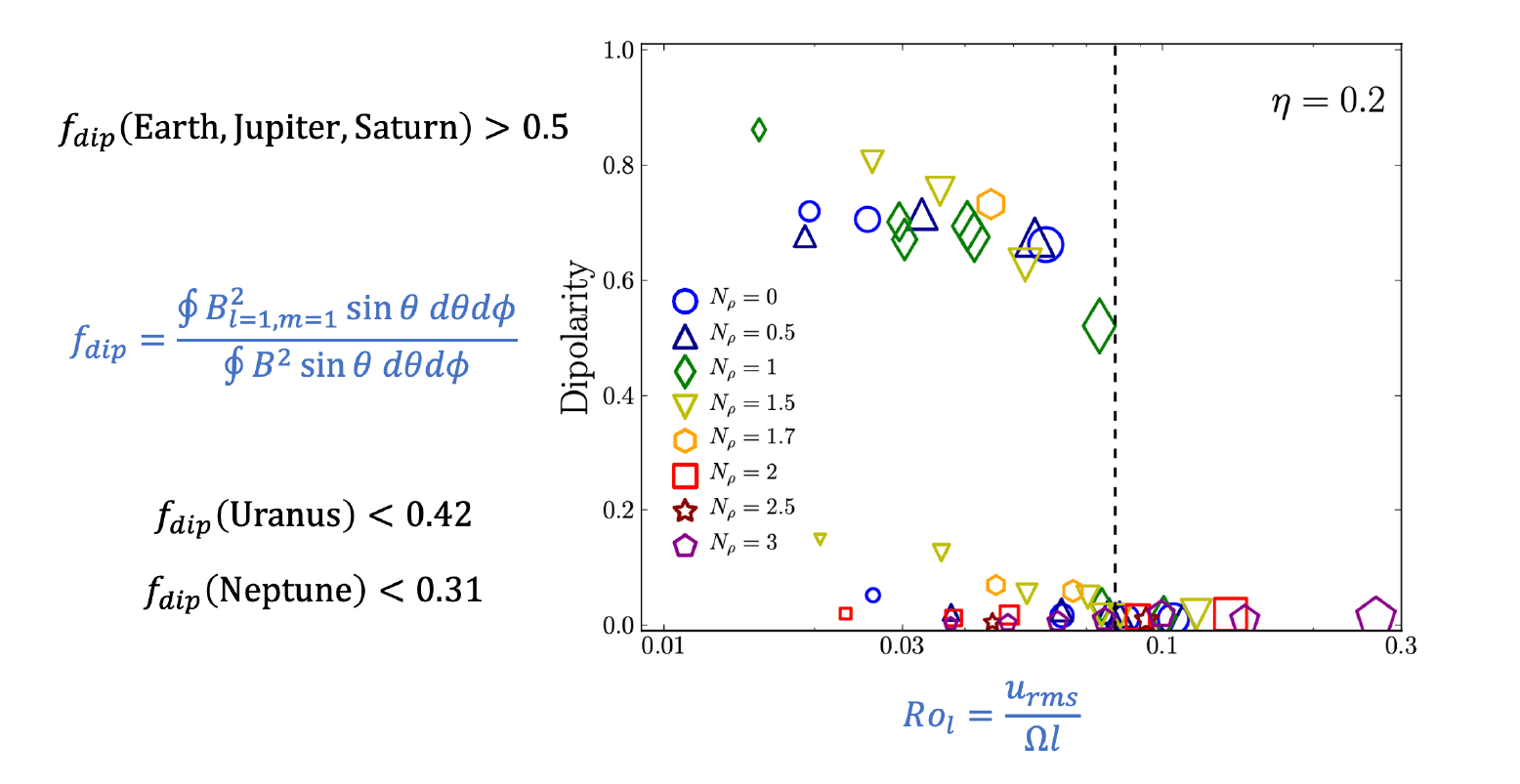}
    \caption{Dipolarity of the dynamo magnetic field as a function of local Rossby number in a suite of 3D numerical dynamo simulations under the anelastic approximation \citep{gastine2012dipolar}. Dipolarity is defined as the ratio of the dipole magnetic energy to the total magnetic energy at the dynamo surface. Local Rossby number, $Ro_l$, is a measure of the influence of background rotation on the convective motion. A small $Ro_l$ indicates that the convective flow is strongly influenced by the background rotation while a large $Ro_l$ indicates that the convective flow is less influenced by the background rotation. The suite of 3D numerical dynamo simulations exhibit two branches: 1) the bi-stable branch for $Ro_l<0.1$, where dipole-dominant and multi-polar solutions co-exist and 2) the multipolar branch for $Ro_l>0.1$ where only multipolar solutions exist.}
    \label{fig: dynamo_dipolarity_Rol}
\end{figure}

Despite being remote from realistic parameter regimes directly applicable to planets, numerical dynamo surveys in the past few decades have significantly advanced our understanding of the MHD dynamo process. Many of these numerical dynamo surveys are generic without efforts to match the detailed background properties of a particular planet. For Uranus, there can be many more complication factors. For example, the radially variable material properties including the electrical conductivity, the existence and properties of stable stratification(s), convective regimes, etc. While it is generally agreed that Uranus’s magnetic field is created by convection and dynamo action in the interior, the mechanisms responsible are not yet understood. As reviewed in \cite{soderlund2020underexplored}, current hypotheses include, but are not limited to: 
\begin{itemize}
    \item {\it Shallow, thin dynamos overlying a fluid region of stable stratification.} \\
    Deep stable layers lead to ice giant-like magnetic fields in all numerical dynamo models of \cite{stanley2006} except for when the convecting layer is very thin ($\chi_s \leq 0.7$ in Fig.~\ref{fig: Uranus_dynamo_models}a).
    \item {\it Shallow, thin dynamos overlying a solid core with lower electrical conductivity.} \\
    Solid inner cores produce multipolar magnetic fields in the numerical dynamo models of \cite{stanley2006} only if the inner core is less electrically conducting than the convecting fluid ($\eta_{io} \geq 10^2$ in Fig.~\ref{fig: Uranus_dynamo_models}b).
    \item {\it Dynamos with convective turbulence that is only moderately constrained by rotation.} \\
    Convection that is characterized by quasi-three-dimensional turbulence, in contrast two quasi-two-dimensional turbulence that is ``vertically stiff" along the direction of the rotation axis, produces ice-giant like magnetic fields for both thick and thin convecting layers in the numerical dynamo models of \cite{soderlund2013} (Fig.~\ref{fig: Uranus_dynamo_models}c).
    \item {\it Bistable dynamos that depend on initial conditions.} \\
    Bistable solutions emerge where both dipolar and multipolar fields are obtained for the same input parameters and boundary conditions depending on the initial conditions in the numerical dynamo models of \cite{gastine2012dipolar} (Fig.~\ref{fig: dynamo_dipolarity_Rol}).
    \item {\it Dynamos with an interplay between electrical conductivity and density stratification.}
    The presence of strong density stratification can inhibit the generation of dipolar magnetic fields due to the concentration of convective features in the lower density region near the outer boundary \citep[e.g.,][]{gastine2012dipolar}. Conversely, if large electrical conductivity variations with radius are present, significant density variations are required for dipolar solutions \citep[e.g.,][]{duarte2013anelastic}.
    \item {\it Double-diffusive dynamos?} \\
    Double-diffusive dynamo models have not yet been evaluated in the context of ice giant planets to our knowledge.
\end{itemize}
  

As illustrated by this sampling of dynamo generation hypotheses, the structure and composition of Uranus's interior are critical to understanding the origin of its highly asymmetric magnetic field. Towards this end, we will next briefly comment on relevant material properties and dynamical end-member models.

\subsubsection{Material properties}

The building blocks of Uranus's interior include chemical constituents such as H$_2$O, CH$_4$, NH$_3$, etc. in the icy layer and MgO, SiO$_2$, FeO, etc. in the rocky layer. Physical properties such as density, electrical and thermal conductivity, and phase equilibrium (phase transformation and melting) of these icy and rocky components are essential parts of any model of Uranus's evolution, dynamics, and structure. 

If Uranus were to separate into distinct water and rock layers, the most abundant component in the icy layer would be H$_2$O. At the top of the icy layer, H$_2$O is likely to be molten, but deeper within the water layer, pressure and temperature conditions could stabilize superionic ices, including both body-centered cubic (BCC) and face-centered cubic (FCC) structures. The exact boundary between these two phases remains controversial.

Due to the relatively high melting temperatures of these superionic ices, the icy layer of Uranus could undergo further separation. It could split into a fluid layer and a frozen layer composed of ice-rich materials. A temperature jump could develop along the boundary between the liquid and solid layers. This thermal boundary layer could play a crucial role in trapping heat in the interior of Uranus, possibly explaining its low luminosity. The size of the frozen core also affects the orbits of Uranus's moons. As the superionic ices melt, their rapidly diffusing protons could contribute to the generation of the strong and complex magnetic fields characteristic of Uranus.

Impurities in the icy layer can also change the nature of the water-rich layer. The addition of NH$_3$ and CH$_4$ could change the stable phase relevant to the water-rich later as well as significantly change the physical properties of the water rich later. For example, superionic NH$_3$ might be almost liquid-like, in contrast to superionic water with finite viscosity. CH$_4$ may significantly depress the melting curve of the water-rich later, and the C-H-O rich fluid would eventually break down to form diamond rain.

A rocky layer consist of silicates may also exist. If we use the simplest analogy, the MgO-SiO system, to present it, for MgO/SiO$_2$ ratios ranging from 1:1 to 2:1 and 1:2, Mg$_2$SiO$_4$ + MgSi$_2$O$_5$, MgSiO$_4$, and MgSi$_2$O$_5$ would be expected at the center of Uranus. One thing to note here is that although properties of the end-member oxides are important, they are only a starting point for understanding solid solution, solubility, and dissociation reactions in plausible phase assemblages that are realistic for the interior of Uranus. For example, a realistic silicate layer of Uranus would have other components such as FeO, but the stable phase or phase assemblages as well as the melting curve of these complex oxide mixtures are unexplored. 

A major question is whether the rock and water would separate into distinct layers, mix into a uniform layers, or mix into a single layer with a compositional gradient inside Uranus. Thermal evidence has shown that the rock and water are likely completely miscible  in the interior if they are liquid, but when they will freeze and what that frozen phase or phase assemblage would look like remains unknown. A gradually mixed rock-water interior for Uranus may prohibit convection and have layered-convection patterns hypothesized for Jupiter with implications for magnetic field generation.






\begin{figure}
    \centering
    \includegraphics[width = 10cm]{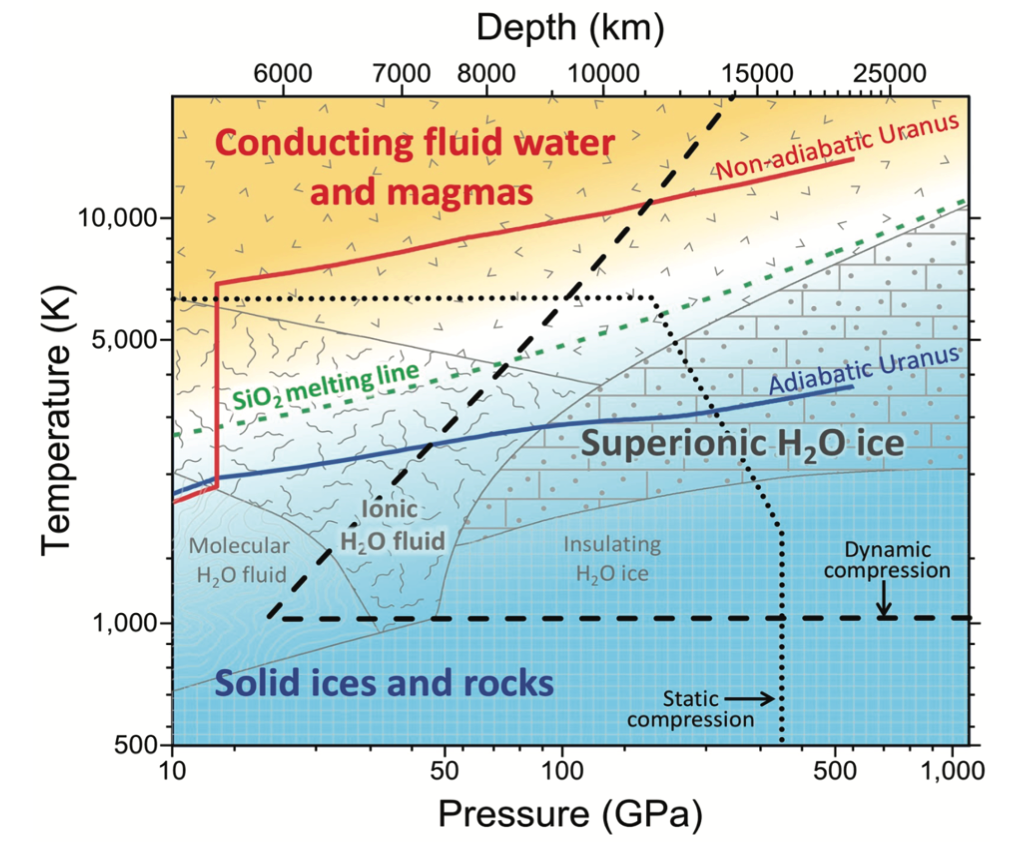}
    \caption{Laboratory experiments and numerical simulations reveal new material behaviors within ice giant interiors. Example profiles of Uranus’ mantle include a hot non-adiabatic model [red line] and a cold adiabatic model [blue line]. The phase diagram of water [gray line] and the melting line of silica [green line], showing the conducting fluid water and magmas and the solid ices and rocks (yellow and blue shaded regions). Range of static and dynamic compression experiments (black dotted and dash lines). Quantum simulations can complement experiments and map material properties over the full range of Pressure-Temperature conditions shown here. From \cite{soderlund2021underexplored}.}
    \label{fig: Uranus_density_profiles}
\end{figure}

\subsubsection{Dynamical end-member models}

\begin{figure}[h]
\centering
\includegraphics[width = 15cm]{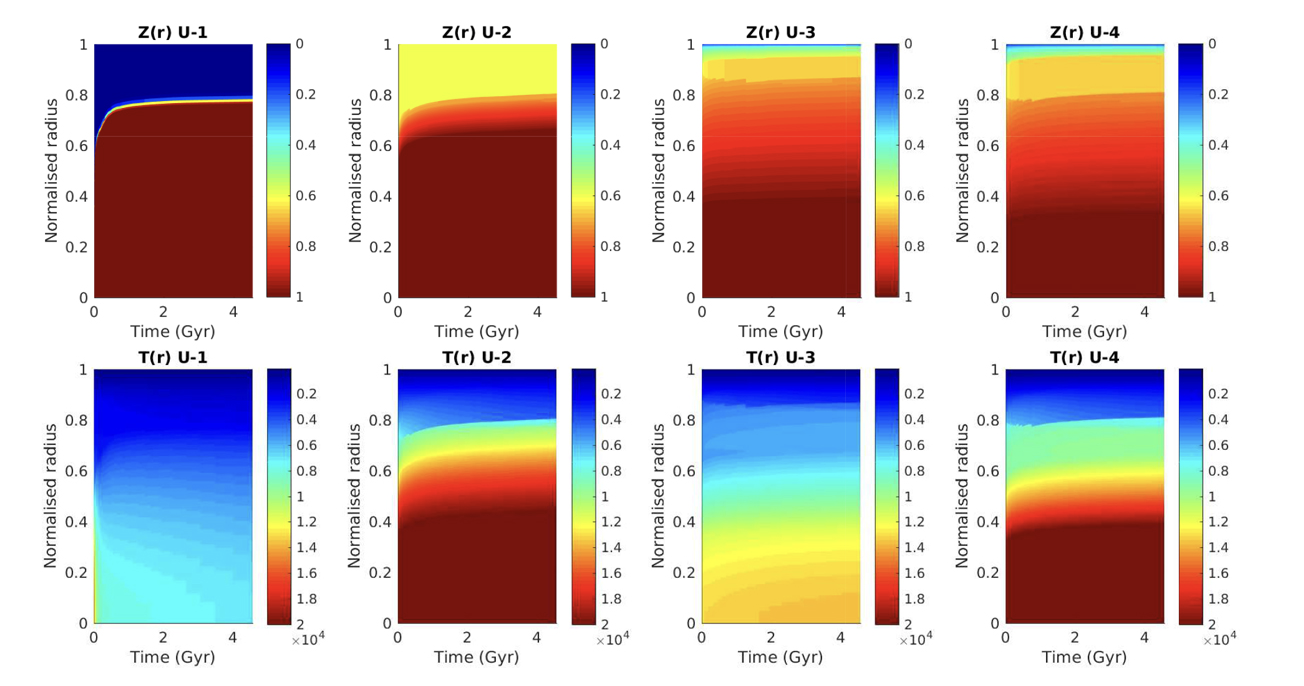}
\caption{Non-adiabatic models of Uranus evolution with time (Gyr) on the x-axis and normalized radius on the y-axis. \textbf{Top row}: the heavy-element mass fraction Z(r). \textbf{Bottom row}: the temperature profile T(r). \textbf{Column 1}: (U-1) two-layer model with a distinct ice-envelope boundary [1:0 ice-rock], \textbf{Column 2}: (U-2) steep gradient model [2:1 ice-rock], \textbf{Column 3}: (U-3) a shallow composition gradient model [2:1 ice-rock], and \textbf{Column 4}: (U-4) rock-rich composition gradient [1:2 ice-rock]. Figure from~\cite{Vazan2020}.}
\label{fig:uranusDynamicModelsAllona2020}
\end{figure}

The uncertainty surrounding Voyager's observations has significant implications for inferring the planetary structure. Dynamic evolution models propose various interior compositions, emphasizing the need for a new mission to break the compositional degeneracy. \cite{Vazan2020} examined evolution models for Uranus and identified four representative compositions for the heavy elements (i.e., rock and/or ice): \textbf{U-1} is all ice, \textbf{U-2} is 2/3 ice and 1/3 rock with a steep compositional gradient, \textbf{U-3} is 2/3 ice and 1/3 rock with a steep compositional gradient, and \textbf{U-4} is 1/3 ice and 2/3 rock (Fig.~\ref{fig:uranusDynamicModelsAllona2020}). These models were found to be consistent with many currently available observations, including that Uranus's magnetic field generation likely extends to relatively shallow depths given the large multipolar contributions. 

\begin{figure}
    \centering
    \includegraphics[width = 14cm]{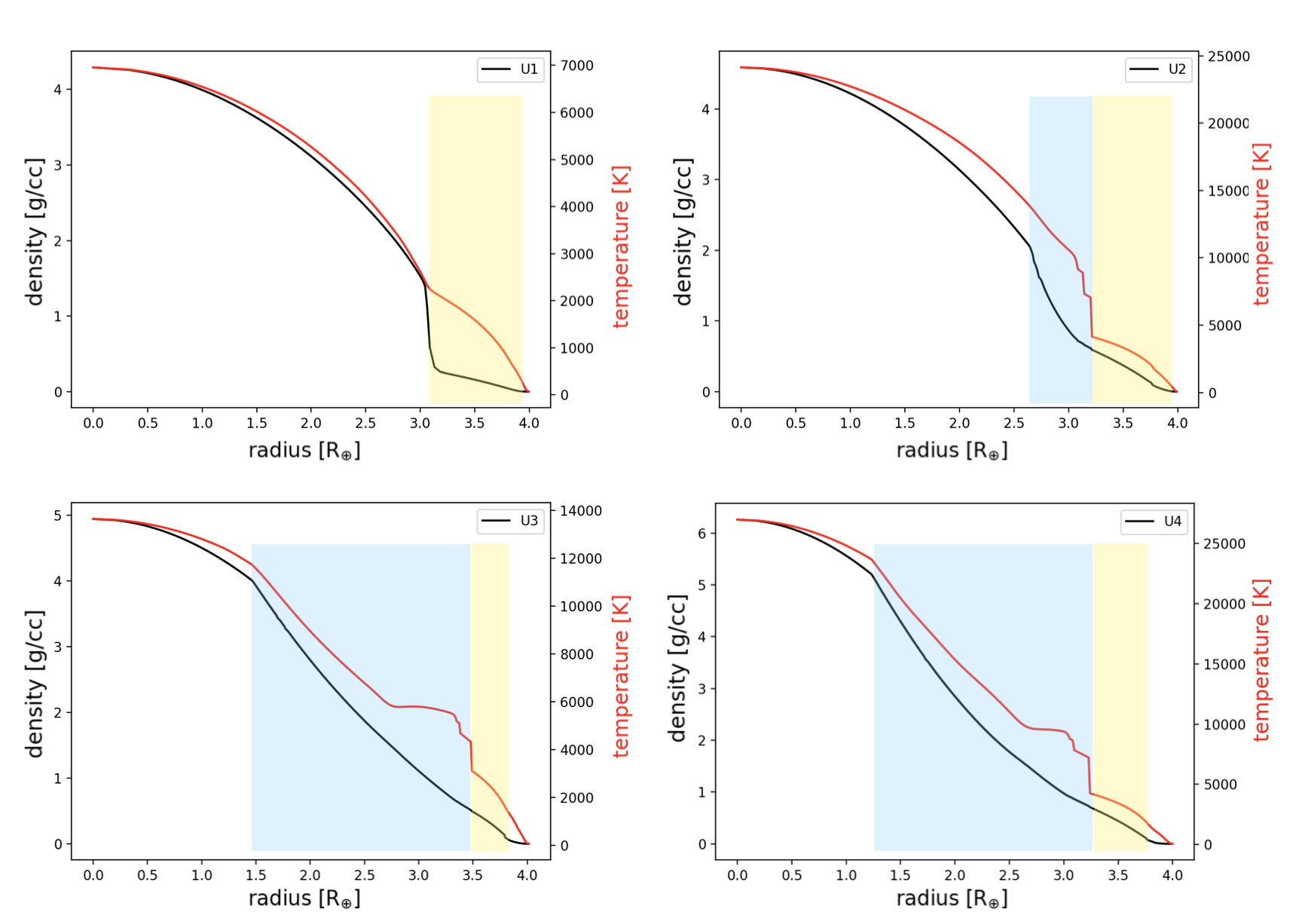}
    \caption{Models of interior profiles of Uranus at current stage, based on models U1--U4 in \cite{vazanhelled20}. Density (black) and temperature (red) are shown. The outer convective shell (yellow) and the stably stratified composition gradient beneath (light blue) are marked on the profiles. 
 }
    \label{fig:ev_models}
\end{figure}

In Figure~\ref{fig:ev_models}, we show density and temperature profiles of Uranus models at present, resulting from thermal evolution models shown in Figure~\ref{fig:uranusDynamicModelsAllona2020}. 
The outer convective shell and the stably stratified composition gradient beneath it are marked in yellow and in light blue, respectively. Model U1 has no heavy element composition gradient, and therefore no stably stratified layers in the interior. Model U2 has an outer convective shell and thin (on the order of 10\% of the planet radius) stably-stratified layer above a convecting deep interior. Models U3 and U4 are similar, but with a much thicker stable layer. These models demonstrate the need to consider the role of double-diffusive convection on Uranus's dynamo. 

